\documentclass[preprint,eqsecnum]{aastexy}

\usepackage{natbib,cases, topcapt, booktabs,longtable,multicol}
\usepackage{fancyhdr}
\fancypagestyle{plain}{ %
\fancyhf{} 
}

\def\degree{\ifmmode {^\circ}\else {$^\circ$}\fi}
\def\mum{\ifmmode {\rm \mu {\rm m}}\else $\rm \mu {\rm m}$\fi}
\def\arcsec{\ifmmode ^{\prime \prime}\else $^{\prime \prime}$\fi}

\def\inch{\ifmmode ^{\prime \prime}\else $^{\prime \prime}$\fi}
\def\arcmin{\ifmmode ^{\prime}\else $^{\prime}$\fi}

\def\mjup{\ifmmode {\rm M_J}\else $\rm M_J$\fi}
\def\rjup{\ifmmode {\rm R_J}\else $\rm R_J$\fi}
\def\mearth{\ifmmode {\rm M_{\oplus}}\else $\rm M_{\oplus}$\fi}
\def\rearth{\ifmmode {\rm R_{\oplus}}\else $\rm R_{\oplus}$\fi}
\def\lstar{\ifmmode {\rm L_{\star}}\else $\rm L_{\star}$\fi}
\def\lsun{\ifmmode {\rm L_{\odot}}\else $\rm L_{\odot}$\fi}
\def\mstar{\ifmmode {\rm M_{\star}}\else $\rm M_{\star}$\fi}
\def\msun{\ifmmode {\rm M_{\odot}}\else $\rm M_{\odot}$\fi}
\def\rstar{\ifmmode {\rm R_{\star}}\else $\rm R_{\star}$\fi}
\def\rsun{\ifmmode {\rm R_{\odot}}\else $\rm R_{\odot}$\fi}
\def\mjupyr{\ifmmode {\rm M_J~yr^{-1}}\else $\rm M_J~yr^{-1}$\fi}
\def\msunyr{\ifmmode {\rm M_{\odot}~yr^{-1}}\else $\rm M_{\odot}~yr^{-1}$\fi}
\def\kms{\ifmmode {\rm km~s^{-1}}\else $\rm km~s^{-1}$\fi}
\def\ms{\ifmmode {\rm m~s^{-1}}\else $\rm m~s^{-1}$\fi}
\def\rhill{\ifmmode R_H\else $R_H$\fi}
\def\vhill{\ifmmode v_H\else $v_H$\fi}
\def\gcm2{g cm$^{-2}$}
\def\2470{[24]--[70]}
\def\sigsb{\sigma_{\rm SB}}

\newcommand{\vc}[1]{\mbox{\boldmath{$#1$}}}

\DeclareMathSymbol{\varOmega}{\mathord}{letters}{"0A}
\DeclareMathSymbol{\varSigma}{\mathord}{letters}{"06}
\DeclareMathSymbol{\varPsi}{\mathord}{letters}{"09}
\def\p{\partial}

\newcommand{\Eq}[1]{eq.\,(\ref{#1})}

\newcommand{\Eqfour}[4]{equations (\ref{#1}), (\ref{#2}), (\ref{#3}) and~(\ref{#4})}

\newcommand{\Fig}[1]{Fig.~\ref{#1}}

\def\cm{{\rm\,cm}}

\def\km{{\rm\,km}}
\def\mm{{\rm\,mm}}
\def\gm{{\rm\,g}}
\def\g{{\rm\,g}}

\def\AU{{\rm\, AU}}
\def\K{{\rm\,K}}  
\def\yr{{\rm\,yr}}
  
\def\mum{\mu{\rm m}}
\def\FeH{\rm [Fe/H]}

\def\degree{\ifmmode {^\circ}\else {$^\circ$}\fi}
\def\Zr{\ifmmode {Z_{\rm rel}}\else {$Z_{\rm rel}$}\fi}
\def\Ri{\ifmmode {{\rm Ri}} \else {${\rm Ri}$}\fi}
\def\Rey{\ifmmode {{\rm Re}}\else {${\rm Re}$}\fi}
\def\ts{t_{\rm s}}
\def\taus{\tau_{\rm s}} 
\def\vk{v_{\rm K}}
\newcommand{\gs}{_{\rm g}}
\newcommand{\ps}{_{\rm p}}

  \renewenvironment{thebibliography}[1]{%
    \begin{oldthebibliography}{#1}%
      \setlength{\parskip}{0ex}%
      \setlength{\itemsep}{0ex}%
  }%
  {%
    \end{oldthebibliography}%
  }

\slugcomment{To appear in {\it Planets, Stars and Stellar Systems},
ed.\ P.\ Kalas \& L.\ French}
\shorttitle{From Disks to Planets}
\shortauthors{Youdin \& Kenyon}

\begin{document}

\title{\LARGE From Disks to Planets}

\author{\Large Andrew N. Youdin and  Scott J. Kenyon}
\affil{Smithsonian Astrophysical Observatory}

\setlength\parskip{0.0ex}%
\tableofcontents
\setlength\parskip{0.5ex}%

\section{INTRODUCTION}
\label{sec:intro}

Theories for the formation and evolution of planets depended 
primarily on geophysical data from the Solar System \citep[see][and references therein]{brush1990}.  
Starting in the 1940's, astrophysical data began to provide new insights. Discoveries of pre-main 
sequence stars in Taurus-Auriga, Orion, and other regions led to the concept that stars form in 
giant clouds of gas and dust \citep[see][and references therein]{kgw2008}.  Because nearly every 
young star has a circumstellar disk with enough mass to make a planetary system, theorists began to 
connect the birth of stars to the birth of planets. Still, the Solar System remained unique until 
the 1990's, when the first discoveries of exoplanets began to test the notion that planetary systems 
are common. With thousands of (candidate) planetary systems known today, we are starting to have 
enough examples to develop a complete theory for the origin of the Earth and other planets.

Here, we consider the physical processes that transform a protostellar disk into a planetary 
system around a single star. Instead of discussing the astrophysical and geophysical data in
detail \citep[e.g.,][]{dauphas2011}, we focus on a basic introduction to the physical steps 
involved in building a planet. To begin, we discuss several observational constraints from the 
wealth of astrophysical and geophysical material in \S\ref{sec:obs}. We then describe the global 
physical properties and evolution of the disk in \S\ref{sec:disk}.  Aside from the special conditions 
required for fragments in the disk to collapse directly into giant planets (\S\ref{sec:planets-bdgg}), 
most planets probably grow from micron-sized dust grains.  Thus, we consider how a turbulent 
sea of grains produces the building blocks of planets, planetesimals (\S\ref{sec:planetesimal}), 
and how ensembles of planetesimals collide and merge into planets which may later accrete a 
gaseous atmosphere (\S\ref{sec:planets}).  We conclude with a brief summary in \S\ref{sec:summary}.

\section{OBSERVATIONAL CONSTRAINTS ON PLANET FORMATION THEORIES}
\label{sec:obs}

We begin with the main observational constraints on planet formation processes, including raw materials, timescales, and outcomes.  
Detailed studies of the Solar System and the disks around the youngest stars yield strict limits on the mass available 
and the time required to make a planetary system. The diverse population of exoplanets illustrate the many 
outcomes of planet formation.

\subsection{Lessons from the Solar System}
\label{sec:obs-ss}

Until the discovery of exoplanets, the Solar System was the only known planetary system.  The Solar System will continue to provide the most detailed data on planet formation, despite obvious issues of statistical significance and anthropic bias.

\subsubsection{The Solar Nebula}
The alignment of major planets in the ecliptic plane suggests that they formed within a flattened disk or ``nebula."   The philosopher Immanuel Kant and the mathematician Pierre-Simon Laplace are often credited for this ``nebular hypothesis."  However, the scientist and theologian Emanuel Swedenbourg first recorded this insight in his 1734 \emph{Principia}.  For a long time the nebular hypothesis competed with the theory, proposed by the naturalist Buffon, that planets were tidally extracted from the Sun during an encounter.    Though Laplace dismissed the encounter theory as being inconsistent with the circular orbits of the planets, it survived to reach peak popularity in the early 20th century as the Chamberlin-Moulton hypothesis \citep{Jef29}.
Once \citet[][1925b]{payne1925a} identified hydrogen as the most abundant element in stars, the nebular hypothesis regained favor. Adding the tidal theory's idea of planetesimals -- small solid particles that condense out of hot gas -- the nebular hypothesis began to develop into a robust theory for planet formation.

The minimum mass solar nebula (MMSN) provides a simple estimate of the mass available in planet forming disks.  The recipe for the MMSN is to distribute the mass currently in the solar system's planets into abutting annuli, adding volatile elements (mainly hydrogen gas) until the composition is Solar.  \citet{kuiper51} and  \citet[see his Table 5]{cameron1962} estimated the mass of the MMSN as a few percent of a solar mass.  \citet{weid1977a} and \citet{hayashi:1981} fit the now canonical $R^{-3/2}$ surface density law,  bravely smoothing the mass deficits in the regions of Mercury, Mars, and the asteroid belt, and the abundance uncertainties for the giant planets.  The roughness of the fit is immaterial: the MMSN is not a precise initial condition, but a convenient fiducial for comparing disk models.  

We use the same MMSN as \citet{chiangyoudin:2010} with disk surface density profiles: 
\begin{eqnarray}
\label{eq_sigmag}
\varSigma\gs = 2200 \,F \left( \frac{R}{\rm AU} \right)^{-3/2} \gm \cm^{-2} \\
\label{eq_sigmap}
\varSigma\ps = 33 \,F\, \Zr \left( \frac{R}{\rm AU} \right)^{-3/2} \gm \cm^{-2} \, ,
\end{eqnarray}
where subscripts ``${\rm g}$" and ``${\rm p}$" respectively denote gas and particles (condensed solids) 
and $R$ is the distance from the Sun.  The parameter $F$ scales the total mass; $F=1$ is a reference MMSN.  
Integrated out to $100 \AU$, the MMSN disk mass is $0.03 M_\odot$.

The parameter $\Zr$ scales the ratio of solids-to-gas, or disk metallicity as
\begin{equation} 
Z_{\rm disk} = \varSigma\ps /\varSigma\gs = 0.015 \Zr\, ,
\end{equation} 
evolves during the planet formation process, as evidenced by the enrichment of heavy elements relative to H in Jupiter and Saturn.  Our fiducial value is normalized to the \citet{lodders2003} analysis of (proto-)Solar abundances, which can be approximately summarized as:
\begin{equation} 
\Zr \simeq \left\{ 
\begin{array}{ccc}
1 &  & T \lesssim 40 \K \\
0.78 & & 40 \K \lesssim T \lesssim 180 \K \\
0.33 & & 180 \K \lesssim T \lesssim  1300 \textrm{ --- } 2000  \K \\
0 & & T \gtrsim 1300  \textrm{ --- }  2000  \K \\
\label{eq:Zrel}
\end{array}
\right.\, .
\end{equation}
The abundance of solids decreases with increasing temperature due to the sublimation of (most significantly) methane ice above $40 \K$, water ice above $180 \K$, and dust over a  range of temperatures from roughly $1300$ --- $2000 \K$, covering the condensation temperatures for different minerals.  Note that this definition of disk metallicity ignores the heavy elements in the gas phase, which are at least temporarily unavailable to produce planetary cores.  For the disk temperature, this work adopts \Eq{eq:irtemp-2}, the result for an irradiated disk with a self-consistently flared surface \citep{chiang:1997}.

In the Solar System today, the temperatures of solids are set by an equilibrium between heating and cooling. 
Although gravitational contraction (Jupiter) and tidal heating (many satellites, including the Moon) contribute
some heating in a few objects, the Sun is the primary source of heating. Radiation from the Sun peaks at a 
wavelength of $\lambda_{max,\odot} \approx$ 0.5~$\mum$. Objects with radius $r$, peak wavelength $\lambda_{max,g} < r$, 
and no atmosphere radiate as nearly perfect blackbodies.  Equating the energy they receive from a central star 
($ \pi r^2 \lstar / 4 \pi R^2$) with the energy they emit ($4 \pi r^2 \sigsb T_{eq,bb}^4$) leads to an 
equilibrium temperature ,
\begin{equation}
T_{eq,bb} = 278 \left ( {\lstar\ \over {\lsun}} \right)^{1/4} \left ( {R \over {\rm AU}} \right )^{-1/2} ~ {\rm K} ~ ,
\label{eq:t-eqbb}
\end{equation}
where \lsun\ is the luminosity of the Sun. Small grains with $r \gtrsim 1~\mum$ and 
$r \lesssim \lambda_{max,g}$ emit radiation inefficiently. In most cases, the radiative efficiency is
\begin{equation}
\epsilon = \left\{ 
\begin{array}{ccc}
1 & & \lambda \le \lambda_0 \\
(\lambda / \lambda_0)^q & & \lambda > \lambda_0 \\
\end{array}
\right.\, ,
\label{eq:rad-eff}
\end{equation}
where $\lambda_0$ is the critical wavelength and $q \approx$ 1--2 depends on grain properties. Usually $\lambda_0$ is roughly equal to the
grain radius $r$.  Because they can only radiate efficiently at short wavelengths, these grains have much larger temperatures. For $q = 1$,
\begin{equation}
T_{eq,s} = 468 \left ( {\lstar\ \over {\lsun}} \right)^{1/5} \left ( {R \over {\rm AU}} \right )^{-2/5} \left ( {\lambda_0 \over {\rm \mu m}} \right )^{-1/5} ~ {\rm K} ~ .
\label{eq:t-eqsm}
\end{equation}
To derive \Eq{eq:t-eqsm}, include the efficiency in the grains' emitted radiation ($\propto 1/\epsilon$) and relate the wavelength of peak emission to the grain temperature with Wein's Law ($\lambda \propto 1/T_{eq,s}$).

Coupled with the condensation temperatures in eq. (\ref{eq:Zrel}), these definitions allows us to identify
the ``snow line."  Also known as the ``frost line," this annulus in the disk\footnote{or, more generally, a spherical 
shell surrounding the central star} which separates an inner region of rocky objects from an outer region 
of icy objects \citep{kk2008}.  For blackbody grains, the water condensation temperature of 180~K implies $R_{snow} \approx$ 
2.7~AU, roughly coincident with the asteroid belt.  Similarly, the methane condensation temperature of 40~K 
yields another region beyond the outer edge of the Kuiper belt at 48~AU where solid objects have a combination 
of water and methane ice.  Inside of $\sim$ 0.1 AU, dust evaporates; rocky grains cannot exist so close to the Sun.

\subsubsection{Isotopic Timescales}

Meteorites delivered to Earth from the asteroid belt provide the most detailed chronology of the early Solar System \citep[e.g.,][]{dauphas2011}.   Primitive meteorites from asteroids that did not undergo differentiation (or other significant alteration) preserve the best record of their formation.  These primitive meteorites are called chondrites because they contain many chondrules.  Chondrules are glassy inclusions, with a typical size $\sim 0.1 - 1 \mm$.  They provide evidence for high temperature melting events in the Solar nebula.  The nature of these melting events is debated and beyond our scope \citep{connolly:messII}.  Calcium Aluminum Inclusions (CAIs) are also present in primitive meteorites.  CAIs experienced even more extreme heating than chondrules.

With ages up to $4567.11 \pm 0.16$ Myr \citep{russ06}, CAIs are the oldest known objects in the Solar System.  This age is consistent with current results for the main sequence age of the Sun \citep{bonn02}.  The absolute ages of CAIs are measured by lead-lead dating, which makes use of half-lives, $t_{1/2}$, of uranium isotopes that are conveniently long.  The decay chain of $^{235}$U $\rightarrow ^{207}$Pb has $t_{1/2} = 0.704$ Gyr, while  $^{238}$U $\rightarrow ^{206}$Pb has $t_{1/2} = 4.47$ Gyr.

Radioisotopes with short half-lives yield accurate relative ages of meteorite components.  These isotopes are ``extinct;" 
they have decayed completely to daughter products whose abundances relative to other isotopes result in an age.  The extinct isotope $^{26}$Al decays to $^{26}$Mg in $t_{1/2} = 0.73$ Myr.  The abundance of $^{26}$Mg relative to $^{27}$Al and to $^{24}$Mg yields an age for the sample. Aside from its use as a chronometer, $^{26}$Al is a powerful heat source in young protoplanets.

Both absolute (lead-lead) and relative ($^{26}$Al) ages support a planet formation timescale of a few Myr.  Most CAIs formed in a narrow window of $1-3 \times 10^5$ yr; chondrule formation persisted for $\sim 4$ Myr or longer.  \citet{russ06} discuss systematic uncertainties.  Here, we emphasize the remarkable agreement of the few Myr formation time derived from primitive meteorite analyses and protoplanetary disk observations (\S\ref{sec:obs-disks}).

The assembly of terrestrial planets from planetesimals requires tens of Myr.  Isotopic analysis of differentiated Solar System bodies (including Earth, Mars and meteorites) probe this longer timescale.   The decay of radioactive hafnium into tungsten, $^{182}{\rm Hf} \rightarrow ^{182}{\rm W}$ with $t_{1/2} = 9.8$ Myr,  dates core-mantle segregation.  Tungsten is a siderophile (prefers associating with metals) while hafnium is a lithophile (prefers the rocky mantle); thus, Hf-W isotope ratios are the primary tool to date differentiation (Chapter by Barlow).  Studies of Hf-W systematics indicate that asteroid accretion continued for $\sim 10$ Myr, Mars' core formed within 20 Myr, and the Earth's core grew over 30--100 Myr \citep{kleine2009}.  Astronomical observations of debris disks (\S\ref{sec:obs-disks}) and dynamical studies of terrestrial planet accretion (\S\ref{sec:planets}) support these longer timescales.

\subsubsection{Water}

After hydrogen, water is the most abundant molecule in disks \citep[e.g.,][]{naj2007}.  Inside the snow line, 
water exists in the gaseous phase, though it dissociates at $T \gtrsim$ 2500 K.  As water vapor interior to the ice line diffuses past the snow line, it condenses into icy grains.  The snow line thus acts as a cold trap, where the enhanced mass in water ice (eq. [\ref{eq:Zrel}]), should accelerate the growth of planetesimals and perhaps gas giant 
planets \citep{stev1988}. 

Water is abundant throughout the Solar System \citep[e.g.,][and references therein]{rivkin2002,jewitt2007}. 
Outside of the Earth, water appears in spectra of comets, Kuiper belt objects, and satellites of giant planets 
(including the Moon)
and bound within minerals on Mars, Europa, and some asteroids. Because they can be analyzed in great detail, 
meteorites from asteroids provide a wealth of information on water in the inner Solar System.  Many meteorites 
show evidence for aqueous alteration prior to falling onto the Earth. Most groups of carbonaceous chondrites 
and some type-3 ordinary chondrites contain hydrated minerals, suggesting association with liquid water.
Despite some evidence that grains might react with water prior to their incorporation into larger solids, most 
analyses of the mineralogy suggest hydration on scales of mm to cm within chondrites and other meteorites 
\citep{zolen1988}.

The water content of Solar System bodies helps trace the evolution of the snow line \citep{kk2008}.  Hydration within meteorite samples demonstrates
that the snow line was at least as close as 2.5--3 AU during the formation of the asteroids \citep{rivkin2002}.
Radiometric analyses suggest hydration dates from 5--10 Myr after the formation of the Sun, close to and perhaps
slightly after the formation of chondrules.
Ice on the Earth and on Mars suggests the possibility that the snow line might 
have been as close as 1 AU to the proto-Sun, a real possibility for passively irradiated disks (\S\ref{sec:irr}).  Within the terrestrial zone, the rise in water abundance from 
Venus (fairly dry) to Earth (wetter) to 
Mars and the asteroids (wetter still) points to processes that either distributed water throughout the inner 
Solar System (with a preference for regions near the snow line) or inhibited accretion of water (either vapor 
or ice) from the local environment. Abundance analyses, including D/H and noble gases, help to probe the (still imperfectly known) history of water in the inner Solar System.

\subsection{Disks Surrounding the Youngest Stars}
\label{sec:obs-disks}

Observations of young stars provide additional constraints on the early evolution of 
planetary systems. In nearby star-forming regions (Orion, Taurus, etc), nearly every star 
with an age of 1 Myr or less has an optically thick circumstellar disk \citep{wil2011}.  
The disk frequency seems independent of stellar mass.  The data suggest the disks are 
geometrically thin, with a vertical extent of roughly 10\% to 20\% of their outer radius.
They are composed of molecular gas and dust grains with sizes ranging from a few microns up 
to several mm.  Around solar-type stars, young disks have typical luminosities of order \lsun\ and 
radii of order 100 AU.

Estimating disk masses is challenging. Aside from a few transitions possible only in warm material 
near the central star, H$_2$ is undetectable. The next most abundant molecule, CO, is optically thick 
and provides a crude lower limit to the total mass. Current estimates rely on a conversion from the 
dust emission at mm wavelengths to a dust mass and then to a mass in gas.  These estimates are highly 
uncertain due to ignorance of dust-to-gas ratios and  grain size distributions.  The typical assumptions give 
disk masses  $\sim$ 0.01~\msun, a factor of 2--4 smaller than the MMSN.  The dispersion for ensembles 
of a few hundred systems is roughly an order of magnitude \citep{and2005,wil2011}.  The size of the typical 
disk is similar to the semimajor axes of orbits in the Kuiper belt.

Current data demonstrate that more massive young stars have more massive disks.  Recent 
observational programs concentrate on whether the ratio of disk mass to stellar mass is 
roughly constant or increases with stellar mass.  Despite the larger scatter in this ratio 
at every stellar mass, the relation is probably linear \citep{wil2011}.

High resolution radio observations reveal interesting limits on the distributions of surface 
density and temperature within the brightest and most massive disks around nearby stars.  
Although disks with a broad range of surface density gradients are observed 
($\varSigma \propto R^{-n}$ with $n \approx $ $-$0.6 to 1.5), most observations indicate a 
typical $n \approx$ 0.5--1.5 \citep{and2005,ise2009,wil2011}. Thus, the surface density 
gradient in the MMSN is steeper than the average protostellar disk but within the range 
observed in disks around other young stars.

The evolution of protostellar disks sets severe limits on the timescale for planet formation. 
Optical and ultraviolet spectra of young stars show that material from the disk flows onto the 
central star. The rate of this flow, the mass accretion rate, drops from well in excess of 
$10^{-8}$ \msunyr\ at 1 Myr to much less than $10^{-11}$ \msunyr\ at 10 Myr \citep{wil2011}.  
Declining mass accretion rates imply much less gas near the young star.  At the same time, 
the fraction of young stars with opaque dusty disks declines from nearly 100\% to less than 1\%. 
Fewer dusty disks implies the solid material has been incorporated into large (km-sized or 
larger) objects, accreted by the central star, or driven out of the system by radiation 
pressure or a stellar wind. Direct constraints on the amount of gas left in 10~Myr old 
systems without opaque disks are limited to a few systems.

Among older stars with ages of 3--10 Myr, many disks have substantial mass beyond 
30--50 AU but have inner holes apparently devoid of much gas or dust.  The frequency of 
these ``transition" disks suggests the evolution opaque disk $\rightarrow$ opaque disk 
with inner hole $\rightarrow$ no opaque disk takes from 0.1--0.3~Myr up to 1--2~Myr 
\citep{CurSic11, EspIng12}.

Once the opaque disk disappears, many pre-main and main sequence stars remain surrounded by 
1--10 $\mu$m dust grains \citep{wya2008}.  This material lies in a belt with radial extent 
$\delta R \approx 0.1 - 0.5 R$ and vertical height $\delta z \approx 0.05 - 0.1 R$, with 
$R \approx$ 1--100 AU. Infrared spectroscopy suggests the grains have compositions similar 
to material in the comets and the dust (Zodiacal Light) of the Solar System. The total mass, 
a few lunar masses, exceeds the mass of dust in the inner solar system by factors of 100--1000.
These properties are independent of stellar metallicity and many other properties of the
central star. However, the frequency of and the amount of dusty material in these dusty disks 
peaks for stars with ages of 10--20 Myr and then declines approximately inversely with time
\citep{currie2008}. 

These disks place interesting limits on the reservoir of large objects around stars with
ages of 10--20 Myr. Among several possible grain removal mechanisms, the most likely are
radiation processes and collisions \citep{back1993}. Grains orbiting the star feel a headwind
from the incoming radiation from the central star, which causes the grain to spiral into 
the star \citep{burns1979}. If the grains have a mass density $\rho_\bullet$, the orbital 
decay time for this Poynting-Robertson drag is
\begin{equation}
t_{pr} = \left ( { 4 \pi r \rho_\bullet \over 3} \right ) \left ( {c^2 R^2 \over \lstar} \right ) = 710 ~ \rho_\bullet ~ \left ( {r \over \mum} \right ) \left ( {R \over \AU} \right )^2 ~ \left ( {\lstar \over \lsun} \right )^{-1} ~ \yr ~ .
\label{eq:tpr}
\end{equation}
The decay time, $t_{pr} \sim$ 1 Myr, is much shorter than the 100 Myr to 10 Gyr main sequence 
lifetime of the central star. The large frequency of dusty disks among 50--500 Myr (1--10 Gyr) 
old A-type (G-type) stars suggests the grains are continually replenished over the main sequence 
lifetime. By analogy with the Solar System, where trails of dust result from collisions of asteroids 
\citep{nesv2003}, high velocity collisions among large (but undetectable) objects can replenish the dust. 
Adopting typical sizes for asteroids, $\sim$ 1--10 km, a reservoir containing $\sim$ 10 \mearth\ of 
material can explain the amount of dust around young stars and the time evolution of the dust emission 
among older stars.  Because this mass is between the initial mass of solids in protostellar disks ($\sim$ 
100--1000~\mearth) and the dust mass in the Solar System ($\lesssim 10^{-4}$ \mearth), these systems 
are often called ``debris disks" \citep[][also Chapter by Moro-Martin]{wya2008}.

\subsection{The Exoplanet Revolution}
\label{sec:exo}

The current pace of exoplanet discovery is extraordinarily rapid.  Here, we highlight the main 
insights exoplanets bring to theories of planet formation.

Within the discovery space accessible with current techniques, exoplanets fill nearly all available phase 
space \citep[e.g.,][and references therein]{cum2008,gould2010,how2011,john2011}.  Among $\sim$ 10\% to 30\% 
of middle-age solar-type stars, exoplanets lie within a few stellar radii from the central star out to several AU.    Because there are many multiplanet systems, the sample of short period planets implies more planets than stars \citep{you11b}.
Though detection becomes more difficult, the frequency of exoplanets seems to grow with increasing distance from the parent star.
The orbits have a broad range of eccentricities, with a peak at $e \sim$ 0.2. Planet masses range from a
rough upper limit at 10--20 \mjup\ to a few Earth masses. Around stars with the same mass, lower mass 
planets are more frequent than higher mass planets.  More massive stars tend to have more 
massive planets.

There is some evidence that exoplanets are more likely around metal-rich stars \citep{gonzalez1997,john2010}.
With the large samples available, this ``planet-metallicity correlation" is now unambiguous for gas giants with
masses ranging from the mass of Saturn up to $\sim$ 10 \mjup. Among lower mass planets, small samples currently prevent 
identifying a clear correlation. Larger samples with the {\it Kepler} satellite will yield a better test of 
the planet-metallicity correlation as a function of planet mass.

The origin of any planet-metallicity correlation establishes some constraints on formation theories.
If the initial metallicity of the disk is identical to the current metallicity of the star, then gas giants --
and perhaps other planets -- form more frequently in more metal-rich disks. However, planets could
pollute the stellar atmosphere after the rest of the gaseous disk disperses, raising the metallicity
of the star above the initial metallicity of the disk.  In this case enhanced metallicity would be a result 
not a cause of planet formation.    Current data contradicts the pollution hypothesis \citep{fv05, pasquini2007}, 
but more study of the diverse exoplanet population is warranted.

To quantify the planet-metallicity correlation, \citet{john2010} fit the frequency, $f$, of giant planets 
as a joint power-law in stellar mass, $M_\star$, and metallicity, $Z_{\rm Fe} \propto \log_{10}(\FeH)$, 
\begin{equation} 
f \propto M_\star^\alpha Z_{\rm Fe}^\beta ~ .
\label{eq:fPMC}
\end{equation}
For giant planets in the California Planet Survey, $\alpha = 1.0 \pm 0.2$ and $\beta = 1.2 \pm 0.2$.
Ignoring the dependence with stellar mass ($\alpha \equiv 0$) introduces a bias, but yields a stronger relation with metallicity,
($\beta = 1.7 \pm 0.3$).  For more massive stars with $M_\star > 1.4 M_\odot$, $\alpha = 1.5 \pm 0.4$ and 
$\beta = 0.73 \pm 0.35$. Thus, the formation of giant planets around more massive stars is more sensitive
to stellar mass and less sensitive to metallicity than for lower mass stars.

With exoplanet samples growing so rapidly, new analyses will change at least some of these conclusions.
The most firm conclusions -- that exoplanetary systems are common and have nearly as much diversity as 
possible -- provides a good counterpoint to the wealth of detail available from {\it in situ} analyses 
of the Solar System.

\section{DISK PROPERTIES AND EVOLUTION}
\label{sec:disk}

Stars form within collapsing clouds of gas and dust.  When a cloud collapses, most infalling material 
has too much angular momentum to fall directly onto the nascent protostar.  This gas forms a 
rotationally supported circumstellar disk \citep{cas1981, tsc1984}. If the angular momentum in
the disk is transported radially outward, gas can accrete onto the central star.
Although jets launched near the protostar \citep{shu2000} or from the disk \citep{pudritz2007} 
or both can remove significant angular momentum from the disk, most analyses 
concentrate on how angular momentum flows through the disk.

At early times, the disk mass $M_{\rm disk}$ is similar to the stellar mass $M_\star$.  
For a circumstellar disk with surface density $\varSigma$ and radial flow velocity $v_R$, 
the rate material flows through the disk as a function of 
radial coordinate $R$ is
\begin{equation} 
\dot{M} = -2 \pi R v_R \varSigma ~ .
\label{eq:mdot}
\end{equation} 
Positive $\dot{M}$ corresponds to gas flowing towards, and eventually draining onto, the central star.  If the mass infall rate from the surrounding molecular cloud $\dot{M}_i$ exceeds $\dot{M}$, the 
disk mass grows.   If $M_{\rm disk}$ exceeds $\sim 0.3 M_\star$, 
gravitational instabilities within the disk can produce a binary companion \citep{ars1989, KraMat10}.  
Smaller instabilities may form brown dwarfs or giant planets (\S\ref{sec:planets-bdgg}).  At 
late times, several processes -- including the clearing action of protostellar jets and 
winds \citep{sal1987} -- stop infall.  Without a source of new material, the disk mass 
gradually declines with time.

In general, all of the physical variables characterizing the cloud and the disk change with 
radius and time.  However, $\dot{M}_i$, $\dot{M}$, and $\varSigma$ often change slowly enough
that it is useful to construct steady disk models with a constant $\dot{M}$ throughout the 
disk. Here, we develop the basic equations governing the evolution of the disk and use
steady disks to show the general features of all circumstellar disks.

An evolving gaseous disk sets the physical conditions in which small particles grow into
planets. Physical conditions within the disk limit how planetesimals can form 
(\S\ref{sec:planetesimal}) and how solid planets grow out of planetesimals (\S\ref{sec:gaseffects}).  
Once solid planets form, the gaseous disk provides the mass reservoir for giant planet 
atmospheres (\S\ref{sec:atmos}) and drives planet migration (see chapter by Morbidelli).

\subsection{Basic Disk Dynamics}
\label{sec:disk-evol}
To introduce basic concepts in disk dynamics we describe orbital motion in a gas disk and the 
radial flow induced by viscosity.  The orbital velocity of the gas $v_{\phi}$, is set by the 
balance of radial forces -- centrifugal, pressure and gravitational\footnote{The radial speeds  associated with accretion produce negligible advective accelerations, $D v_R/Dt \sim v_R^2/R$.} -- as  
\begin{equation}
- \frac{v_{\phi}^2}{R} + \rho^{-1} \frac{\partial P}{\partial R} + {G M_\star \over R^2} = 0 ~ ,
\label{eq:disk-radmom}
\end{equation}
where $\rho$ is the gas density and $P$ is the gas pressure.   Away from the immediate vicinity of 
protoplanets, gravity from the central star typically dominates.  Even a self-gravitating thin disk has 
$M_{\rm disk} \ll M_\star$.  For $P = P_0 (R/R_0)^{-n}$ (the index $n$ need only be locally valid), the orbital motion is
\begin{equation}
v_{\phi} = v_K \left ( 1 - {n c_s^2 \over \gamma v_K^2} \right )^{1/2} ~ ,
\label{eq:disk-gasvel}
\end{equation}
in terms of  the Kepler velocity, $v_K = \sqrt{G M_\star/R}$, and the sound speed, $c_s = \sqrt{ \gamma P/\rho}$, 
with $\gamma$ the adiabatic index.  Plausible disk models have an outwardly decreasing pressure ($n > 0$) 
almost everywhere.  Thus pressure support typically gives sub-Keplerian rotation, but the correction is 
quite small, $v_\phi - v_K \sim - 10^{-3} v_K$.   It is often safe to ignore the pressure correction to orbital motions, 
but not when studying the motion of solids relative to gas \citep[\S\ref{sec:radialdrift}]{weid1977b, houches10}.

\begin{figure}
\vskip 5ex
\includegraphics[width=6.0in]{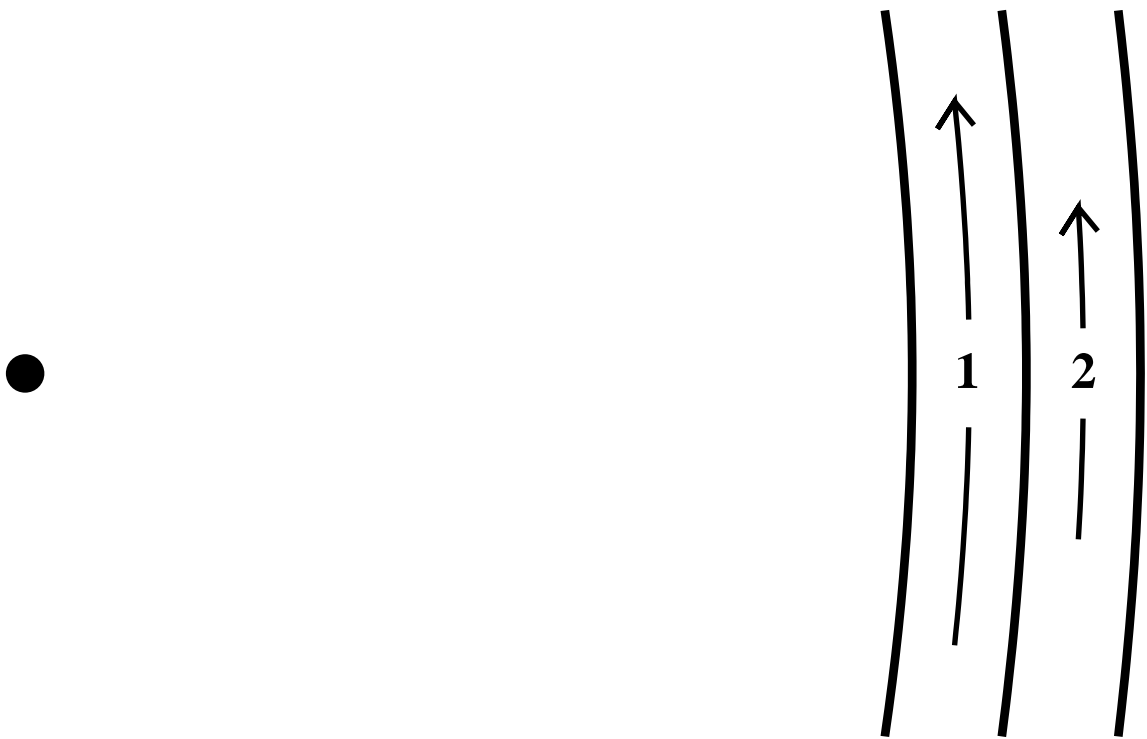}
\vskip 5ex
\caption{
Schematic view of two adjacent annuli in a disk surrounding a star (black point at left).
Annulus 1 lies inside annulus 2; material in annulus 1 orbits the star more rapidly than
material in annulus 2 ($\Omega_1 > \Omega_2$).
\label{fig: disk1}
}
\end{figure}

Accretion disks also have radial inflow, which is constrained by the laws of mass, angular momentum and energy
conservation.  Indeed accretion disks can be considered as machines that radiate energy as they transport mass inwards and 
angular momentum outwards.  The viscous disk model offers the simplest means to understand how a disk 
manages this feat.  Consider a thin ring with two adjacent annuli at 
distances $R_1$ and $R_2$ from a star with mass \mstar\ (Fig. \ref{fig: disk1}). Material
orbits the star with angular velocities, $\Omega_1 = \sqrt{G M_{\star} / R_1^3}$ and 
$\Omega_2 = \sqrt{G M_{\star} / R_2^3}$. If the gas has viscosity, the differential 
rotation, $\Omega_1 - \Omega_2 < 0$ for $R_2 > R_1$, produces a frictional
(shear) force that attempts to equalize the two angular velocities.  The resulting torques
produce an outward flow of angular momentum.   By moving a small amount of disk material onto 
distant, high-angular momentum orbits, large amount of mass can fall inwards to low angular 
momentum orbits.  Mass accretion is biased towards inflow because specific (i.e.\ per unit mass) 
angular momentum can increase to very large values, but cannot fall below zero.
The heat generated by viscous dissipation affects the disk temperature and the predicted spectra 
as described in \S\ref{sec:disk-viscous}.

\begin{figure}
\includegraphics[width=6.5in]{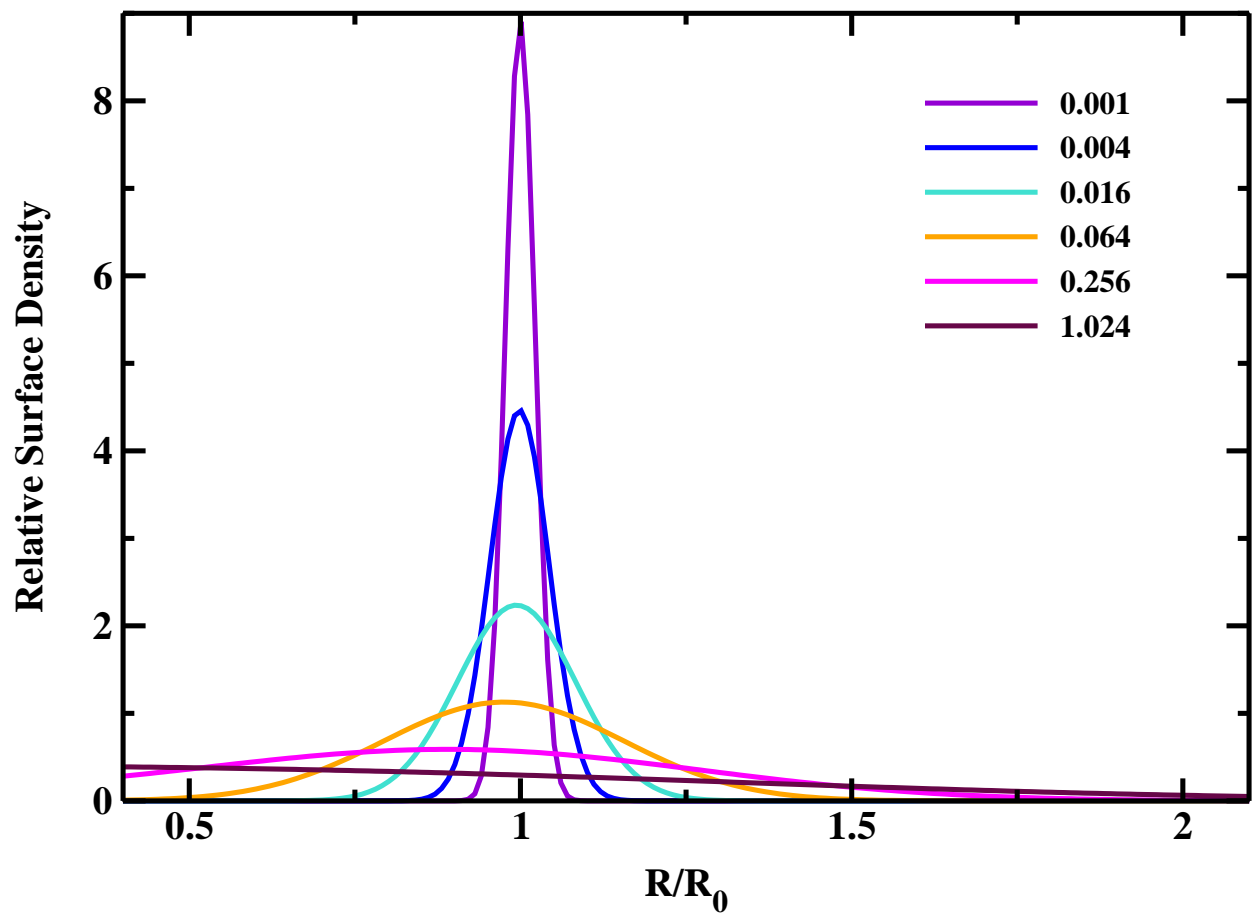}
\vskip 3ex
\caption{
Time evolution of the surface density for a ring with constant viscosity (eq. [\ref{eq:sigma-ring}]).
Over time, viscous diffusion spreads the ring into a disk. The legend indicates the scaled time, $\tau$,
for each curve.
\label{fig: disk2}
}
\end{figure}

For a disk with surface density $\varSigma$ and viscosity $\nu$, mass continuity and 
conservation of angular momentum lead to a non-linear diffusion equation for $\varSigma$ 
\citep[e.g.,][]{lbp1974,pri1981},
\begin{equation}
\frac{\partial \varSigma}{\partial t} = 3 R^{-1}~\frac{\partial}{\partial R} ~ \left ( R^{1/2}~\frac{\partial}{\partial R} ~ \{ \nu \varSigma R^{1/2} \} \right ) ~ + ~ \dot{\varSigma}_{ext}.
\label{eq:disk-diffusion}
\end{equation}
The first term on the right hand side corresponds to viscous evolution; the second is a source term 
which is positive for infall from the cloud ($\dot{\varSigma_{\rm ext}} = \dot{\varSigma_i}$) and negative 
for mass loss due to  photoevaporation, disk winds, or accretion onto giant planets.  
Consider a simple model with $\nu$ = constant, $\dot{\varSigma}_{\rm ext} = 0$, and $\varSigma(R, t = 0)$ = $m \delta(R - R_0) / 2 \pi R_0$. i.e.\ the initial mass $m$ is in a narrow 
ring at radius $R_0$.  The time evolution of the surface density is
\begin{equation}
\varSigma(x,\tau) = \frac{m}{\pi R_0^2} ~ \tau^{-1} ~ x^{-1/4} ~ e^{-(1 + x^2)/\tau} ~ I_{1/4}(2x / \tau) ~ ,
\label{eq:sigma-ring}
\end{equation}
where the scaled distance and time are $x = R/R_0$ and $\tau = t/t_0 = 12 \nu t / R_0^2$, 
and $I_{1/4}$ is the modified Bessel function.  Fig. \ref{fig: disk2} shows how this viscous solution 
asymptotically transports all of the mass to $R$ = 0 and all of the angular momentum to infinity.  This 
evolution occurs on the viscous timescale $\tau$.

We now consider the evolution of disks with a more general viscosity law, $\nu = \nu_0 (R/R_0)^{\beta}$.  
With a constant powerlaw 
$\beta$, exact similarity solutions to eq.\ (\ref{eq:disk-diffusion}) exist  \citep{lbp1974}.   To develop 
intuition, we instead physically derive the approximate solution.  After several viscous timescales 
have passed, the disk ``forgets" the initial conditions (as seen in Fig. \ref{fig: disk2}).  
Conservation of angular momentum alone (without applying mass conservation needed to derive 
eq.\ [\ref{eq:disk-diffusion}]) gives
\begin{equation}
{\p \varSigma \over \p t} = {1 \over R^{3/2}} {\p \over \p R} \left[{\dot{M} \sqrt{R}\over 2 \pi} - {3 \over 2} \nu \varSigma  \sqrt{R} \right] \label{eq:disk-ang-mom}\, .
\end{equation} 
The two terms in square brackets represent the advection of angular momentum and viscous torques.  
For a steady disk with $\p \varSigma /\p t = 0$, the term in square brackets must equal a constant 
(independent of $R$).  This constant represents the torque at the inner boundary, 
which can be neglected far from that boundary (as we show explicitly in eq.\ [\ref{eq:disk-nusigma}]). 
Thus angular momentum conservation gives
\begin{equation} 
\varSigma \simeq {\dot{M} \over 3 \pi \nu} \propto R^{-\beta} ~ ,
\label{eq:sigma1}
\end{equation} 
where the final proportionality assumes the powerlaw viscosity and  constant $\dot{M}$.  From eq.\ (\ref{eq:mdot}) the accretion speed follows as
\begin{equation}\label{eq:vR}
v_R = -{3\nu \over 2R} \, .
\end{equation} 

To derive the evolution of disk mass (and $\dot{M}$) in this limit, we adopt $R_0$ as the outer edge 
of the disk.  The outer radius changes on the local viscous timescale
\begin{equation} 
R_0 \sim \sqrt{\nu t}  \sim (\nu_0 t)^{1 /( 2-\beta)}\, .
\end{equation} 
To conserve angular momentum and energy, the disk expands, requiring
$\beta < 2$.  Conservation of angular momentum, $J \sim M_{\rm disk} \Omega(R_0) R_0^2$,
drives the evolution of the disk mass as
\begin{equation} 
M_{\rm disk} \sim {J \over \sqrt{G M_\star}} (\nu_0 t)^{-1 /( 4-2\beta)}\, .
\label{eq:Mdisk-J}
\end{equation} 
The accretion rate through the disk is
\begin{equation} 
\dot{M} = {d M_{\rm disk} \over dt} \propto t^{-(5-2\beta)/(4-2\beta)}\, .
\end{equation} 
This physical derivation emphasizes the role of conservation laws in setting the global
viscous evolution of the disk. The results are consistent with similarity solutions derived 
using Green's functions \citep{lbp1974}. 

\subsection{Transport Mechanisms and the $\alpha$ Disk Model}
\label{sec:disk-transport}

The source of a physical mechanism to drive disk accretion is a longstanding problem.
The molecular viscosity, $\nu_{\rm mol} = c_{\rm s} \lambda$ --- the product of sound speed, 
$c_{\rm s}$, and collisional mean free path, $\lambda$ --- is far too small.  For the MMSN, 
$\lambda = \mu m_H c_s / \Omega \varSigma \sigma_{H_2} \approx 0.8 R^{11/4}$ cm, where $\mu$ = 2.4 is 
the mean molecular weight, $m_H$ is the mass of a hydrogen atom, and $\sigma_{H_2}$ is the collision 
cross-section for the dominant constituent of the gas, $H_2$.  The resulting accretion timescale 
\begin{equation} 
t_{\rm acc, mol} = {M_{\rm disk} \over \dot{M}} \sim {R^2 \over  \nu_{\rm mol}} \sim 7 \times 10^{13} \left(R \over {\rm AU}\right)^{-4/7}~{\rm yr} 
\end{equation} 
vastly exceeds the age of the universe.  Viable mechanisms for angular momentum transport are sometimes 
identified as ``anomalous" sources of viscosity.  Because long range interactions are often important,
the analogy between a transport mechanism and the local viscosity is inexact.  For sufficiently small 
scale fluctuations, however, even self-gravitating disks are well approximated by the viscous 
prescription \citep{lodato2004}.

Anomalous viscosity is most often associated with turbulence in the disk.  
Because the size of turbulent eddies can greatly exceed $\lambda$, the effective turbulent viscosity can vastly exceed 
the molecular one, even for slow, subsonic turbulence. However, turbulence by itself does not explain 
accretion, velocity fluctuations must correlate for angular momentum to be transported outwards.  Moreover 
it is difficult (if not impossible) for Keplerian shear to drive turbulence; the angular momentum gradient in disks is 
quite stable according to the Rayleigh criterion \citep{sb96}.

Over the past 50 years, theorists have considered a wide range of transport mechanisms: convective eddies, 
gravitational instabilities,  internal shocks, magnetic stresses, orbiting planets, sound waves, spiral 
density waves, and tidal forces.   Currently the ``magneto-rotational instability" (MRI) is the leading 
candidate for a transport mechanism in low mass disks \citep{balb1991,papa2003}.  In this mechanism, 
 modest magnetic fields thread ionized material orbiting the central star. An outward (or inward) perturbation of
 fluid stretches and shears the magnetic fields.  The resulting torque amplifies the original perturbation and, crucially, 
transports angular momentum outwards.  Although this mechanism is attractive, the low 
ionization fraction of protostellar disks restricts the MRI to surface layers of the disk at many radii.  
Disks may then contain extensive ``dead zones" \citep{gammie96}, where levels of transport and turbulence 
are reduced.  In massive protostellar disks, gravitational waves and gravitoturbulence are another likely 
source of angular momentum transport \citep{lin1987}.

To sidestep fundamental uncertainties of transport mechanisms, it is convenient to adopt a simple
viscosity model \citep{ss1973,lbp1974}. Setting $\nu = \alpha c_s H$ -- where $H = c_s/\varOmega$ is the vertical 
scale height of the disk -- leads to the popular ``$\alpha$-disk" model.  The dimensionless parameter 
$\alpha < 1$ (even $\ll 1$) since large values would lead to rapid shock dissipation and/or gravitational 
fragmentation.   Similar to the mixing length parameterization of convection, $\alpha$-disk models allow 
progress despite ignorance of the underlying dynamics.  Detailed simulations typically quote measured 
transport rates in terms of effective $\alpha$ values.

This definition allows us to define the three important timescales in a viscous disk 
\citep[e.g.,][]{lbp1974,pri1981}. The shortest disk timescale is the dynamical 
(orbital) timescale, $t_d \sim \Omega^{-1}$:
\begin{equation}
t_d \approx {\rm 0.17~yr}~ \left ( \frac{M_{\star}}{{\rm M_{\odot}}} \right )^{-1/2} ~ \left ( \frac{R}{\rm AU} \right )^{3/2} ~ .
\label{eq:t-dyn}
\end{equation}
The disk establishes hydrostatic equilibrium in the vertical direction on the same timescale 
$t_v \approx H / c_s \approx \Omega^{-1} \approx t_d$. 

The disk cooling time, $t_c \approx U/D$ is the ratio of the thermal energy content (per unit area), 
$U = C_v \varSigma T$, to the energy generation or dissipation rate, $D$.  For a viscous disk 
$D = 9 \nu \varSigma \Omega^2 / 4$ \citep{pri1981}; $C_v$ is the specific heat at constant volume.
With 
$C_v T = (\gamma - 1)^{-1}P/\rho$ for an ideal gas, the cooling time is
\begin{equation}
t_c \approx \frac{4}{9 \gamma (\gamma - 1)} ~ \alpha^{-1} ~  \Omega^{-1} ~ . 
\label{eq:t-c1}
\end{equation}
In this classical result, the thermal timescale depends only on the local 
dynamical timescale, the dimensionless viscosity ($\alpha$), and the equation of state ($\gamma$).  For 
a molecular gas, $\gamma \approx$ 7/5.  The cooling time is a factor of roughly $\alpha^{-1}$
larger than the dynamical time:
\begin{equation}
t_c \approx {\rm 0.08~yr} ~ \alpha^{-1} ~ \left ( \frac{M_{\star}}{{\rm M_{\odot}}} \right )^{-1/2} ~ \left ( \frac{R}{\rm AU} \right )^{3/2} ~ .
\label{eq:t-c2}
\end{equation}

The viscous timescale -- $t_{\nu} = R^2 / 3 \nu$ -- measures the rate matter diffuses through 
the disk. Using our expressions for the sound speed and the viscosity, 
$t_{\nu} \approx (\alpha \Omega)^{-1} (R/H)^2$ .  Thus, the viscous timescale is 
\begin{equation}
t_{\nu} \gtrsim {\rm 0.17 ~ yr} ~ \alpha^{-1} \left ( \frac{R}{H} \right )^2 ~ \left ( \frac{M_{\star}}{{\rm M_{\odot}}} \right )^{-1/2} ~ \left ( \frac{R}{\rm AU} \right )^{3/2} ~ .
\label{eq:t-visc}
\end{equation}
Typically the disk is thin, $H/R \approx$ 0.03--0.1; thus, the viscous timescale is 100--1000 times
longer than the cooling time.

The radial velocity in \Eq{eq:vR} becomes $v_R  \approx \alpha c_s H / R \approx 0.1 \alpha (H/R)^2 v_{\phi}$. 
With $\alpha < 1$ and $H/R < 1$, $v_R$ is much smaller than both the 
orbital velocity and the sound speed. 

\subsection{Viscously Heated Disks}
\label{sec:disk-viscous}

As shown above, $t_{\nu} \gg t_c > t_d$ for $\alpha <$ 1, so the thermal properties of the disk adjust rapidly to changes 
in the surface density distribution.  This property is very useful in describing the thermal properties of many astrophysical 
disks, including active galactic nuclei, interacting binaries, and pre-main sequence stars \citep{pri1981}.  We now describe the basic energetics of accretion disks with constant $\dot{M}$

Since gas remains on Keplerian orbits as it accretes, the specific energy release in falling from $R + dR$ to $R$ is 
$G M_\star dR/(2 R^2)$, since half the potential energy goes into the increase in kinetic energy.  The energy release 
per unit area for a disk accreting at $\dot{M}$ is $F_K(R) = G \dot{M} M_\star/(4 \pi R^3)$.  This result is completely 
independent of how the energy was released.

The energy released by viscous dissipation does not simply match the local change in kinetic energy.   To fully describe the energetics of steady viscous disks, we must keep the integration 
constant when integrating eq. (\ref{eq:disk-ang-mom}) over radius to get,
\begin{equation}
\nu \varSigma = \frac{\dot{M}}{3 \pi} \left(1 - \sqrt{R_J  \over  R}\right) ~ .
\label{eq:disk-nusigma}
\end{equation}
The integration constant, $R_J$, represents the torque $\dot{J}$ exerted at the inner boundary, $R_{in}$ as $\dot{J} =  \dot{M} \sqrt{G M_\star}(\sqrt{R_{in}} - \sqrt{R_J})$.   The standard choice $R_J = R_{in}$ is a zero torque boundary condition.  Negative torques are not allowed for steady disks, as $R_J > R_{in}$ would require $\nu \varSigma < 0$.   For $R_J = 0$, the maximum torque $\dot{J}_{max} =  \dot{M} \sqrt{G M_\star R_{in}}$ matches the flow of angular momentum past the inner boundary.

%

The laws of fluid dynamics in cylindrical coordinates \citep{shuv2} give the viscous dissipation as\footnote{Note that eq.\ 3.10 in \citet{pri1981} has a factor of two typo in the intermediate result (involving $\nu$) but reaches the correct final result (in terms of $\dot{M}$).}
\begin{equation}
D(R) = \nu \varSigma  \left ( R \frac{\partial \Omega}{\partial R} \right )^2 = 
\frac{3 G M_{\star} \dot{M}}{4 \pi R^3} \left(1 - \sqrt{R_J \over R}\right) ~ .
\label{eq:disk-diss}
\end{equation}
As advertised, this expression does not simply match the local release of  Keplerian orbital energy; far from the boundary ($R \gg R_J$), $D(R) \approx 3 F_K(R)$.  Viscous disks transport energy (in addition to angular momentum) from the inner disk to the outer disk.  Nevertheless, the rapid falloff with $R$ means that most energy is dissipated close to the disk's inner edge.

Now consider the total energy release from large $R$ to the inner boundary.  The Keplerian energy release is just $L_K = G M_\star \dot{M}/(2 R_{in})$.   The total viscous luminosity is 
\begin{equation}
L_d = 2 \pi \int_{R_{in}}^\infty D(R) 2 \pi R dR = {3 \over 2} {G M_\star \dot{M} \over R_{in}} \left( 1 - {2 \over 3} \sqrt{R_J \over R_{in}} \right)
\label{eq:Ld}
\end{equation}
For the zero torque boundary condition ($R_J = R_{in}$) the luminosity simply matches the release of  Keplerian energy.  However the disk's luminosity increases due to work done by torques at the inner edge, up to $L_d = 3 L_K$ for $R_J = 0$. 
For typical parameters in protostellar disks
\begin{equation}
L_d = {f_d G M_\star \dot{M} \over 2 R_{in}} \approx 0.16 f_d L_\odot \left ( \frac{\dot{M}}{\rm 10^{-8}~M_\odot~yr^{-1}} \right ) ~ \left ( \frac{M_\star}{M_\odot} \right ) ~ \left ( \frac{R_{in}}{\rm R_{\odot}} \right )^{-1} ~ ,
\label{eq:l-disk2}
\end{equation}
where $f_d$ ranges from 1 (no torque) to 3 (maximum torque).

The maximum disk luminosity occurs for a disk that extends to the stellar stellar surface, $R_{in} = R_\star$.  The total accretion luminosity, 
\begin{equation}
L_{acc} = {f_\star G M_\star \dot{M} \over 2 R_\star} \, ,
\label{eq:Lacc}
\end{equation} 
with $1 \lesssim f_\star \lesssim 2$, includes all the energy loss needed to come to rest on the rotating stellar surface.  For a star rotating at breakup $f_\star \approx 1$.  For a slowly rotating star the damping of the orbital kinetic energy gives twice the energy release $f_\star \approx 2$.  Any difference $L_{acc} - L_d \geq 0$ is emitted at the stellar surface.   This difference must be positive (accretion should not cool the star), further constraining $f_d$.  As a consistency check, note that a disk with an inner boundary at the surface of a star, $R_{in} = R_\star$, that rotates at breakup, must satisfy the zero torque boundary condition to avoid $L_{acc} < L_d$. 


In many cases, the accreting star has a magnetosphere that truncates the disk at $R_{in} > R_{\star}$ 
\citep[e.g.,][]{ghosh1979}. Material then flows onto the star along magnetic field lines, collimated onto 
 hot spots, which are hot because the accretion energy is emitted from a small fraction of the stellar photosphere. 
 In most young stars, $R_{in} \approx$ 
3--5 \rstar\ \citep[e.g.,][]{kenyon1996,bouv2007}.  Thus the hot spot luminosity
\begin{equation}
L_{hot} = L_{acc} - L_d = L_{acc} \, \left(1 - {f_d R_\star \over f_\star R_{in}}\right)
\end{equation} 
can easily reach 60-80\% of the total accretion luminosity.  For star that corotates with the disk's inner edge,  $f_d \approx f_\star \approx 1$ is expected  \citep{ShuNaj94}.
 
To calculate the temperature structure of viscously heated disks, note that the upper and lower halves of the disk each radiate half of $D(R)$.  If the vertical optical depth $\tau > 1$, the disk photosphere then has effective temperature $T_e$,
\begin{equation} 
\sigsb T_e^4 = \frac{3 G M_{\star} \dot{M}}{8 \pi R^3} \left(1 - \sqrt{R_J \over R}\right) ~ .
\label{eq:vtemp1}
\end{equation}
Though $T_e$ is calculated as if the disk radiates as a blackbody, the disk's atmosphere will radiate as a stellar atmosphere  with spectral lines, especially when $T_e \gtrsim $ 1000--1500 K.  The effective temperature declines as 
$T_e \propto R^{-3/4}$ for $R \gtrsim$ a few $R_{\star}$:
\begin{equation}
T_e = {\rm 85 ~ K} \left ( \frac{\dot{M}}{\rm 10^{-8}~M_\odot~yr^{-1}} \right )^{1/4} ~ \left ( \frac{M_\star}{M_\odot} \right )^{1/4} ~ \left ( \frac{R}{\rm AU} \right )^{-3/4} ~ (1 - \sqrt{R_J/R}) ~ .
\label{eq:vtemp2}
\end{equation}
In a simple, grey-atmosphere approach, the midplane temperature $T_d$ is a factor of $\tau^{1/4}$
larger than $T_e$ and is used to derive the scale height $H$, the viscosity $\nu$, and other 
physical variables. More rigorous approaches calculate $T_e$ and $T_d$ using a self-consistent
prescription for the opacity throughout the disk.

The temperature distribution in eq. (\ref{eq:vtemp1}) allows us to derive the surface density of an $\alpha$ disk. 
Assuming the midplane temperature scales like the effective temperature 
($T_d \approx T_e \propto R^{-3/4}$), $\nu \propto c_s^2 \Omega^{-1} \propto r^{3/4}$ for the $\alpha$ prescription.  Then \Eq{eq:sigma1} gives \begin{equation}
\varSigma(R) = \varSigma_0 \left ( \frac{R}{\rm AU} \right )^{-3/4}
\label{eq:sigma-v}
\end{equation}
where $\varSigma_0$, the surface density at 1 AU, depends on the mass accretion rate. Solving for the midplane 
temperature with realistic opacities yields $\varSigma \propto R^{-\beta}$ with $\beta \approx$ 0.6--1 \citep[e.g.,][]{step1998,cha2009}.

\subsection{Steady Irradiated Disks}
\label{sec:irr}
Although viscous dissipation drives the overall evolution, radiation from the central star also heats 
the disk \citep{fried1985,adams1986,kh1987}. If the disk is an infinite, but very thin, sheet, it absorbs 
roughly 25\% of the light radiated by the central star.  For a 1 \lsun\ central star and disk accretion 
rates $\dot{M} \lesssim 10^{-8}$ \msunyr, emission from irradiation exceeds emission from viscous 
dissipation (eq. [\ref{eq:l-disk2}]).  If the disk re-radiates this energy at the local blackbody 
temperature, the radial temperature gradient of a flat, irradiated disk follows the gradient for a viscous 
disk, $T_d \propto R^{-3/4}$ \citep{fried1985,adams1986}. 

Disks with vertical scale height $H$ absorb and re-radiate even more starlight \citep{kh1987}. 
\citet{chiang:1997} derive a general formalism for $H$ and $T_d$ in a ``passive" disk with negligible
$\dot{M}$. Defining $\theta$ as the grazing angle that starlight hits the disk, the temperature of a 
disk that emits as a blackbody is 
\begin{equation}
T_d \approx \left ( \frac{\theta}{2} \right )^{1/4} ~ \left ( \frac{R_\star}{R} \right )^{1/2} ~ T_{\star} ~ ,
\label{eq:irtemp-1}
\end{equation}
where $T_{\star}$ is the stellar temperature. The grazing angle is 
\begin{equation}
\theta \approx 0.4 \frac{R_{\star}}{R} + R \frac{d}{dR} \left ( \frac{h}{R} \right ) ~ ,
\label{eq:theta-1}
\end{equation}
where $h$ is the height of the photosphere above the disk midplane. For a blackbody disk in 
vertical hydrostatic equilibrium, the grazing angle is the sum of a nearly flat component 
close to the star and a flared component far from the star:
\begin{equation}
\theta \approx 0.005 \left ( \frac{R}{\rm AU} \right )^{-1} + 0.05 \left ( \frac{R}{\rm AU} \right )^{2/7} ~ .
\label{eq:theta-2}
\end{equation}
The disk temperature beyond a few tenths of an AU is then
\begin{equation}
T_d \approx {\rm 155~K} ~ \left ( \frac{R}{{\rm AU}} \right )^{-3/7} \left ( \frac{R_\star}{R_\odot} \right )^{1/2} \left ( \frac{T_{\star}}{T_\odot} \right ) ~ .
\label{eq:irtemp-2}
\end{equation}
For $\dot{M} \approx 10^{-8}$ \msunyr, the irradiated disk is roughly twice as hot as a viscous 
accretion disk.

This temperature relation leads to a steeper surface density gradient in $\alpha$ disks. With $\nu \propto c_s^2 \Omega^{-1}$
and $c_s^2 \propto T_d$, $\nu \propto r^{15/14}$. Using this viscosity in eq. (\ref{eq:sigma1}),
\begin{equation}
\varSigma(R) = \varSigma_{0,irr} \left ( \frac{R}{\rm AU} \right )^{-15/14} ~ , 
\label{eq:sigma-irr}
\end{equation}
where
\begin{equation}
\varSigma_{0,irr} = \frac{\rm 2~g~cm^{-2}}{\alpha} \left ( \frac{\dot{M}}{\rm 10^{-8}~M_{\odot}~yr^{-1}} \right ) ~ . 
\label{eq:sigma0-irr}
\end{equation}
The surface density gradient for an irradiated disk is steeper than the gradient for a viscous disk
and is reasonably close to the gradient for the MMSN.

For identical $\alpha$, hotter irradiated disks have larger viscosity and smaller surface density than
cooler viscous disks.
Integrating eq. (\ref{eq:sigma-irr}) over radius, the mass of an irradiated disk is 
\begin{equation}
M_d \approx \left ( \frac{10^{-4} M_{\odot}}{\alpha} \right ) ~ \left ( \frac{\dot{M}}{10^{-8} ~ M_{\odot} ~ {\rm yr^{-1}}} \right ) ~ \left ( \frac{R_d}{\rm 100~AU} \right )^{15/14} ~ .
\label{eq:mdisk-irr}
\end{equation}
When $\alpha \approx 10^{-3} - 10^{-2}$, this estimate is close to the observed masses of protostellar
disks around young stars.

\subsection{Time Dependence}

Deriving more robust estimates of disk evolution requires a direct solution of eq. (\ref{eq:disk-diffusion}).
This exercise requires a model for the viscosity and a prescription for the thermodynamics and opacity of disk
material. Analytic approaches assume a constant mass accretion rate through the disk. If $\alpha$ and 
$\tau$ are simple functions of the local variables $\varSigma$ and $T$, then the diffusion equation can
be solved exactly for $\varSigma(t)$, $\dot{M}(t)$, and other disk properties \citep{step1998,cha2009}. 
Numerical approaches allow $\alpha$ and $\dot{M}$ to vary with radius. Some solutions consider iterative 
solutions to the temperature and vertical structure \citep[e.g.][]{hues2005}; others solve for the vertical 
structure directly using techniques developed for the atmospheres of stars \citep[e.g.][]{bell1994,DAless1998}.

To compare these approaches, we consider a simple model for a viscous disk irradiated by a central star.
We assume that the optical depth of cool disk material is dominated by dust grains with a constant opacity 
$\kappa_0$; warmer dust grains evaporate and have a smaller opacity:
\begin{equation}
\kappa = \left\{ \begin{array}{cll}
      \kappa_0 ~ , &    & T_d \le T_{evap} \\
      \kappa_0 \left ( \frac{T_d}{T_{evap}} \right )^n,  &    & T_d > T_{evap} \\
      \end{array} \right.
\label{eq:kappa}
\end{equation}
For material with roughly solar metallicity, typical values are $\kappa_0 \approx 2~{\rm cm^2/g}$,
$T_{evap} \approx$ 1380 K, and $n = -14$ \citep[e.g.,][]{cha2009}. With this opacity,
we derive a self-consistent disk temperature and scale height \citep[as in][]{hues2005} and solve for 
the time evolution of $\varSigma$ using an explicit solution to the diffusion equation 
\citep[as in][]{bath1982}. 

\begin{figure}
\includegraphics[width=6.5in]{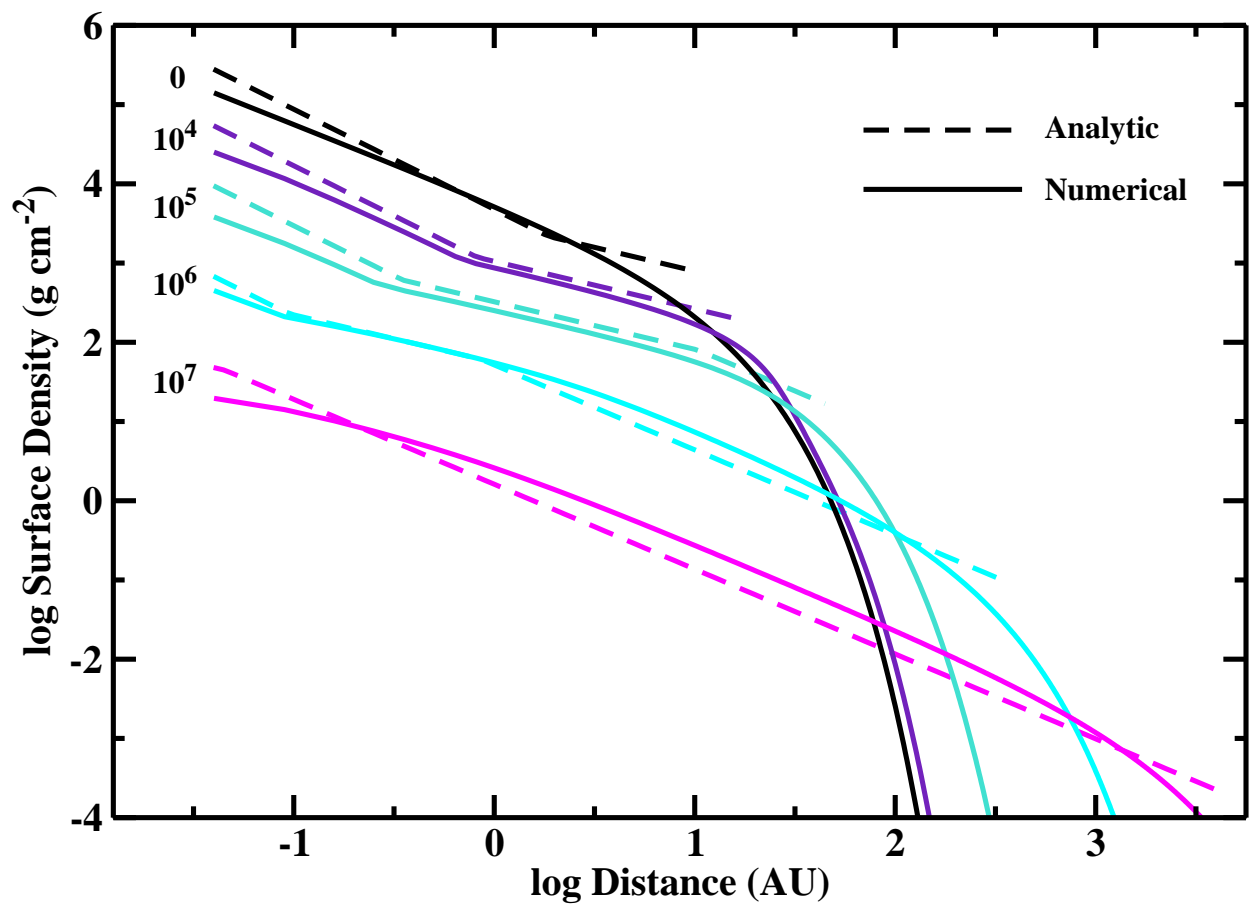}
\vskip 3ex
\caption{
Time evolution of the surface density of a gaseous disk surrounding
a 1 \msun\ star.  Dashed lines show results for the analytic disk model
of \citet{cha2009}; solid lines show results for our numerical solution
of the diffusion equation. Depsite small differences in the initial
conditions, the numerical solution tracks the analytic model.
\label{fig: disk3}
}
\end{figure}

Fig. \ref{fig: disk3} compares analytic and numerical results for a disk with
$\alpha = 10^{-2}$, initial mass $M_{d,0}$ = 0.04~\msun, and initial radius $R_0$ = 10~AU 
surrounding a star with \mstar\ = 1~\msun. The numerical solution tracks the analytic model 
well.  At early times, the surface density declines steeply in the inner disk 
($\varSigma \propto R^{-1.2}$; where dust grains evaporate) and more slowly in the outer disk 
($\varSigma \propto R^{-0.6}$; where viscous transport dominates).  At late times, irradiation 
dominates the energy budget; the surface density then falls more steeply with radius, 
$\varSigma \propto R^{-1}$.

Other approaches lead to similar time evolution in the surface density. Early on, a massive
disk is dominated by viscous heating.  For these conditions, the simple analytic estimate of 
the surface density yields $\varSigma \propto R^{-n}$ with $n = 3/4$ (eq. [\ref{eq:sigma-v}]), 
close to results for the numerical solution ($n = 0.6$) and other analytic and numerical
($n = 0.6 - 1$) approaches \citep{bath1982,lin1990,step1998,cha2009,alex2009}.  As the disk 
ages, it evolves from a viscous-dominated to an irradiation-dominated system. Thus, the 
exponent $n$ in the surface density relation approaches the limit ($n = 15/14$) derived in 
eq. (\ref{eq:sigma-irr}). 

\begin{figure}
\includegraphics[width=6.5in]{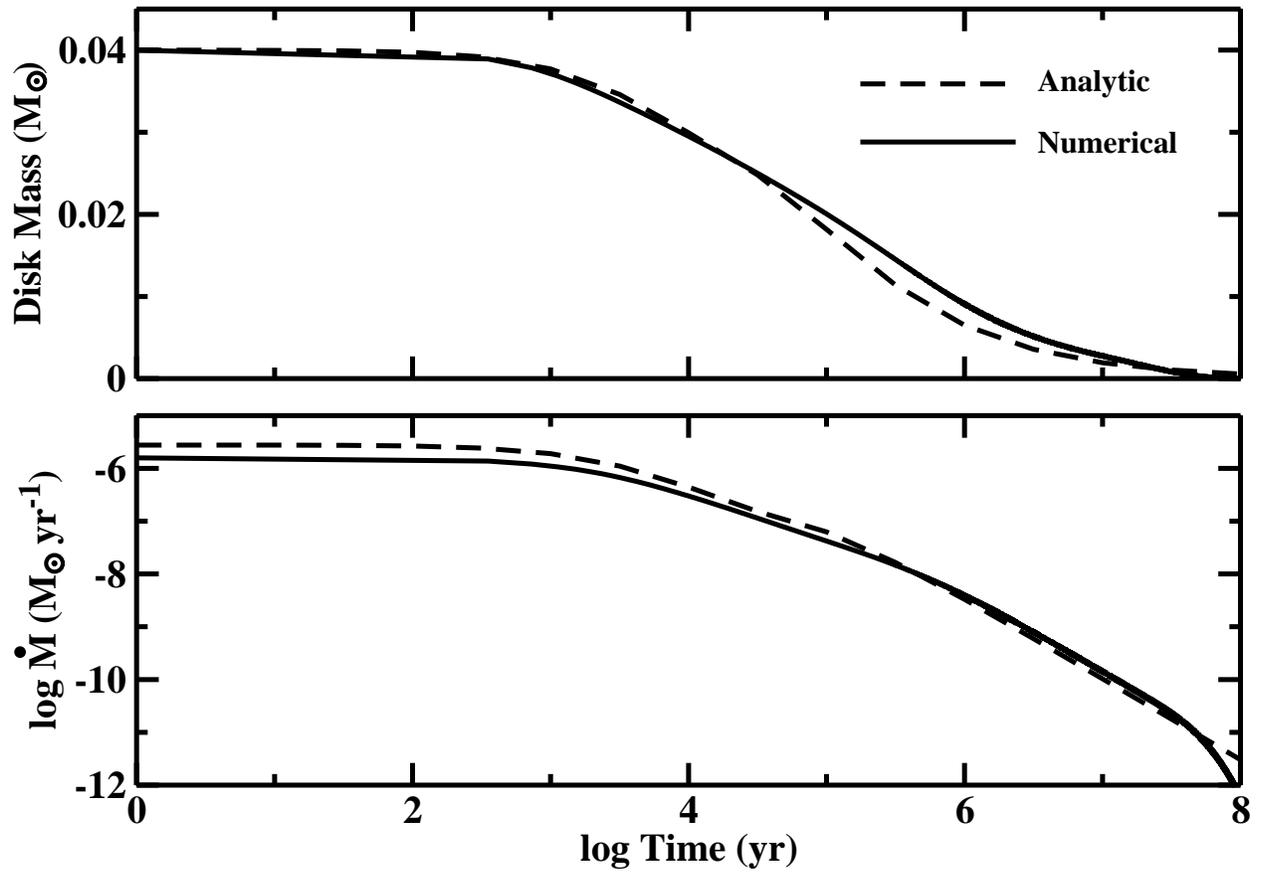}
\vskip 3ex
\caption{
Time evolution of the disk mass (upper panel) and disk accretion rate onto
the central star (lower panel) for the analytic and numerical solutions in
Fig. \ref{fig: disk3}.
\label{fig: disk4}
}
\end{figure}

Fig. \ref{fig: disk4} compares the evolution of the disk mass and accretion rate
at the inner edge of the disk. In both solutions, the disk mass declines by a factor 
of roughly two in 0.1 Myr, a factor of roughly four in 1 Myr, and a factor of roughly 
ten in 10 Myr. Over the same period, the mass accretion rate onto the central star
declines by roughly four orders of magnitude.

\subsection{Disk Instabilities and Fragmentation}\label{sec:gravity}

In addition to evolution on the viscous timescale shown in Fig. \ref{fig: disk3}, all disks vary their energy 
output on much shorter timescales. In compact binary systems, these fluctuations range from small, 10\%--20\%, 
amplitude flickering on the local dynamical time scale to large-scale eruptions, factors of 10--100, 
that can last for several times the local viscous time scale \citep{warner1995}. Although many pre-main sequence
stars also display distinctive brightness variations \citep{joy1945,herbig1962}, the FU Ori variables provide 
the cleanest evidence for large-scale variations of the disk, rather than the environment or the central star
\citep{hk1996}.

Theory suggests several types of instabilities in viscous disks \citep[see][]{pri1981}. In standard
derivations of the structure of steady disks, radiative cooling balances heating from viscous stresses.
However, radiative losses are set by local disk parameters; local parameters and an input accretion rate
set viscous energy generation. Usually radiative losses can keep up with changes in disk structure; 
sometimes, radiation cannot balance viscous stresses, leading to a thermal instability. 
A limit cycle arises, where regions of the disk 
alternate between states where radiative losses exceed (and then fall below) the viscous energy input. This
mechanism may produce FU Ori and other eruptions in the disks of pre-main sequence stars \citep{hk1996}.

Viscous instabilities occur when changes in the local surface density do not produce parallel increases 
in the local mass transfer rate. From eq. (\ref{eq:sigma1}), $\dot{M} \propto \nu \varSigma$.
In a steady disk, $\nu$ is fairly independent of $\varSigma$; thus, $\dot{M}$ changes in step with $\varSigma$. 
For the MRI viscosity mechanism, however, larger $\varSigma$ leads to larger optical depths, less ionization,
and smaller $\alpha$. Thus, an MRI disk with growing (falling) surface density can produce a smaller 
(larger) viscosity, leading to an ever greater over- or under-density in the surface density. 

Although thermal and viscous instabilities change the temperature and surface density throughout the disk, 
they evolve on timescales much longer than the local orbital period.  Massive disks can evolve 
on shorter timescales.  If the local gravity in a region with size $\lambda$ overcomes rotational support
($G \varSigma \gtrsim \Omega^2 \lambda$) and  thermal support ($G \varSigma \gtrsim c_s^2/\lambda$),
this region can (begin to) collapse \citep{saf60,toom64,gold65,pac78}.  Together these conditions require $c_s^2/(G \varSigma) \lesssim \lambda \lesssim G \varSigma/\Omega^2$.   Collapse at any wavelength requires the disk satisfy the ``Toomre instability criterion," 
\begin{equation} 
Q \equiv \frac{c_s \Omega}{\pi G \varSigma} \lesssim 1 ~ .  
\label{eq:Q} 
\end{equation} 
Setting the disk mass $M_d \approx \varSigma R^2$, a stable disk has 
\begin{equation} 
c_s \gtrsim \frac{M_d}{M_{\star}} ~ v_{\phi} ~ .  
\label{eq:v-sound-q} 
\end{equation}
When the disk first forms, $M_d \approx M_{\star}$.  Such ``disks" cannot be thin because $H/R \sim c_s/v_\phi \gtrsim 1$.

In a viscous accretion disk, the stability criterion can be re-written in terms of the accretion rate
\citep{gamm2001}. With $\dot{M} = 3 \pi \nu \varSigma$ and $\nu = \alpha c_s^2 \Omega^{-1}$ 
an unstable disk has $\dot{M_Q} \gtrsim 3 \alpha c_s^3 / G$. To evaluate the temperatures of unstable
disks, we use $c_s = (\gamma k T / \mu m_H)^{1/2}$ and set $\gamma = 7/5$ and $\mu$ = 2.4 for molecular
gas:
\begin{equation} 
\dot{M_Q} \gtrsim 2.4 \ 10^{-4} ~ \alpha T^{3/2} ~ \msunyr ~ ,
\label{eq:mdot-q} 
\end{equation}
with $T$ in Kelvins.  For the observed accretion rates in very young stars, $\dot{M} \sim 10^{-7} ~ \msunyr$, unstable disks 
have $\alpha T^{3/2} \lesssim 0.4$. If $\alpha$ is large ($10^{-2}$), only very cold disks are unstable 
(T $\sim$ 10--15~K); smaller $\alpha$ (e.g., $10^{-3}$) allows instability in warmer disks 
(T $\sim$ 50--60~K). 

\section{FROM DUST TO PLANETESIMALS}\label{sec:planetesimal}

The accumulation of dust grains into planetesimals --- solids greater than a kilometer in size --- is the first step in the formation of terrestrial planets and giant planet cores.  Several observational and theoretical reasons suggest the formation of planetesimals is a separate step.  Observationally, remnant planetesimals in the Solar System and in extrasolar debris disks shows that growth sometimes stalls before planets accumulate all planetesimals. Comets from the Oort cloud also suggest an intermediate stage between dust grains and planets (see Chapter by Moro-Martin).

Theoretically, the physical processes responsible for the growth of planetesimals differ from those relevant to the final stages of planet formation.  As Section 5 describes, few-body gravitational encounters --- both scattering and gravitationally focused collisions --- establish the rates of growth for icy and terrestrial planets.  By contrast, the sticking of dust grains involves electrostatic forces.  During planetesimal formation, particle dynamics is qualitatively different.  Drag forces exerted by the gas disk dominate the motions of small solids.  Though not negligible, the drag exerted on km-sized or larger planetesimals is weaker than gravitational interactions (\S5.2). 

While planetesimal formation is a common occurrence in circumstellar disks, understanding how it happens has proved elusive.  Observations of planetesimal formation in action are indirect.  Particles beyond cm-sizes contribute negligibly to images and spectra of circumstellar disks.  Primitive meteorites record the conditions during planetesimal formation, but the implications for formation mechanisms are difficult to interpret --- we need a better instruction manual.

Especially beyond millimeter sizes, experiments show that particle collisions often result in bouncing or breaking instead of sticking \citep{bw00,zsom10,weidling2012}.  Inefficient growth by coagulation is further complicated by the rapid infall of centimeter to meter sized solids into the star.  These difficulties --- often termed the ``meter-sized barrier" --- are explained in more detail in \S\ref{sec:meterbarrier}.

Gravitational collapse is one way to overcome the growth barrier.  The mutual gravitational attraction of a collection of small solids could lead to a runaway collapse into planetesimals --- even when sticking is inefficient and radial drift is fast.   While appealing, this path encounters theoretical difficulties when stirring by turbulent gas is included.  Section \ref{sec:ptmlGI} describes the current status of the gravitational collapse hypothesis.

Even when self-gravity is weak, aerodynamic effects can concentrate solids in the disk.  Particles tend to seek high pressure regions in the disk.  This tendency causes the inward drift mentioned above.  Particles can also concentrate in localized pressure maxima.  Predicting the sizes and lifetimes of pressure maxima is a difficult (and currently unsolved) problem of disk meteorology.  

In addition to the passive response of solids to the gas disk, active particle concentration occurs when particles cause their own clumping by altering the flow of gas.
Instabilities caused by gas drag, notably the streaming instability, provide a clumping mechanism that is both powerful and amenable to study by direct numerical simulations.  The strong clumping driven by the streaming instability is capable of triggering gravitational collapse into $\sim 100$~km planetesimals.  Both passive and active particle concentration mechanisms are reviewed in \S\ref{sec:concentrate}.

Theories based on complex non-linear dynamics must be tested against, and refined by, observations.  We discuss observational consequences of planetesimal formation models in \S\ref{sec:ptmlobs}.  Unless stated otherwise, the numerical estimates in the section use the passively heated MMSN disk (\S3).  For more detailed reviews of planetesimal formation, see \citet{chiangyoudin:2010} and \citet{houches10}.  For a thorough review of collision experiments and their relevance to planetesimal formation, see \citet{bw08}.

\subsection{The ``Meter-Size" Barrier}\label{sec:meterbarrier}
We discuss in more detail the two components of the ``meter-size" barrier to planetesimal formation.  The review of radial drift timescales in \S\ref{sec:radialdrift} also serves as an introduction to the dynamics of solids in a gas disk.  The discussion of collisional growth and destruction in \S\ref{sec:sticking} couples  dynamical models of collision speeds to the complex physics of contact mechanics.  

\subsubsection{Radial Drift and the Basics of Disk Aerodynamics}\label{sec:radialdrift}
The aerodynamic migration of small solids imposes the most stringent timescale constraint on planet formation: $\approx 100$ years.  Aerodynamic radial drift arises because solids encounter a headwind as they orbit through the gas disk (eq. [\ref{eq:disk-gasvel}]). This headwind removes angular momentum from particle orbits, causing their inspiral.  Infall speeds are fastest for solids near roughly meter sizes.   The critical size is actually below a meter in standard disk models --- especially in their outer regions.  So the ``meter-size" barrier is a slight misnomer, but it has a better ring than the ``millimeter-to-tens-of-centimeters-size" barrier.

Radial pressure gradients in gas disks set the speed of the headwind. Plausible disk models are hotter and denser in the inner regions; on average, the radial pressure gradient is directed outwards.  If the radial pressure gradient is directed inwards, a tailwind --- and outward particle migration --- results.  

We express the headwind speed as the difference between the  Keplerian velocity, $\vk = \sqrt{G M_\star/R} = \varOmega R$, and the  orbital speed of the gas, $v_{{\rm g},\phi}$: 
\begin{equation}\label{eq:etavK}
\eta \vk \equiv \vk - v_{{\rm g},\phi}  \approx - \frac{\p P /\p \ln R}{2 \rho\gs \vk} 
\approx 25 \left(\frac{R}{\rm{AU}}\right)^{1/14} ~{\rm m~s}^{-1}\, ,
\end{equation}
where $P$ and $\rho\gs$ are the pressure and density of the gas, and $\eta \sim c_{\rm s}^2/\vk^2 \sim (H/R)^2 \sim 10^{-3}$ is a dimensionless measure of pressure support.   In disks hotter than our passive model,  headwinds and drift speeds are faster.

To derive \Eq{eq:etavK}, compute radial force balance assuming (correctly) that the radial pressure acceleration, $f_{P,R} = -\rho\gs^{-1}\partial P/\partial R$, is weak compared to the centrifugal acceleration.  Equivalently we can reproduce  \Eq{eq:etavK} by balancing the pressure and Coriolis forces,  $f_{P,R} + f_{{\rm Cor},R} = 0$ with  $f_{{\rm Cor},R} = - 2\varOmega \eta \vk$.

Drag forces set the response of particle orbits to the gas headwind.  We express the drag acceleration felt by a solid particle as
\begin{equation} \label{eq:fdrag}
\vc{f}_{\rm drag} = - {\Delta \vc{v} / \ts}\, ,
\end{equation} 
where  $\Delta \vc{v}$ is the particle velocity relative to the gas and $\ts$ is the aerodynamic damping timescale for this relative particle motion.  

The value  of $\ts$ depends on particle  properties  -- such as the internal density, $\rho_\bullet$, and spherical radius, $s$  --- and on properties of the gas disk ---  $\rho\gs$ and $c_{\rm s}$ -- as:
\begin{subnumcases}
{\label{eq:ts} \ts = }
\ts^{\rm Ep} \equiv \rho_\bullet s / (\rho\gs c_{\rm s}) & if  $s < {9 \lambda/4}$\label{Ep} \\
\ts^{\rm Stokes} \equiv \ts^{\rm Ep} \cdot  4s /(9 \lambda) & if  ${9 \lambda/4} < s < \lambda /(4 {\rm Ma})$ \label{St}\\
\ts^{\rm int}\cdot \left(s / \lambda\right)^{3/5} {\rm Ma}^{-2/5}/4 & if $\lambda /(4 {\rm Ma}) < s <200 \lambda/{\rm Ma}$ \label{mix} \\
\ts^{\rm turb} \equiv \ts^{\rm Ep} \cdot 6 /{\rm Ma} & if  $s > 200 \lambda/{\rm Ma}$ \label{turbdrag}
\end{subnumcases}
where ${\rm Ma} \equiv |\Delta\vc{v}|/c_{\rm s}$, $\lambda \propto 1/\rho\gs$ is the gas mean free path, and  ${\rm Re} \equiv 4 s {\rm Ma}/\lambda$ is the Reynolds number of the flow around the particle.  The cases are written in order of increasing particle size: Epstein's Law of drag from molecular collisions, Stokes' Law for viscous drag when ${\rm Re}  < 1$, an approximate intermediate ${\rm Re}$ case, and the drag from a fully developed turbulent wake for ${\rm Re} > 800$.  The turbulent drag force is more relevant for fully formed planetesimals and is commonly expressed as
\begin{equation} \label{eq:turbdragforce}
F_{\rm drag} = -m {|\Delta \vc{v}| \over \ts^{\rm turb}} = - {C_D \over 2} \pi s^2 \rho\gs |\Delta \vc{v}|^2 \, ,
\end{equation} 
where the drag coefficient, $C_D \approx 0.44$ \citep{ada76,weid1977b}.

\begin{figure}[t]
\includegraphics[width=6.5in]{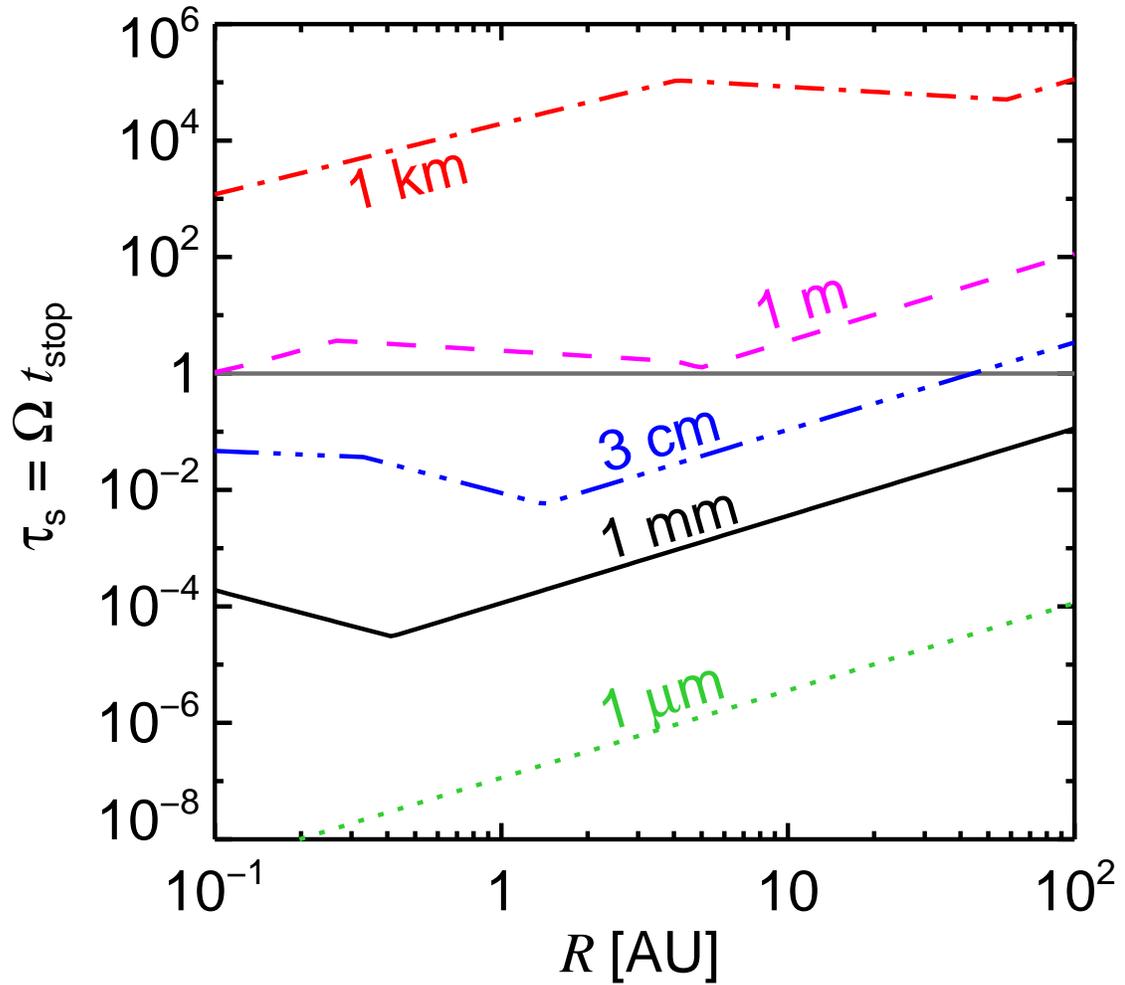}
\vskip 3ex
\caption{Aerodynamic stopping time normalized to the Keplerian orbital frequency for a range of particle sizes in our reference minimum mass disk model.  Small (large) values of $\taus$ indicate strong (weak) coupling of solids to the gas disk.  The breaks in the curves are due to transitions between different drag laws, as described by \Eq{eq:ts}.  An internal density of $\rho_\bullet = 1 \g \cm^{-3}$ is assumed for the solids.
}  
\label{fig:ts}
\end{figure}

The dynamical significance of drag forces is measured by comparing the stopping time and the orbital frequency, via the parameter
\begin{equation} \label{eq:taus}
\taus \equiv \varOmega \ts\, .
\end{equation} 
For $\taus \ll 1$, particles are carried along with the gas; for $\taus \gg 1$, gas drag is a small correction to Keplerian orbits.  Fig.\ \ref{fig:ts} plots $\taus$ for a range of particle sizes in our passively heated MMSN.  At least in the inner disk, objects near meter-sizes have $\taus \approx 1$.  In the outer disk, where gas densities are lower, smaller solids have the critical $\taus = 1$.   As we now show, $\taus = 1$ solids have the fastest radial drift speeds.

To derive the particle drift caused by the gas headwind, we consider the equations of motion for a particle in cylindrical coordinates, $r$ and $\phi$,
\begin{eqnarray}
\ddot{R} - R\dot{\phi}^2 = -\vk^2/R - \dot{R}/\ts  \label{eq:eomr}\\
R\ddot{\phi} + 2 \dot{R}\dot{\phi} = - (R\dot{\phi} - v_{{\rm g},\phi} )/\ts \,. \label{eq:eomphi}
\end{eqnarray}
To find the steady drift solutions, we make several approximations that are valid when drag forces are strong.  We neglect the radial inertial acceleration, $\ddot{R}$, and express the azimuthal motion as a small deviation from the Keplerian frequency, $\dot{\phi} = \varOmega +  \delta v_\phi/R$ where $|\delta v_\phi| \ll \varOmega R$.  The azimuthal acceleration is then $\ddot{\phi} \approx \dot{\varOmega} \approx -3\varOmega \dot{R}/ (2R)$.

\begin{figure}[t]
\includegraphics[width=6.5in]{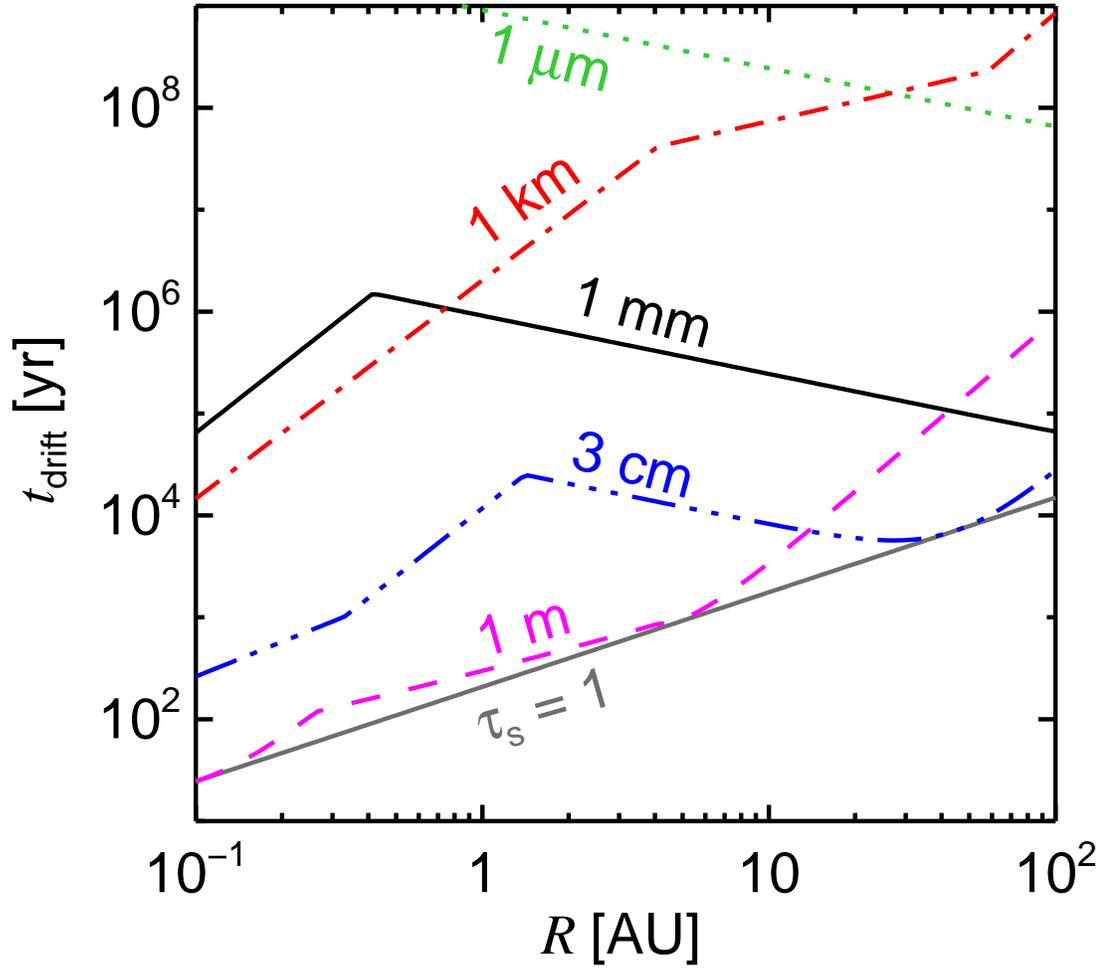}
\vskip 3ex
\caption{Radial drift timescales, $R/\dot{R}$, for the same disk models and particle sizes as in \Fig{fig:ts}.  The fastest drift timescale is for the particle size that has $\taus = 1$ as indicated by the grey curve.  Drift timescales are much faster than disk lifetimes of a few Myr, especially near the ``meter-sized" barrier.
}  
\label{fig:tdr}
\end{figure}

The radial drift speed follows from \Eqfour{eq:etavK}{eq:taus}{eq:eomr}{eq:eomphi} as
\begin{equation} 
\dot{R} \approx -2 \eta \vk \left( \frac{\taus}{1+\taus^2} \right) \label{eq:rdot} \, . \\
\end{equation} 
Solids with $\taus = 1$ have the fastest infall speed, $-\dot{R} = \eta \vk$.  The corresponding timescale
\begin{equation} 
\min (t_R) \sim (\eta \varOmega)^{-1} \sim 200 (R/\rm{AU})^{13/14} ~{\rm yr}
\end{equation} 
is a very strong constraint on growth.  This constraint is the main element of the meter-sized growth barrier.  \Fig{fig:tdr} plots the radial drift timescales for a range of particle sizes.

To complete this brief introduction to particle aerodynamics, we give the azimuthal drift speed of solids through the (sub-Keplerian) gas as 
\begin{equation} 
R \dot{\phi} - v_{{\rm g},\phi} = \delta v_\phi + \eta \vk = \eta \vk {\taus^2 \over 1 + \taus^2}\, .
\end{equation} 
Large, $\taus \gg 1$ solids experience the full $\eta \vk$ headwind, yet their radial drift is slow because their inertia is so large.  Small, $\taus \ll 1$ solids are dragged by the gas and only feel a mild headwind.  We thus see why  radial drift is fastest near $\taus \approx 1$.  For these intermediate sizes, drag forces are strong enough to overcome particle inertia, but not so strong as to cause perfect coupling.

These idealized calculations explain the basics of radial drift.  A pressing question is whether ignored effects could mitigate the radial drift problem.  The existence of a headwind is the most crucial assumption, and it can vanish in localized pressure maxima as addressed below.  Even when these maxima exist, headwinds still prevail in the majority of the disk.  We also assume that aerodynamic drag only affects the solids, and not the gas component of the disk.  When the distributed mass density of solids $\rho\ps$ becomes comparable to the gas density $\rho\gs$, then it is no longer acceptable to ignore the feedback of drag forces on the gas.  \citet{nsh86} showed how drift speeds become slower when $\rho\ps \gtrsim \rho\gs$.  This feedback is also the source of powerful drag instabilities --- both shearing and streaming --- that we address below.  Thus there is no simple way to ignore the radial drift problem --- its resolution has consequences for how planetesimals form.

\subsubsection{Early Collisional Growth}\label{sec:sticking}
Planetesimal formation begins with the collisional agglomeration of dust grains into larger solids, a process that is observed to proceed up to mm-sizes in T Tauri disks \citep{wil2011}.  The conceptually simplest mechanism to form planetesimals is for this collisional growth to proceed past kilometer sizes.   However both direct experiment and theoretical arguments show that coagulation beyond mm-sizes is inefficient at best.
This inefficiency is particularly problematic due to the timescale constraints imposed by radial drift.  The combination of inefficient sticking and rapid infall together comprise the formidable ``meter-size" barrier.

Although collision rates do not rule out rapid growth, they place tight constraints on the sticking efficiency  To make this conclusion, we approximate the mean collision time as $t_{\rm coll} \sim 1/(\varOmega \tau)$, where $\tau \sim \varSigma\ps/(\rho_\bullet s)$ is the vertical optical depth.   This approximation for the collision rate is good when $\taus \gg 1$, and we show below that it suffices for $\taus = 1$.  The ratio of collision to drift timescales for $\taus = 1$ solids is thus roughly
\begin{equation} \label{eq:collvdrift}
\left.{t_{\rm coll} \over t_{\rm drift}}\right|_{\taus = 1} \sim {\eta \over Z} \sim 0.3 \left(R \over 10\AU\right)^{4/7}\left({0.01 \over Z}\right)\, ,
\end{equation} 
where we assume Epstein drag, appropriate for the outer disk.  We use the result of hydrostatic balance that $\varSigma\gs \sim \rho\gs c_{\rm s}/\varOmega$.  When the collision time exceeds the drift time, collisional growth is ruled out.  Even when the two are close, growth requires an efficient rate of sticking per collision.  This constraint is most severe in the outer disk.

While turbulent motions  increase collision speeds, they do not increase the collision rate above the geometric estimate used to derive \Eq{eq:collvdrift}.  The reason is that turbulence also increases the particle layer thickness, $H\ps$, thereby decreasing the mean particle density, $\rho\ps$.   We can compute the collision rate due to turbulence as $t_{\rm coll}^{-1} \sim n\sigma v$.  Here the particle number density is $n \sim \varSigma\ps/(H\ps m\ps)$, where $m\ps$ is the particle mass.  The particle layer thickness due to turbulent stirring is\footnote{To accommodate the even mixing of small grains (not our current concern) we require $H\ps \leq H\gs$.}
\begin{equation} \label{eq:halpha}
H\ps = H_\alpha = \sqrt{\alpha_D \over \taus} H\gs\, .
\end{equation} 
This well-known result \citep{cuzzi:1993,cfp06,yl07} normalizes the turbulent diffusion, $D$, to the dimensionless parameter, $\alpha_D \equiv D/(c_{\rm s} H\gs)$.   The cross section is $\sigma \sim s^2$ and the relative velocity due to turbulent motions is, $v \sim \sqrt{\alpha \taus/(1+\taus^2)} c_{\rm s}$ \citep{mmv91, chiangyoudin:2010}.  For $\taus > 1$, the collision rate necessarily agrees with the optical depth estimate.  For $\taus < 1$, the collision rate is also independent of the strength of turbulence as $t_{\rm coll}^{-1} \sim Z \varOmega$.  These cases agree at $\taus \sim 1$ and confirm the constraint set by \Eq{eq:collvdrift}.

Collision rates are not the only concern.  Collisions can also result in bouncing or fragmentation that stalls, or even reverses, growth.  Below speeds of $\sim 1 ~{\rm m~s}^{-1}$, small dust grains stick efficiently as a consequence of van der Waals interactions and the efficient dissipation of kinetic energy \citep{cth93, bw00}.  As particles grow and as collision speeds increase, the collisional kinetic energy increases.  Short range sticking forces cannot match this increase in kinetic energy, because they are surface area limited \citep{you04}.   Experimental work confirms that collisions between equal mass objects do not produce growth beyond $\sim$ mm-sizes \citep{bw08}.

Collisions between lower mass projectiles and higher mass targets offer another route to growth.  In this scenario impact speeds exceed the meters-per-second value expected to produce growth.  As explained above, since small solids are tied to the gas flow, they impact larger solids (which decouple from the gas) at the full headwind speed, $\eta \vk \gtrsim 25~{\rm m~s}^{-1}$.  Indeed when the latest experimental results are combined with dynamical estimates of collision speeds for a  dispersion of particle sizes, growth stalls at only millimeter sizes \citep{zsom10}.

Based on observed SEDs, disks likely find a way to overcome these obstacles and grow solids beyond mm-sizes \citep{wil2011}.  The mechanisms responsible for enhanced coagulation remain unclear.  The particle concentration mechanisms discussed below could augment particle sticking.  As shown explicitly in \citet{jym09}, collision speeds are reduced in dense particle clumps.

Most experimental work on grain-grain collisions uses porous silicates.  If ices are stickier, growth beyond mm-sizes is possible. In the low pressure of disks, ices sublimate; there is no liquid available to make the equivalent of wet snow.  Saturn's rings are an excellent laboratory to explore the outcomes of gentle, $\sim$ mm s$^{-1}$, collisions between ices \citep{youdin:2002}.  Here, sticking forces are constrained by their inability to overcome tidal shear and produce growth beyond $\sim 5$ m objects.  Terrestrial experiments on low temperature ices suggest that cm-sized frosty objects stick at collision speeds below $\sim 0.1 ~{\rm m~s}^{-1}$ \citep{sup97}. While possibly a crucial ingredient, this limit appears insufficient to allow icy surfaces to bridge the meter-size barrier.

\subsection{Gravitational Collapse of Solids into Planetesimals}\label{sec:ptmlGI}

Self-gravity provides a qualitatively different route to the formation of planetesimals.  Instead of bottom-up growth, the gravitational instability (GI) hypothesis of \citet{saf1969} and \citet{gold1973} offers a top-down approach.  In this theory, a sea of small solids collapses coherently into a gravitationally bound planetesimal.  This collapse does not rely on sticking forces, proceeds faster than radial drift, and bypasses the meter-size barrier.

The gravitational collapse hypothesis encounters several theoretical difficulties.  The crucial issue is the ability of turbulent gas to prevent collapse \citep{weid1995}.  Until recently, these theoretical obstacles seemed insurmountable.  Progress in coupled particle-gas dynamics has led to a revival \citep{youdin:2002,jkh06,nature07, chs08,you11a}.  Some of these mechanisms use aerodynamic concentration as the initial concentration mechanism (S\ref{sec:concentrate}) but all eventually rely on self-gravity for the final collapse to solid densities.

We focus in this subsection on ``pure" gravitational collapse from a relatively smooth background.  Although separating these processes from aerodynamic concentration is artificial, this historical approach allows us to isolate the main issues of each mechanism.  We first discuss the standard model of gravitational collapse of a disk of solids, which has many similarities to gravitational instabilities in a gas disk (\S5). We then briefly describe how gas drag changes this standard picture, a research area where progress is still being made.

The simplest criterion for gravitational collapse requires self-gravity to overcome the tidal distortion of the central star.  This condition is met when the particle density exceeds the Roche limit,
\begin{equation} \label{eq:Roche}
\rho\ps > \rho_{\rm R} \simeq 0.6 {M_\star \over R^3} \simeq 130 {\sqrt{m_\star} \over F}\left(R \over {\rm AU}\right)^{-3/14} \rho\gs \, ,
\end{equation} 
where $m_\star = M_\star/M_\odot$ and $F$ is the mass enhancement factor for the MMSN from \S2.
\citet{sek83} derives this result for the case of a disk midplane with solids perfectly coupled to an incompressible gas, making use of the powerful formalism of \citet{gold65}.  The relation of this critical density to the \citet{toom64} $Q$ criterion for GI is discussed (in the context of planetesimals) by \citet{chiangyoudin:2010} and \citet{you11a}.

For a given particle surface density, \Eq{eq:Roche} implies that planetesimals form with a mass $M_{\rm ptml} \sim \varSigma\ps^3/\rho_{\rm R}^2$.  After contraction to solid densities, the planetesimal size is
\begin{equation} 
R_{\rm ptml} \sim {\varSigma\ps \over \rho_\bullet^{1/3} \rho_{\rm R}^{2/3}} \approx 5 {F Z_{\rm rel} \over m_\star} \sqrt{R \over {\rm AU}}~{\rm km}\, .
\end{equation} 
Though the current relevance is not so clear, this kind of estimate played a key role in defining the canonical planetesimal size to be near a kilometer.

To satisfy the density criterion of \Eq{eq:Roche}, solids must settle vertically to a midplane layer with thickness $H_{\rm R} = \varSigma\ps/\rho_{\rm R}$.  Even the faintest whiff of turbulence probably produces a much thicker layer.  Although the disk midplane could be a ``dead zone" devoid of magnetized turbulence \citep{gammie96}, interactions among particles can drive enough turbulence to halt settling \citep{houches10}.

Vertical shear instabilities usually prevent the sedimentation of small particles into a layer thinner than $H_\eta \simeq \eta R$ \citep{weidenschilling:1980,youdin:2002}.   As particle inertia in the midplane increases from sedimentation, solids begin to drag the midplane gas towards the full Keplerian speed.  As in the Kelvin-Helmholz instability, the vertical shear with the overlying particle-poor gas drives overturning. With $H_\eta/H_{\rm R} \sim 200$, GI seems ruled out.

The revival of the GI hypothesis requires abandoning two faulty assumptions.  The surface density of solids can increase above MMSN --- or any initial --- values.  The evolution of solid and gas components decouples due to drift motions \citep{sv96}.  The radial drift of small solids from the outer disk generically leads to ``particle pileups," a snowplow effect that increases the surface density in the inner disk \citep{youdin:2002,yc04}.  The local concentration mechanisms discussed in \S\ref{sec:concentrate} can be even more powerful.

The critical Roche density is also too stringent.  Planetesimal formation can be triggered when $\rho\ps \gtrsim \rho\gs$, a criterion about a hundred times less severe than the Roche limit in \Eq{eq:Roche}.  Several interesting effects arise when the particle density approaches the gas density.  Vertical shear instabilities lose their ability to overturn a layer that is so heavy \citep{sek98, youdin:2002}.  When disk rotation is included, the case for particle inertia halting vertical overturning is less clear \citep{go05, lee10b}.  However, when perfect coupling is relaxed, and streaming instabilities appear, the relevance of $\rho\ps \gtrsim \rho\gs$ reemerges as the threshold for strong clumping, as described below.

When gas drag is present,  the Roche density is not the relevant criteria for GI. \citet{war76, war00} investigated a dissipative mode of GI that has no formal stability threshold.  Collapse always occurs in principle, but it becomes slower and spans a larger radial extent for small particles.  \citet{you11a} included radial turbulent diffusion, and showed that radial spreading --- not vertical stirring --- is the dominant stabilizing influence for dissipative GI.   When vertical stirring is accounted for, it turns out that $\rho\ps \gtrsim \rho\gs$ is typically required for dissipative GI to proceed faster than radial drift.  The result that dissipative GI depends so simply on particle inertia is mostly a numerical coincidence and relies on the fact that $Q\gs \sqrt{2\eta} \sim 1$ in the MMSN, see eq.\ (55) of \citet{you11a}.  The important point is that the relevance of particle inertia --- specifically the $\rho\ps \gtrsim \rho\gs$ criterion --- has been established for a range of mechanisms.

Although this lesser degree of particle settling is substantial, it may require a local enrichment of the disk metallicity, $Z = \varSigma\ps/\varSigma\gs$.  If the particle scale-height is set by particle-driven turbulence to $H_\eta$, then the particle density exceeds the gas density if
\begin{equation} \label{eq:Zcrit}
Z > {\eta R \over \sqrt{2 \pi} H\gs} \simeq 0.014 {1 \over \sqrt{m_\star}} \left(R \over {\rm AU}\right)^{2/7} \, ,
\end{equation} 
again with $m_\star = M_\star/M_\odot$.  The near agreement with Solar abundances is remarkable and could be related to the correlation of giant planets with host star metallicity \citep{youdin:2002}.  Assuming that the stellar photospheres reflect the abundance of solids in the disk \citep{fv05}, the early formation of planetesimals could be a crucial factor in the formation of gas giants \citep{jym09}.

Though poorly constrained, the role of external (not particle-driven) midplane turbulence may be interesting.  For small solids, constraints on the level of turbulence that allows settling to $\rho\ps \gtrsim \rho\gs$ are quite stringent.  
Using \Eq{eq:halpha}, sedimentation to $\rho\ps > \rho\gs$ requires that midplane turbulence satisfy
\begin{equation} 
\alpha_D \lesssim 2 \pi Z^2 \taus \approx 10^{-4} Z_{\rm rel}^2 {s \over {\rm cm}} \left(R \over 10~{\rm AU}\right)^{3/2} \, .
\end{equation} 
 Thus when trying to form planetesimals via GI it helps to have some combination of weak turbulence, particle growth and enriched metallicity ($\varSigma\ps/\varSigma\gs$).  These requirements become more stringent towards the inner disk \citep{you11a}.
 
Thus even in the GI hypothesis, particle growth by coagulation plays a crucial role.  Particles must grow until they decouple from the gas.  Provided this growth occurs, GI --- likely aided by other concentration mechanisms --- provides a plausible way past the meter-size barrier. 

\subsection{Aerodynamic Particle Concentration}\label{sec:concentrate}
We now consider aerodynamic processes that can concentrate particles even when self-gravity is negligible.  Many of these processes rely on the presence of turbulence in the disk.  This connection raises a general question: does turbulence help or hinder planetesimal formation?  By stirring particles, turbulence increases their collision speeds which can lead to more destructive collisions.  Furthermore, the diffusive effects of turbulence oppose particle settling and concentration.  On the other hand, turbulence can concentrate particles in a variety of ways.  Which tendency wins depends on details, notably particle size.  Since small solids with $\taus \ll 1$ drift and settle slowly, they require much weaker turbulence to participate in aerodynamic concentration.

Localized pressure maxima are very powerful particle traps.  When the pressure bump takes the form of an axisymmetric ring, the trap is very effective  \citep{whipple1972}.   Solids migrate into these rings and accumulate at the center where they encounter no headwind.  The MRI naturally produces axisymmetric pressure bumps, via the generation of zonal flows that are somewhat analogous to the surface winds of Jupiter \citep{jyk09, fs09}.   The relevance of MRI-induced pressure maxima is  subject to two caveats:  turbulent stirring associated with MRI may lead to destructive collisions and the disk midplane may be insufficiently ionized for the MRI to operate \citep{tcs10}. 

Non-axisymmetric pressure maxima can also trap particles.  When the disk is young and massive, spiral arms in the gas probably provide an important source of turbulence \citep{ric06}.  However, this phase of disk evolution may be too turbulent and/or brief for significant planetesimal formation.  Isolated pressure maxima take the form of anticyclonic vortices \citep{chav00}.  Vortices are embedded in, and thus flow with, the sub-Keperian gas \citep{houches10}.  Although the vortex center is not a stable point for particle concentration, a point upstream (in the direction of orbital motion) is.  The implications for vortex size is discussed in \citet{houches10}.  The formation and survival of vortices is a topic of ongoing research \citep{lith09}.

We have so far focused on particle concentration over many orbits where disk rotation and Coriolis forces play a central role.  Small turbulent eddies have short turnover times, $t_{\rm eddy} \ll 1/\varOmega$, and are unaffected by rotation.  In this regime pressure maxima occur not at the centers of anti-cyclonic vortices, but between vortices of either sign.  The concentration of heavy particles in these regions of low enstrophy (vorticity-squared) was first described in the fluid dynamics community \citep{max87}.

\citet{cuzzi:2001} applied small-scale concentration to protoplanetary disks.  They showed that $\sim$ 1~mm solids --- specifically the chondrules that are discussed in \S\ref{sec:ptmlobs} --- can concentrate at the ``inner" or dissipation scale of turbulence.  These are the smallest eddies that have the shortest turnover time $t_{\rm i}$.   Particles with a matching stopping time, $\ts, \sim t_{\rm i} \sim 30$~s, are preferentially flung from the eddies and concentrated between them.  Chondrules can plausibly satisfy this condition.  The ability to concentrate such small particles makes this mechanism unique.  The relevance of such brief concentrations is unclear.  The characteristic mass involved is also quite small, at most that of a 10 cm rock \citep{chiangyoudin:2010}.

To overcome these issues, \citet{chs08} developed a model that concentrates chondrules on larger scales that contain enough mass to form $\sim 100$ km-planetesimals.  This model involves a somewhat speculative extrapolation.   In particular, it assumes that all scales of a turbulent cascade contribute equally to the concentration of chondrule-sized particles.  This assumption is a significant deviation from the original mechanism that requires eddy and stopping times to match.  See \citet{chiangyoudin:2010} for further discussion, which concludes that more study of this intriguing mechanism is required.

Clearly particle concentration mechanisms are fraught with uncertainties in the detailed dynamical behavior of gas in protoplanetary disks.  Some --- certainly not all --- of these uncertainties are overcome by the realization that particles can cause their own concentration by collectively altering the gas dynamics \citep{gp00}.  In the streaming instability of \citet{yg05}, particle concentrations arise spontaneously from radial drift motions.  As described in \S\ref{sec:radialdrift} these drift motions are an inevitable consequence of pressure support in disks.  The linear growth of streaming instabilities is strongest for $\rho\ps > \rho\gs$, because particle inertia must be large for drag feedback to influence gas motions.  When $\rho\ps < \rho\gs$, growth is fastest for $\taus \approx 1$, when drift speeds are fastest \citep{yj07}.  While streaming instabilities involve complex dynamics --- 3D motions of both the gas and solid components --- simplified toy models \citep{gp00, chiangyoudin:2010} and considerations of geostrophic balance \citep{jbl11} help explain how particle density perturbations self-reinforce.

Numerical simulations show that the non-linear clumping from the streaming instability can be quite strong \citep{jy07,jym09,bal09,bs10a}.  Particle densities $\gtrsim 10^3 \rho\gs$ are achieved in the absence of self-gravity, and clumping tends to increase with numerical resolution.  The conditions for strong clumping are similar to those giving rapid linear growth: partial decoupling, $\taus \gtrsim 0.1$, and large particle inertia $\rho\ps \gtrsim 0.2 \rho\gs$. 

When vertical stratification is included,  the midplane particle density evolves consistently due to settling and stirring by both streaming and vertical shearing instabilities.  In these simulations, there is a critical disk metallicity for particle clumping which is slightly super-Solar \citep{jym09, bs10a}, consistent with \Eq{eq:Zcrit}.  This metallicity varies with the radial pressure gradient, $\eta$; smaller gradients promote clumping \citep{nature07, bs10b}.  

In \S\ref{sec:ptmlGI}, we noted that GI depends on particle growth by coagulation.  Since particle sedimentation to $\rho\ps \gtrsim \rho\gs$ is a crucial prerequisite,  particle growth remains essential when the streaming instability provides the initial particle concentration.    However growth need not result in a single particle size, or a very narrow size distribution.  Though the smallest solids participate less in clumping by streaming instabilities, including a dispersion in particle sizes does not prevent strong clumping \citep{nature07}.

The particle concentration produced by the streaming instability is more than sufficient to trigger gravitational collapse.  \citet{nature07} formed gravitationally bound objects equivalent to $\sim 500$ km planetesimals within only a few orbits of initial collapse.  More recent simulations suggest the formation of lower mass objects with equivalent $\sim 100$ --- $200$ km radii \citep{jym09}.  The crucial differences are the inclusion of the MRI in the earlier study and smaller particle sizes in the second.  A more thorough investigation of parameter space, combined with resolution studies, is required.  These results exceed the standard estimate of km-sized planetesimals because gravitational collapse occurs not from a smooth background, but from aerodynamically concentrated clumps.

\subsection{Observational Constraints on Planetesimal Formation}\label{sec:ptmlobs}

We now discuss how observations constrain dynamical theories of planetesimal formation.  The Solar System provides the most detailed information on planetesimals, and allows comparison between the inner asteroidal reservoir and the Kuiper belt objects and comets of the outer Solar System.  The crucial issue is to what extent today's planetesimals reveal the clues of their formation, especially after $\sim 4$ Gyr of dynamical, collisional and thermal evolution.

Primitive, undifferentiated meteorites give us a hands-on view of the composition of planetesimals.  The most common of these are the aptly named ``ordinary chondrites."  With filling factors up to 90\%, they are primarily composed of 0.1-1~mm chondrules.  Chondrules are glassy spheres, poetically referred to as ``fiery drops of rain"  \citep{sorby1863}.  The origin of chondrules --- in particular their source of heating ---  is debated and beyond our scope, see \citet{hewins96}.  The prevalence of chondrules in ordinary chondrites strongly motivates further investigation of the mechanisms that could concentrate solids this small \citep{cuzzi:2001,chs08}.

Despite this attractive conclusion, chondrules may not be the universal building blocks of all planetesimals.  Because their abundances most closely match Solar, the CI class of chondrites is considered the most primitive \citep{lodders2003}.  Yet CI chondrites contain no chondrules.  It is also unclear whether chondrules were present in the first generation of planetesimals. Most chondrules formed at least 1.5 Myr and up to 4 Myr after the rarer CAIs (Calcuim-Aluminum Inclusions; \citealp{con07, krot07}).  Thus, planetesimals probably formed before major chondrule forming events, especially the planetesimals that formed the cores of Jupiter and Saturn.  Since planetesimals that form early will trap more radioactive heat and differentiate, it seems likely that the undifferentiated chondrites represent a later phase of planetesimal formation \citep{tk05}.  The relation between chondrules, meteorites and planetesimal formation continues to be the focus of intense interdisciplinary research.

Planetesimals that remain in the asteroid belt can also provide clues to their formation.  The radially banded zonation of different spectral classes of asteroids is well known \citep{gradie:1982}.  This observation suggests separate formation epochs, with each event creating a ``clan" of chemically and spectrally similar planetesimals.  \citet{you11a} proposed large-scale, drag mediated GI as the cause of these events.

The size distribution of objects within planetesimal belts provides other clues to their formation.  Breaks in the size distribution --- i.e.\ changes in its powerlaw slope --- point to shifts in formation and/or erosion processes.  The asteroid belt has a break near a radius $\sim 50$~km.  \citet{bornbig} argue that the asteroids with radii $\gtrsim 50 $~km reflect their initial sizes.  Specifically they assert that the largest asteroids have undergone minimal collisional evolution and could not have formed via collisional growth of smaller planetesimals.  Since most of the mass is contained in the largest asteroids, their model plausibly produces the numerous small objects below the break via collisional disruption.
That interpretation places GI as the preferred formation mechanism.  By including streaming instabilities, the simulations of  \citet{nature07} predicted that large initial sizes were possible.   Conclusively proving that a size distribution is unobtainable by collisional growth is rather difficult. \citet{stu10} contends that collisional growth of asteroids can be accomplished starting with 0.1~km planetesimals --- which themselves presumably grew by coagulation past the meter-size barrier.

Curiously, the Kuiper belt also has break in its size distribution at $\sim 50$ km radii \citep{bernstein:tnodist}.  This break is not measured directly; a combination of an observed luminosity distribution and an estimate of the albedo yields the distribution of radii \citep{PetitSSBN}. Ongoing surveys of the Kuiper belt seek to provide constraints on the size distribution for the various components of the Kuiper belt.

Understanding the origin of the break requires a model for KBO formation and dynamical interactions with gas giants. Reproducing the observed size distribution with collisional growth models requires an initially massive Kuiper belt followed by dynamical depletion; a break occurs when depletion excites erosive collisions among KBOs with radii below the break \citep{kb2004c}.  The break radius depends on excitation; more (less) excitation by more (less) massive gas giants yields a break at larger (smaller) radii. Matching the location of the break and the apparent slope of the KBO size distribution below the break requires numerical calculations with growth and depletion, which are an active area of research \citep{kbod2008,morbiSSBN}.

The relative roles of collisional and dynamical depletion affect the interpretation of the KBO size break.  \citet{pan:2005} argue that the break is not primordial but due to ongoing collisional erosion that continues to push the break to larger sizes.  However \citet{nesv11} claim that this collisional history is ruled out on two grounds.  First the collisional strengths required for such destruction are too weak.  Second, such an intense collisional bombardment would destroy the observed Kuiper belt binaries.  

The observed binary fraction in the cold, classical Kuiper belt is $\gtrsim 20\%$ \citep{nollSSBN}.   The colors of the two components of Kuiper belt binaries are nearly identical, a fact interpreted as representing a common chemical composition \citep{ben09}.  This observation provides the most compelling support for the GI hypothesis in the outer Solar System, or perhaps anywhere.   Gravitational collapse can naturally produce binary planetesimals as a consequence of angular momentum conservation during the contraction of a swarm of small solids \citep{nyr10}.  Binaries --- and higher-order multiples ---  formed this way should have the same chemical composition since they formed from the same well-mixed clump of small solids.  While mechanisms for the dynamical capture of KBO binaries are well developed \citep{GolLit02, nollSSBN}, these models do not obviously explain matching colors.  Moreover the physical conditions require for capture, make the collisional survival of these binaries questionable \citep{nesv11}, especially for wide binaries \citep{ParKav12}.

Outside the solar system, exoplanets and debris disks inform the prevalence and consequences of planetesimal formation.  The higher incidence of giant planets around stars with super-Solar metallicities (discussed in \S\ref{sec:exo}) might be tied to planetesimal formation.  As shown in \Eq{eq:Zcrit}, this connection is strongly suggested by the super-Solar \emph{disk} metallicity threshold for strong clumping by streaming instabilities.  Since the disk metallicity can increase over time \citep{yc04}, this threshold does not imply that lower metallicity stars can never form planetesimals. 

Indeed, the streaming instability/GI model explains why  lower metallicity and lower mass stars should form less massive planets \citep{jym09}.  Either directly or by the passage of time, enriching the disk metallicity involves the loss of gas. Thus the initially low metallicity systems that require enrichment are less likely to form giant planets.  This conclusion is especially true in the lower mass disks thought to surround lower mass stars.  These general trends are revealed by radial velocity surveys \citep{sousa2008,john2010}.   The \emph{Kepler} transit survey will test these trends, since it is finding striking numbers of small, short period planets \citep{how2011, you11b}.  Characterization of the \emph{Kepler} stars will thus powerfully constrain planetesimal formation models.

\section{PLANETESIMALS TO PLANETS}
\label{sec:planets}

Once planetesimals become larger than a few kilometers --- potentially they are born much larger as discussed above --- gravitationally focused collisions dominate growth into protoplanets.    The size when ``planetesimals" become ``protoplanets" is vague.  Although we use the terms interchangeably, $\sim$1000~km is a useful threshold.  Depending on location and gas temperature, $\sim$1000~km protoplanets are the smallest planets capable of binding disk gas into an atmosphere.

We describe the accretion of solid protoplanets in \S\ref{sec:solidprotoplanets}.  We start by discussing the processes that operate in a gas-free disk, including gravitationally focused collisions (\S\ref{sec:growth}), velocity excitation (\S\ref{sec:stirring}) and collisional fragmentation (\S\ref{sec:frag}).  Wh then describe planetesimal interactions with the 
gaseous disk (\S\ref{sec:gaseffects}).
Section \ref{sec:TPsims} describes simulations of terrestrial planet formation that put these ingredients 
together. The accretion of a gaseous atmosphere (\S\ref{sec:atmos}) affects planetesimal accretion (\S\ref{sec:atmos-planetesimal})  
and transforms a planetary core into a gas giant (\S\ref{sec:atmos-CI}, \S\ref{sec:atmos-accr}). 
We discuss numerical simulations combining the growth of giant planet cores and 
atmospheres  in \S\ref{sec:atmos-GGsims}.  Finally, a young, 
massive gas disk might fragment directly into a gas giant or a brown dwarf.  Section \ref{sec:planets-bdgg} describes this formation channel and whether it might explain some exoplanets, especially massive giants at large radial distances.

\subsection{Growth of Solid Protoplanets}
\label{sec:solidprotoplanets}

Unlike planetesimal formation, it is easy to understand why planetesimals grow into larger protoplanets, even if the details are complicated.  For the largest  planetesimal in any region of the disk, collisions essentially always result in growth.  Planetesimal velocities cannot be locally excited above the escape speed of the largest protoplanet.  Consequently, the kinetic energy of collisions does not exceed the gravitational potential at the surface of the largest protoplanet.    Collisions dissipate a fraction, sometime quite large, of the impact kinetic energy.    Even if the impacting planetesimal shatters, growth is assured. 

Collisions among smaller planetesimals, however, often lead to erosion or catastrophic fragmentation.  When an external,
massive perturber stirs a belt of planetesimals, planetesimals collide at velocities larger than their escape velocity.
These high velocity collisions tend to erode or completely shatter planetesimals.  The dust and changes in the planetesimal size distribution that result from these collisions are relevant for debris disks and for asteroids and Kuiper Belt objects.

Unlike the planetesimal formation phase, aerodynamic drag no longer plays a starring role in protoplanet growth.  However drag can still help regulate planetesimal velocities.  The accretion of atmospheres (see \S\ref{sec:atmos}) also affects planetesimal capture.

Deriving the precise evolution of a swarm of planetesimals is a complex numerical problem being attacked from several angles (\S\ref{sec:TPsims}).  However, we can develop a reasonably accurate picture of the evolution with the ``two groups approximation" reviewed in greater detail by \citet{gold2004}.  This approximation considers interactions between small, low mass planetesimals with mass $m_s$ and larger, more massive planetesimals with mass $m_l$.  
The planetesimal masses $m_{s,l} = 4 \pi r_{s,l}^3 \rho_\bullet / 3$ are related to their radii $r_{s,l}$ and internal mass density, which we fix at  $\rho_\bullet = 2 \g \cm^{-3}$ unless stated otherwise.  Neighboring planetesimals have similar  semimajor axes, $a$, and orbital frequencies, $\Omega$.  Though detailed treatments need not make this approximation, we equate orbital eccentricities $e_{s,l}$ and inclinations (in radians), but allow $e_s$ and $e_l$ to differ.  The random velocities relative
to a circular orbit are thus $v_{s,l} \approx e_{s,l} \Omega a$ and the vertical scale heights of the planetesimal disks are $H_{s,l} \approx v_{s,l}/\varOmega$.  When the nature of the planetesimal is unspecified, we drop the $s$ and $l$ subscripts.

\subsubsection{Basic Length and Velocity Scales}
A useful scale for studying interactions of planetesimals and protoplanets is the \citet{hill1878} radius
\begin{equation}
R_H = \left ( \frac{m}{3 M_\star} \right )^{1/3} a ~ .
\label{eq:rhill}
\end{equation}
Planetesimals separated by $ \lesssim R_H$ are within the Hill sphere where their mutual gravitational attraction dominates the tidal gravity from the central star.   While a mutual Hill radius can be defined, in practice it suffices to consider the more massive planetesimal.  

The size of a planetesimal in Hill units defines the parameter
\begin{equation}
\psi \equiv {r \over R_H} =  \left ( {3 \rho_{\star} \over \rho_\bullet} \right )^{1/3} {R_{\star} \over a} \simeq 6 \times 10^{-3} \left ( {\mstar \over \msun} \right )^{1/3} \left ( \frac{\rm AU}{a} \right ) ~ ,
\label{eq:psi-star}
\end{equation}
where $\rho_{\star}$ is the mean mass density of the central star.  Since $3 \rho_{\star} / \rho_\bullet \sim 1$, the parameter $\psi \sim R_\star/a$ is roughly the angular size of the central star as observed from the planetesimal.  The smallness of $\psi$ represents the fact that physical collisions are rare compared to gravitational scattering.

At the Hill radius, the orbital speed about the protoplanet is the Hill velocity
\begin{equation}
v_H = \Omega \rhill = \left ( \frac{m}{3 M_\star} \right )^{1/3} v_K  ~ .
\label{eq:vhill}
\end{equation}
 When planetesimal random velocities exceed $v_H$, two-body encounters are ``dispersion-dominated," negligibly affected by orbital shear.  Random speeds below $v_H$ cause ``shear-dominated" encounters that involve (restricted) three-body dynamics.  We describe below how the Hill velocity divides different accretion regimes.

The outcome of a shear dominated encounter between planetesimals depends on the difference in semimajor axes $\delta R$, relative to the Hill radius \citep{petit1986}. 
When $\delta R \lesssim$ 1--2 \rhill, the two planetesimals are deflected on a horseshoe orbit.  More distant encounters with 
$\delta R \gtrsim 2 \sqrt{3} \rhill$ result in  small angle scattering. For intermediate separations, 1--2 \rhill\ $\lesssim \delta r \lesssim 2 \sqrt{3} \rhill$, 
planetesimals enter the Hill sphere, experience chaotic 
deflections, and (if no collision occurs) leave the Hill sphere with typical relative velocity \vhill. 

In Hill units, the escape speed from the surface of a protoplanet, $v_{esc} = [2 G m /r]^{1/2}$, is $v_{esc} \sim v_H/\psi^{1/2}$.  Planetesimal velocities can be gravitationally excited up to the escape speed of the largest protoplanet.  To estimate when a massive protoplanet might eject nearby planetesimals (or protoplanets), we compare the escape velocity of the protoplanet to the orbital escape velocity, $v_{esc,\star} \approx \sqrt{2} v_K$.  When
\begin{equation}
{v_{esc} \over v_{esc,\star}} \approx 0.15 \left ( {m \over \mearth} \right )^{1/3} \left ( {a \over {\rm AU}} \right )^{1/2} \left ( {\rstar \over \rsun} \right )^{-1/2} ~ 
\label{eq:escvel-ratio}
\end{equation}
exceeds unity, a planet of mass $m$ can eject other nearby protoplanets.  Terrestrial planets like Earth are too low mass and too close to the Sun to eject objects.  The four Solar System giants can all eject planetesimals; Jupiter is the most efficient at ejecting comets (and spacecraft) from the Solar  System \citep{fernandez:1984}.  Aside from collisional grinding, \Eq{eq:escvel-ratio} implies that planetesimal formation is more efficient closer to a star.  

The concepts of the Hill sphere and the Roche lobe are identical, though often used in different contexts.  Both describe the region where the gravity of an object exceeds the tidal perturbation from its companion.  Formally, both volumes are defined by the critical equipotential containing the $L_1$ and $L_2$ Lagrange points.  The Roche lobe is more distorted than a sphere when it describes binary stars that are similar in mass.   The Roche radius or Roche limit  describes the distance from a primary object at which the secondary becomes tidally disrupted and might form planetary rings.   Aside from order unity corrections due to fluid effects or internal strength, the concept of individual vs.\ tidal gravity is again identical.  To summarize, a secondary is at the Roche limit from the primary when it fills its own Hill sphere (or Roche Lobe).  For an ensemble of very small planetesimals trying to become a much larger planetesimal, the Roche limit sets the critical density for gravitational collapse (eq. [\ref{eq:Roche}]).

\subsubsection{Gravitationally Focused Collisions}\label{sec:growth}

In this section, we describe the growth rates of large protoplanets (subscripted by $l$) accreting either large protoplanets or smaller planetesimals (unsubscripted).  Gravitational focusing is the most important aspect of growth. Smaller random velocities for accreted planetesimals yield larger gravitational focusing factors and shorter growth times.  We defer to later sections the self consistent calculation of planetesimal velocities and assume the standard case, $v_l < v_s$. \citet{green1984}, \citet{weth1993}, \citet{kb2008}, and references in each paper describe more detailed expressions for growth rates.

We begin with the dispersion-dominated regime where $v > v_{H,l}$.  The mass accretion rate results from the usual isotropic expression as $\dot{m}_l = m n \sigma v$.  Adopting the surface mass density of planetesimals, $\varSigma$, the number density of
planetesimals, $n$, is
\begin{equation} \label{eq:nvsigma}
m n v = \varSigma \varOmega ~.
\end{equation} 
The cross section
\begin{equation}
\sigma = \pi (r_l + r)^2 f_{G,disp}  ~ ,
\label{eq:cross}
\end{equation}
is the product of the geometric area and the gravitational focusing factor $f_{G,disp}$.  If the velocity of the incoming planetesimal at infinity is $v > v_{esc,l}$, there is no gravitational focusing and $f_{G,disp} = 1$.  When $v \ll v_{esc,l}$, the speed on impact is roughly $v_{esc,l}$.  Angular momentum conservation during a two body encounter sets 
the impact parameter for a grazing collision as $r_l v_{esc,l} / v$.   This expression yields $f_{G,disp} \approx (v_{esc,l}/v)^2$.  Including energy conservation gives both cases simultaneously as
\begin{equation}
\label{eq:fg-disp}
f_{G,disp} = 1 +  \beta (v_{esc,l} / v)^2 ~ ,
\end{equation}
where $\beta = 1$ for a pure two body interaction and $\beta \approx 2.7$ accounts for anisotropic effects introduced by orbital dynamics \citep{green1990,spaute1991,weth1993}.

Putting these results together and ignoring order unity coefficients gives the dispersion-dominated growth timescale $m_l/\dot{m}_l$ as
\begin{equation}
\label{eq:t-disp}
t_{disp} \approx { \rho_\bullet r_l \over \Sigma \varOmega f_{G,disp} } ~ .
\end{equation}
This result is just the geometric collision time over the focusing factor.

For shear-dominated encounters with $v < v_{H,l}$, collision rates are affected by chaotic trajectories inside the Hill sphere \citep{green1991,DonTre93}.  In this regime, planetesimal disks are thinner than the Hill radius, $H \sim v/\Omega < R_{H,l}$.  Thus planetesimals enter the Hill sphere at the 2D mass accretion rate, $\dot{m}_H \sim \varSigma R_{H,l}^2 \varOmega$.   The probability, $P$, of a collision within the Hill sphere has two cases.  Both use the maximum impact parameter for gravitationally focused collisions $b_{\rm max} \sim r_l v_{esc,l}/v_{H,l} \sim \psi^{1/2} R_{H,l}$.  
If the scale height of the disk is (relatively) thick with $H \sim v/\varOmega > b_{\rm max}$, the collision probability  $P \sim b_{\rm max}^2/(R_{H,l} H)$ is the ratio of the collision cross-section to the area of the accreting disk of planetesimals.  For a thinner planetesimal disk, $P \sim b_{\rm max}/R_{H,l}$ is the ratio of the impact parameter to the Hill radius. 

Combining the mass flow rate through the Hill sphere  with both limits of the collision probability yields the shear dominated growth timescale, $m_l/(P\cdot \dot{m}_H)$, as  
\begin{equation}
t_{shear} \sim { \rho_\bullet r_l \over \Sigma \Omega f_{G,shear}} ~ .
\label{eq:t-shear}
\end{equation}
This timescale is again expressed as the product of the geometric collision time and a gravitational focusing factor
\begin{equation}
\label{eq:fg-shear}
f_{G,shear} \sim   \left(\psi {v \over v_{H,l}} + \psi^{3/2}\right)^{-1} ~.
\end{equation}
Thus for $v < \psi^{1/2} v_{H,l}$,  gravitational focusing reaches its maximum value of $f_{G} \sim \psi^{-3/2}$, resulting in the fastest possible growth rate.  Because inclination excitation is weak in the shear dominated regime, this fastest thin-disk accretion rate likely applies for all large protoplanets with $v_l < v_{H,l}$ \citep{gold2004}.  Aside from this issue of anistropic velocities, the dispersion and shear dominated focusing factors match at $v \sim v_{H,l}$ with $f_G \sim 1/\psi$.

For numerical estimates of growth timescales we consider three cases.  For the slow case we take $v > v_{esc,l}$ and no gravitational focusing.  For the intermediate case, we identify $v \approx v_{H,l} \approx \psi^{1/2} v_{esc,l}$ as the transition between shear- and dispersion-dominated.  The fast case considers the maximum focusing factor $f_G \approx \psi^{-3/2}$ appropriate for $v \lesssim \psi^{1/2} v_{H,l} \approx \psi v_{esc,l}$ (and possibly for higher speeds when planetesimal $i \ll e$).  Using \Eq{eq_sigmap} for the surface density of planetesimals, the growth times become
\begin{subeqnarray}
t_{\rm slow} \approx  { \rho_\bullet r_l \over \Sigma \varOmega}  &\approx& 10^7 \left ( \frac{m_l}{\mearth}\right )^{1/3}  \left ( \frac{1}{F \Zr} \right) \left ( { a \over {\rm AU} } \right )^3 {\rm yr} ~ .
\label{eq:t-slow}\\
t_{\rm int} \approx   { \rho_\bullet r_l \over \Sigma \varOmega}\psi   &\approx& 5\times 10^4 \left ( \frac{m_l}{\mearth}\right )^{1/3} \left ( \frac{1}{F \Zr} \right) \left ( { a \over {\rm AU} } \right )^2 {\rm yr} ~ .
\label{eq:t-int}\\
t_{\rm fast} \approx  { \rho_\bullet r_l \over \Sigma \varOmega}  \psi^{3/2}  &\approx& 4000 \left( \frac{m_l}{\mearth}\right )^{1/3}  \left ( \frac{1}{F \Zr} \right) \left ( { a \over {\rm AU} } \right )^{3/2} {\rm yr} ~ .
\label{eq:t-fast}
\end{subeqnarray} 
These mass doubling times increase with protoplanet mass.  Thus if gravitational focusing stays fixed or decreases (far from a certainty) these estimates also give the total accumulation time.  These expressions omit the dependence on stellar mass and planet density for clarity but the growth times scale $\propto \rho_\bullet^{2/3}/\mstar^{1/2}$, $\rho_\bullet^{1/3}/\mstar^{1/6}$, and $\rho_\bullet^{1/6}\mstar^{0}$ for the three cases, respectively.  Though stellar mass is not a major dynamical effect,  it could correlate with disk mass or metallicity (here meaning planetesimal to gas ratio), normalized above by $F$ and $\Zr$, respectively.  Higher density protoplanets have smaller cross-sections and grow more slowly, but this effect becomes much less significant as gravitational focusing increases.

Gravitational focusing dramatically speeds up the growth of protoplanets, especially in the outer disk.  Without focusing, planets accumulate in $t_{\rm slow} \sim$ tens of Myr inside a few AU and more than 1 Gyr outside 5 AU.  While a long growth time for terrestrial planets is acceptable, gas giants must form within a few Myr. Thus, formation of giant planet cores in the outer disk requires strong focusing, when the growth time for a $10~\mearth$ core at 50~AU, $t_{\rm fast} \sim 3$ Myr, is close to the lifetime of most gas disks.  If strong focusing occurs, the formation of distant gas giants depends on the ability of cores to accrete gas \citep{Raf11}.

Protoplanet accretion also depends on how the velocity distribution evolves.  In dispersion-dominated 
gravitational focusing, the growth time $t_{disp} \propto 1/r_l$; larger planetesimals grow faster than 
smaller ones.  Although it is not the fastest regime, this ``runaway" growth requires that gravitational 
focusing factors remain in the dispersion-dominated regime.  
With $v_{esc}/v_H \sim \psi^{-1/2} \propto a^{1/2}$, runaway growth persists longer in the outer disk 
\citep{green1990}. 

When the largest protopanets enter the shear-dominated regime, runaway growth ends. 
For the thick disk case, $t_{shear} \propto r_l^0$ is independent of size.  With either no focusing, 
$t_{\rm slow} \propto r_l$, or the fastest (thin disk) shear-dominated accretion $t_{\rm fast} \propto r_l$, 
smaller protoplanets grow faster and can catch up to the larger ones. In this ``oligarchic" growth, many 
oligarchs compete to accrete small planetesimals, leading to an ensemble of oligarchs throughout the disk.

As the largest protoplanets grow, they try to accrete all solid material in their vicinity. Two planets 
on circular orbits separated by a little more than 2$\sqrt{3}$ \rhill\ are stable \citep{Gla93}. However, a fairly 
stable system with more planets requires larger separations, $\sim B \rhill$ with $B$ = 7--10 
\citep{liss1987,kok1998}. Planets that accrete all material within $B \rhill$ are ``isolated." 
Setting $m_{iso} = 2 \pi \Sigma a B \rhill$ leads to the isolation mass,
\begin{equation}
\label{eq:miso}
m_{iso} = { (2 \pi B \Sigma)^{3/2} \over (3 \mstar)^{1/2} } a^3 \approx 0.08 \left ( {B \over 7} \right )^{3/2} \left ( { F \Zr \over 0.33 } \right )^{3/2} \left ( { \mstar \over \msun } \right )^{-1/2} \left ( { a \over {\rm AU} } \right )^{3/4} ~ \mearth\ ~ .
\end{equation}
With $F$ = 1 and $\Zr = 0.33$, isolated objects in the terrestrial zone have masses comparable to Mercury and Mars. 
The MMSN has room for 30--50 isolated objects between the orbits of Mercury and Mars. Because their escape velocities
are much smaller than their orbital velocities (eq.  [\ref{eq:escvel-ratio}]), isolated protoplanets eventually
collide and merge to form Earth-mass planets (Fig. \ref{fig:hybrid_evol}).

Outside the snow line, $\Zr = 0.78$ at 5 AU yields an isolation mass of roughly 1 \mearth. As we show later, 
this mass is too small to bind the gas required for a gas giant. Increasing the mass of the MMSN ($F \approx$ 5) 
increases the isolation mass to the `typical' core mass of 10 \mearth\ needed for a massive atmosphere. Thus, 
the MMSN is fine for the terrestrial planets, but it is not massive enough to allow formation of gas giants 
Simialr to Jupiter and Saturn.  The extra mass required is consistent with observations of disks around the
youngest stars (\S2).

\subsubsection{Planetesimal Velocity Evolution}\label{sec:stirring}

As the previous section makes clear, the evolution of planetesimal velocities establishes the rate protoplanets accrete smaller planetesimals.  Gravitational scattering is more common than physical collisions; thus, planetesimal velocities rapidly adjust as large protoplanets grow.

Several processes modify the random velocities of planetesimals.  The source of random kinetic energy is known as viscous stirring.  This process  uses  planetesimal encounters --- predominantly gravitational scattering ---  to extract energy from orbital shear.  Dynamical friction redistributes kinetic energy among planetesimals of different masses, pushing them towards equipartition.  Thus, smaller (larger) planetesimals damp 
(excite) the random velocities of the larger (smaller) planetesimals \citep{weth1989, kok1995, kl1998}.   Ignoring ejections and gas drag, physical collisions are the only source of kinetic energy damping.  Collisional damping is especially effective for small planetesimals,
$r \lesssim$ 1--100 m, that collide frequently \citep{oht1992,kl1998}.  When collisions produce small fragments that collide even more frequently, damping is very efficient.
\citet{gold2004} discuss order-of-magnitude derivations of these processes.  As with accretion, behaviors vary between the dispersion- and shear-dominated regimes.  It is common to refer to the excitation and damping of planetesimal velocities as ``heating" and ``cooling," respectively.

The main goal of this introduction to velocity evolution is to show that planetesimals cannot be heated above --- and can sometimes be cooled significantly below --- the escape velocity of the large protoplanets.   We focus on dispersion-dominated encounters to explain this result, which is crucial for ensuring the gravitationally focused collisions required to make planets on reasonable timescales.  

We first consider the simple case where all planetesimals have the same size.   When $v < v_{esc}$, viscous stirring is dominated by gravitational scattering and occurs on the scattering timescale.  This heating timescale is well approximated by the two body relaxation time from stellar dynamics \citep{binn2008}.  For the $n \sigma v$ estimate of the gravitational scattering rate, we use \Eq{eq:nvsigma} and compute the cross section $\sigma \sim b_{\rm scatt}^2$ from the impact parameter for strong gravitational scattering, $b_{\rm scatt} \sim Gm/v^2$.  Together, these give the viscous stirring timescale \citep{IdaMak93}
\begin{equation}
t_{stir,disp} \simeq C_1 {v^4 \over G^2 m \Sigma \Omega } \sim {\rho_\bullet r \over \Sigma \varOmega} \left(v \over v_{esc}\right)^4 ~ ,
\label{eq:t-vs}
\end{equation}
where the constant $C_1 \approx 1/40$ arises from a more detailed analysis \citep{oht2002}, and is similar to the Coulomb logarithm, $\ln \Lambda$, in stellar dynamics (and plasma physics).  The final approximate expression in \Eq{eq:t-vs} facilitates comparison with the collision rates.

The cooling rate is the gravitationally focused collision rate, which follows from \Eq{eq:t-disp} as
\begin{equation} 
t_{cool,disp} \sim {\rho_\bullet r \over \Sigma \varOmega} \left(v \over v_{esc}\right)^2 ~ ,
\label{eq:t-colldisp}
\end{equation} 
Balancing the stirring and cooling rates implies $v \sim v_{esc}$.  While the correct answer, the reasoning is incomplete. 
Gravitational focusing is weak for $v \gtrsim v_{esc}$.  In this regime, stirring and collisional cooling rates are comparable.  
A slight imbalance in favor of heating could lead to runaway growth of $v$ and an eventual collisional cascade.  This 
runaway requires nearly elastic physical collisions, as in the collision of two basketballs.
In the idealized model of \citet{gt78}, collisions among planetesimals with coefficients of restitution $\gtrsim 0.63$ 
(comparable to a baseball, but smaller than a basketball or a table tennis ball) bounce often enough to lead to runaway 
heating. Coefficients of restitution for planetesimals are probably much smaller than 0.5 \citep{porco2008}; the velocity 
runaway is unlikely.  Similarly sized planetesimals will excite random velocities to the surface escape speed, $v \sim v_{esc}$.

Returning to the two groups approximation, we consider stirring of small planetesimals by large protoplanets.  Dynamical friction ensures $v_s > v_l$ (confirmed below); planetesimals dominate the encounter speed.  The stirring of small planetesimals by larger ones then occurs on a timescale
\begin{equation} 
t_{\rm stir, disp} \sim {\rho_\bullet r_l \over \Sigma_l \varOmega} \left(v_s \over v_{esc,l}\right)^4 ~ .
\label{eq:t-vs2}
\end{equation}
Comparison with \Eq{eq:t-vs} shows that large planetesimals dominate the stirring of small planetesimals when $\varSigma_l m_l > \varSigma_s m_s$.  Initially, $\varSigma_s > \varSigma_l$; small planetesimals contain enough mass to affect growth.  To dominate stirring, however, large planetesimals can contain a minority of the mass.  

Due to stronger stirring by large protoplanets, small planetesimals are excited to $v_s > v_{esc,s}$.  At these speeds,  collisions between small planetesimals generally cause collisional fragmentation or erosion.  The resulting smaller planetesimals then collisionally cool more efficiently.  Even without this extra cooling, gravitational focusing arises.  When $v_s > v_{esc,s}$, small planetesimals cool by colliding with other small planetesimals on the geometric timescale, $t_{cool} \sim \rho_\bullet r_s/(\varSigma_s \varOmega)$.  Balancing these heating and cooling rates gives 
\begin{equation} 
v_s \sim \left({\varSigma_l \over \varSigma _s}{r_s \over r_l}\right)^{1/4}  v_{esc,l}~.
\end{equation} 
When $\varSigma_s > \varSigma_l$, $v_s \ll v_{esc,l}$; small planetesimal accretion is strongly gravitationally focused.

We now consider whether the growth of large protoplanets is dominated by the accretion of small planetesimals or other large protoplanets.  Planetesimals with the larger product of surface density and gravitational focusing, $\varSigma f_G$, drive the fastest growth (eq. [\ref{eq:t-disp}--\ref{eq:t-shear}]).  A balance of viscous self-stirring and cooling by dynamical friction against small planetesimals then sets the velocity dispersion of large protoplanets.  For dispersion-dominated encounters this balance gives \citep[for details, see][]{gold2004}
\begin{equation} 
{v_l \over v_s} \sim \left({\varSigma_l \over \varSigma_s}\right)^{1/4}~.
\end{equation} 
Since $\varSigma f_{G,disp} \propto \varSigma/v^2$, small planetesimals contribute more to the growth of large protoplanets, by a factor $(\varSigma_s/\varSigma_l)^{1/2} > 1$.

This introduction only begins to touch on the complexities of planetesimal velocity evolution.  However even these simple considerations show that gravitationally focused accretion of small planetesimals by large protoplanets is likely.  Earlier, we explained that collisional erosion plays a key role in cooling small planetesimals to $v_s > v_{esc,s}$. Now, we turn to even more violent encounters, with $v_s \gg v_{esc,s }$, which lead to catastrophic disruption.

\subsubsection{Fragmentation}\label{sec:frag}

As large planetesimals grow, they stir up the velocities of smaller planetesimals to the disruption velocity. 
Instead of mergers, collisions then yield smaller planetesimals and debris.  Continued disruptive collisions 
lead to a collisional cascade, where leftover planetesimals are slowly ground to dust 
\citep{dohn1969,will1994,koba2010}. Radiation pressure from the central star ejects dust grains with 
$r \lesssim$ 1--10 $\mu$m; Poynting-Robertson drag pulls larger grains into the central star 
\citep{burns1979,arty1988,take2001}.  Eventually, small planetesimals are accreted by the large 
planetesimals or ground to dust.  

To understand the origin of the collisional cascade, we consider the outcome of a head-on collision between
two identical planetesimals. During the impact, some kinetic energy heats up the planetesimals; the rest
goes into the internal energy of material in the planetesimals. When the impact energy is small, the extra
internal energy is small compared to the binding energy of either planetesimal; the two objects merge into
a single, larger planetesimal. When the impact energy is larger than the binding energy, the collision
shatters the planetesimals into a few smaller planetesimals and a lot of dust.

Estimating the binding energy of planetesimals relies on two approaches 
\citep{davis1985,housen1990,housen1999,hols1994,benz1999,lein2008,lein2009}.  
Sophisticated collision experiments yield the internal strengths of small rocky and icy objects,
$r \lesssim$ 10--100 cm.  Theoretical investigations derive the strength from analytic or numerical 
models of the crystalline structure and the equation-of-state of the material. In both cases, 
investigators derive the energy $Q_D^*$ required to remove half of the combined mass of two colliding 
planetesimals and eject this mass to infinity. Although more sophisticated approaches include the
impact velocity in $Q_D^*$, we focus on a simpler expression that depends only on radius,
\begin{equation}
Q_D^* = Q_b r_s^{\beta_b} + \rho Q_g r_s^{\beta_g} ~ .
\label{eq:qdis}
\end{equation}
Here $Q_b r_s^{\beta_b}$ is the bulk (tensile) component of the binding energy 
and $\rho Q_g r_s^{\beta_g}$ is the gravity component of the binding energy.

\begin{figure}[t]
\includegraphics[width=6.5in]{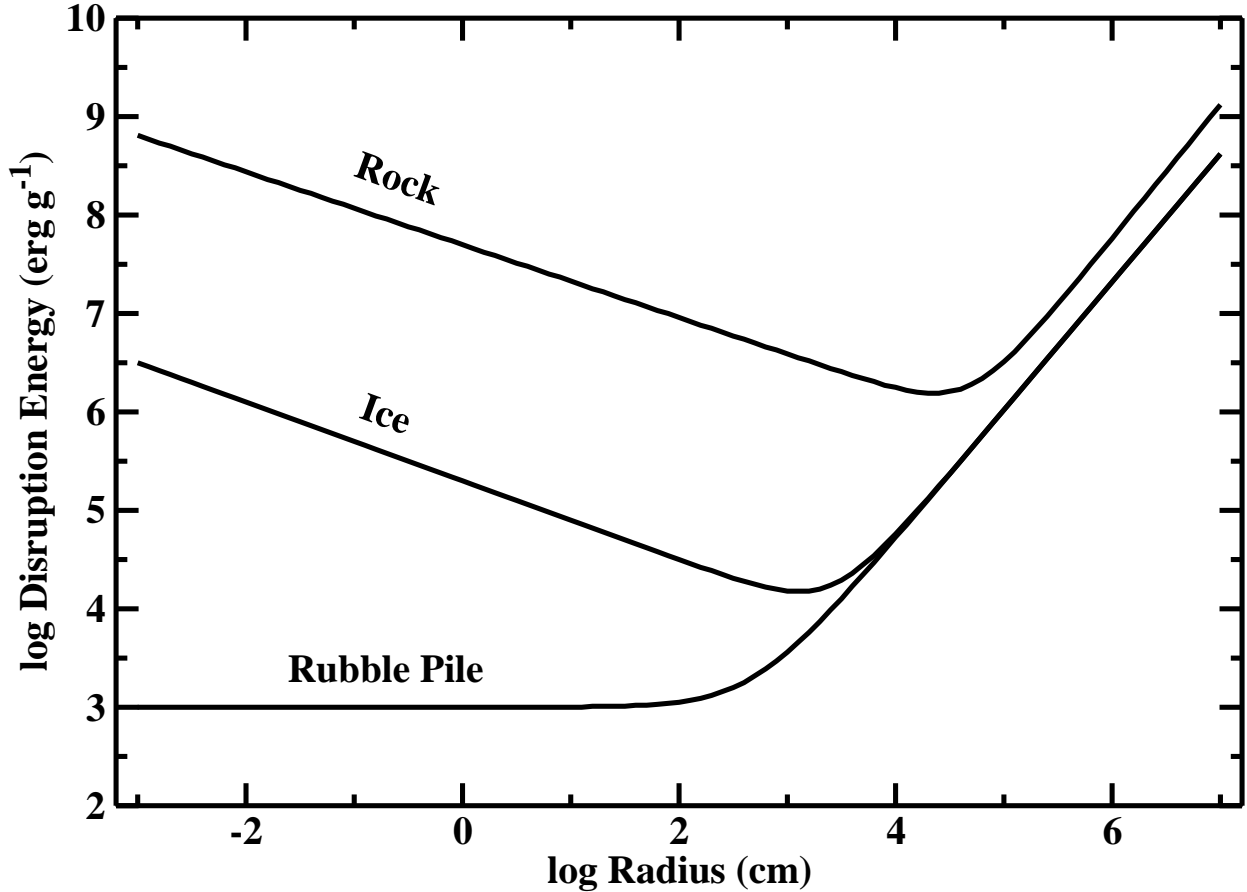}
\vskip 3ex
\caption{
Disruption energy, $Q_D^*$, for icy objects. The solid curves plot typical results 
derived from numerical simulations of collisions \citep[e.g.,][]{benz1999,durda2004,lein2009}
that include a detailed equation of state for basalt (rock) or crystalline ice (ice).
In the strength regime ($r \lesssim 10^2$--$10^4$~cm), smaller particles are stronger.
In the gravity regime ($r \gtrsim 10^5$~cm), larger objects are stronger.
The ``Rubble Pile" curve shows results consistent with model fits to comet breakups 
\citep[e.g.,][]{asp1996}.
}  
\label{fig:qdis}
\end{figure}

Laboratory experiments and detailed numerical collision simulations yield a broad range
of results for $Q_D^*$ \citep[Fig. \ref{fig:qdis};][]{housen1990,benz1999,hols2002,lein2008}. 
In the strength regime at small sizes, the binding energy of a planetesimal depends on
the number of flaws -- cracks, fissures, etc -- in the material. Larger planetesimals
have more flaws and smaller strengths. In the gravity regime at large sizes, the binding
energy depends on the internal pressure. Larger planetesimals have larger internal pressures
and larger strengths.  The lower density and weaker crystalline structure of ice leads to
smaller strengths than basalts and other rocks. 

Models for the breakup of comet Shoemaker-Levy 9 suggest a smaller component of the bulk 
strength \citep{asp1996}, implying small disruption energies for small planetesimals 
(Fig. \ref{fig:qdis}; ``Rubble Pile"). A low strength is consistent with numerical 
simulations of ``rubble piles", structures with countless flaws held loosely together.
This structure probably results after icy or rocky planetesimals suffer numerous impacts 
which disrupt the internal structure (removing most of the tensile component of the
binding energy) but do not destroy the object.

The collisional cascade begins when the impact energy of colliding small planetesimals equals $Q_D^*$.  
Because the random velocities of small planetesimals equal the escape velocities of large planetesimals, 
the impact energy depends only on the mass of a large planetesimal. Equating this energy to $Q_D*$
allows us to derive the ``disruption mass," the mass of a large planetesimal at the onset of the
collisional cascade.  With $v_{esc,s} \ll v_{esc,l}$, the impact energy per unit mass in the 
center-of-mass frame is roughly $v^2 / 8 \approx v_{esc,l}^2/8$.  Setting this energy equal to $Q_D^*$, 
we solve for the disruption mass:
\begin{equation}
m_d = \left ( {3 \over 4 \pi \rho_\bullet} \right ) ^{1/2} \left ( {8 Q_D^* \over G} \right )^{3/2} \sim 3.5 \times 10^{-6} \rho_\bullet^{-1/2} \left ( \frac{Q_D^*}{10^7 ~ \rm erg~g^{-1}} \right )^{3/2} ~ \mearth.
\label{eq:mdis}
\end{equation}
When small planetesimals have sizes exceeding $\sim$ 1~km, $Q_D^*$ is fairly independent of their
composition. For typical $Q_D^* \approx 10^7 - 10^9$ erg g$^{-1}$, the disruption mass is roughly
0.003--3 Pluto masses. Collisional cascades begin well before planets reach their final masses.

Once disruption commences, the final mass of a planet depends on the timescale for the 
collisional cascade \citep{kb2004a,kb2008,lein2005}.  If disruptive collisions produce 
dust grains much faster than planets accrete planetesimals, planets cannot grow much 
larger than the disruption radius and have a maximum mass $m_{l,max} \approx m_d$.  However, 
if planets accrete grains and leftover planetesimals effectively, planets reach the isolation 
mass before collisions and radiation pressure remove material from the disk 
\citep[$m_{l,max} \approx m_{iso}$;][]{gold2004}. 

In a gas-free environment, larger $m_d$ in the inner disk enables planets to accrete much of 
the debris before destructive collisions and radiative processes remove it.  Rocky planet 
masses then approach the isolation mass. In the outer disk, smaller planets cannot accrete 
debris before it is lost. Icy planets cannot grow much larger than $m_d$.

\subsubsection{Planetesimal Accretion with Gas Damping}
\label{sec:gaseffects}

Gas slows the random velocities of smaller planetesimals.   Larger protoplanets are less affected by drag and are
damped by dynamical friction.  The 
drag force, $F_D$, of \Eq{eq:turbdragforce} damps the kinetic energy of planetesimals (now with size $r_s$, not the size $s$ of dust, pebbles, and boulders) at a rate
 $t_{gas} \equiv v_s (dv_s/dt)^{-1} = v_s (F_D/m_s)^{-1}$.  With $\rho\gs \sim \varSigma\gs \varOmega / c_{\rm s}$, where $c_{\rm s}$ is the gas sound speed,
\begin{equation}
t_{gas} \sim {1 \over C_D}{\rho_\bullet r_s \over \varSigma\gs \varOmega}{c_{\rm s} \over v_s} ~ .
\label{eq:t-gas}
\end{equation}

To understand the impact of damping, we consider an ensemble of small planetesimals stirring themselves 
(eq. [\ref{eq:t-vs}]).  Without gas, small planetesimals excite their velocities to $v_s \sim v_{esc,s}$.  
If $t_{gas} < t_{coll} \sim \rho_\bullet r_s /(\varSigma_s \varOmega)$, drag exceeds collisions as the 
dominant cooling mechanism.  This switch happens when 
\begin{equation} 
r_s \gtrsim {1 \over C_D}{\varSigma_s \over \varSigma\gs} {c_{\rm s} \over \sqrt{G \rho_\bullet}} \sim 30 \left(R \over {\rm AU}\right)^{-3/14}\Zr ~~ \km ~ . 
\label{eq:rgas}
\end{equation} 
To be damped by gas drag, planetesimals must exceed this minimum size.  This somewhat counterintuitive result (drag is often more significant for smaller bodies) arises from (i) non-linear drag laws and (ii) a velocity scale, $v_{esc,s}$, that increases with size.   The numerical value of the size threshold decreases if $\Zr$ is reduced due to an inefficiency of turning dust into planetesimals.

When \Eq{eq:rgas} holds, the stronger damping of self-stirred planetesimals ensures $v_s < v_{esc,s}$.  Collisions are gravitationally focused and runaway growth begins earlier than in a disk without gas.

As growth proceeds, larger protoplanets dominate the stirring of smaller planetesimals.   With stronger stirring, smaller and smaller planetesimals are damped by non-linear gas drag instead of collisions.  Even if collisions are initially more significant, gas drag becomes the dominant coolant as growth proceeds.  To compute the random speeds of small planetesimals, we assume dispersion-dominated encounters and balance the heating of \Eq{eq:t-vs2} with the cooling of \Eq{eq:t-gas} to get
\begin{equation} 
{v_s \over v_{esc,l}} \sim \left({1 \over C_D }{r_s \over r_l}{\varSigma_l \over \varSigma\gs} \right)^{1/5} ~.
\end{equation} 
With more mass in small planetesimals $\varSigma_l < \varSigma_s \ll \varSigma\gs$, gravitational focusing, $f_{G, disp} \sim (v_{esc,l}/v_s)^2$, becomes strong.  Accretion of small planetesimals can become shear-dominated.  As described in \S\ref{sec:growth}, runaway growth transitions to oligarchy (but does not slow down) in the transition to shear-dominated accretion.  A self-consistent analysis of these processes is facilitated by the numerical calculations summarized in \S\ref{sec:TPsims}.

Details aside, the large oligarchs stir small planetesimals past their escape speed and up to the disruption velocity (\S5.1.3). 
Disruptive collisions among small planetesimals produce a collisional cascade, which grinds planetesimals
into smaller and smaller objects.  Without gas, planetesimals are ground into small dust grains which are dragged 
into the star by Poynting-Robertson drag or ejected from the planetary system by radiation pressure.  With gas damping, 
the collisional cascade halts at some intermediate size  ($\sim$ 10 cm to 10 m), depending on factors such as the mass of the oligarchs, gas density, material strength, and orbital distance.
The damped velocities are then slow enough that the oligarchs accrete these small rocks rapidly.  This rapid accretion enables oligarchs to reach the isolation mass on short timescales, even in the outer disk \citep{raf2004,kb2009}. 

Collisional grinding a set of small planetesimals into small dust grains requires a very depleted gas disk.  For the cascade to proceed down to  1--10 $\mu$m particles, these grains must decouple from the gas (on an orbital timescale, we assume for simplicity).  Epstein drag applies for low gas densities (and long gas mean free paths).  From \Eq{Ep}, particles with sizes $s \gtrsim \varSigma\gs / \rho_\bullet$ decouple from the gas.  From \Eq{eq_sigmag} the depletion factor (relative to the MMSN) required to avoid entrainment is
\begin{equation}
F \lesssim 10^{-7} {s \over \mu{\rm m}} \left(R \over {\rm AU}\right)^{3/2}  ~ .
\label{eq:disk-depletion}
\end{equation}
This low mass disk may not last very long.  Nevertheless, current observational limits only constrain gas surface 
densities in debris disks to $F \lesssim 1\%$ of the MMSN, not yet sufficient to asses the dynamical significance 
of gas.  ALMA should should place much tighter constraints (see chapter by Moro-Martin).


\subsubsection{Numerical Simulations of Low Mass Planet Formation}
\label{sec:TPsims}

Analytic estimates provide a good understanding of each piece of the planet formation process. However,
putting the whole set of processes into a coherent theory requires numerical calculations.  Clusters of 
computers can now finish an end-to-end calculation in a reasonable amount of time. Several groups are 
building towards this simulation, but no complete calculation exists.

Constructing numerical simulations of planet formation involves identifying and solving a set of 
coupled differential equations which describe the evolution of the gaseous disk and the masses and 
orbital properties of solid objects. Selecting the proper approach depends on the nature of the problem. 
Hydrodynamics codes address the evolution of the gaseous disk and how planets accrete material and migrate 
within the disk \citep{dangelo2003,nelson2004}.
Smooth particle hydrodynamics allows detailed solutions to outcomes of binary collisions between large 
protoplanets \citep[e.g., Earth-Moon and Pluto-Charon formation;][]{canup2008,canup2011}.
Solving the coagulation equation with a fragmentation algorithm yields the mass and time evolution of 
solid particles ranging in size from 1 $\mu$m up to roughly 1000~km \citep{saf1969,weth1993,kl1999a,birn2010}.
To treat the dynamical evolution of large planets,  $N$-body treatments provide accurate and often
fast solutions \citep{chambers1998,chambers2001a,raymond2004,nagasawa2005,kokubo2006}.

Most investigations of terrestrial planet formation employ a coagulation code or an $N$-body code.  An $N$-body code 
cannot possibly follow the trajectories of the $\gtrsim 10^{12}$ small planetesimals expected in a MMSN. 
Coagulation models, which treat planetesimals as a statistical ensemble of objects with a distribution
of $e$ and $i$, can solve for the time evolution of their masses and orbits throughout runaway and 
oligarchic growth \citep{weth1993}. 
Once most of the solid mass is in a few protoplanets, the statistical approach fails. $N$-body codes can
then follow the evolution during the late stages of oligarchic growth and throughout chaotic growth.

Several hybrid codes combine aspects of both approaches \citep{spaute1991,weid1997,bk2006,charnoz2007,raymond2011}.  
To follow the evolution of the gaseous disk together with the solids, \citet{kb2011} solve the radial diffusion 
equation for the gaseous disk (eq. [\ref{eq:disk-diffusion}]) and employ a merged coagulation + $N$-body 
code for the solids.  
In these treatments, the coagulation code follows solids with masses smaller than the promotion mass, 
$m_{pro}$; the $N$-body code tracks protoplanets with $m > m_{pro}$. Comparisons with other simulations 
and with analytic theory provide tests of these techniques \citep[e.g.,][]{kl1998,fraser2009,morby2009}.

To illustrate the formation process, we summarize results for several calculations of terrestrial planets
and gas giant cores.  Because this aspect of this field is growing so rapidly, we focus on a few
simple examples. 

Coagulation codes begin with an ensemble of planetesimals in place at $t$ = 0 \citep{kb2010}. Planetesimals 
are placed in concentric annuli according to a fixed initial surface density relation. These planetesimals
often have a single size of 1--100~km; sometimes calculations begin with a distribution of sizes.  Because 
dynamical friction efficiently damps the velocities of the largest planetesimals, planets grow faster in 
calculations with a size distribution of planetesimals. Starting with an ensemble of small planetesimals 
leads to faster growth than an ensemble of large planetesimals. The initial surface density sets the 
growth time. Planets grow faster in more massive disks. In many calculations, the planetesimals evolve 
in a gaseous disk which also evolves in time; the disk evolution may be proscribed in advanced or calculated 
along with the planetesimals.

\begin{figure}[t]
\begin{center}$
\begin{array}{cc}
\includegraphics[width=3.5in]{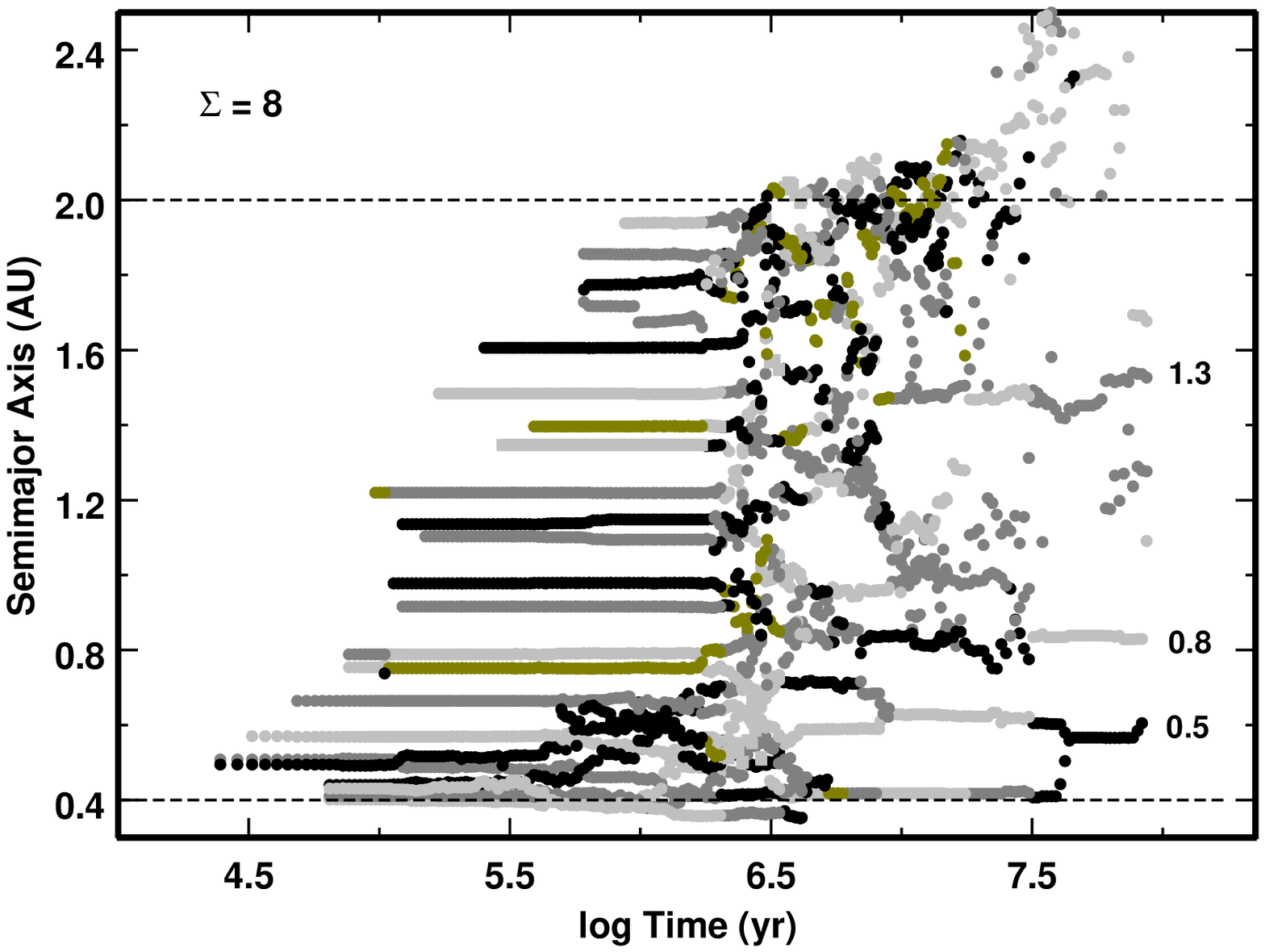} &
\includegraphics[width=3.5in]{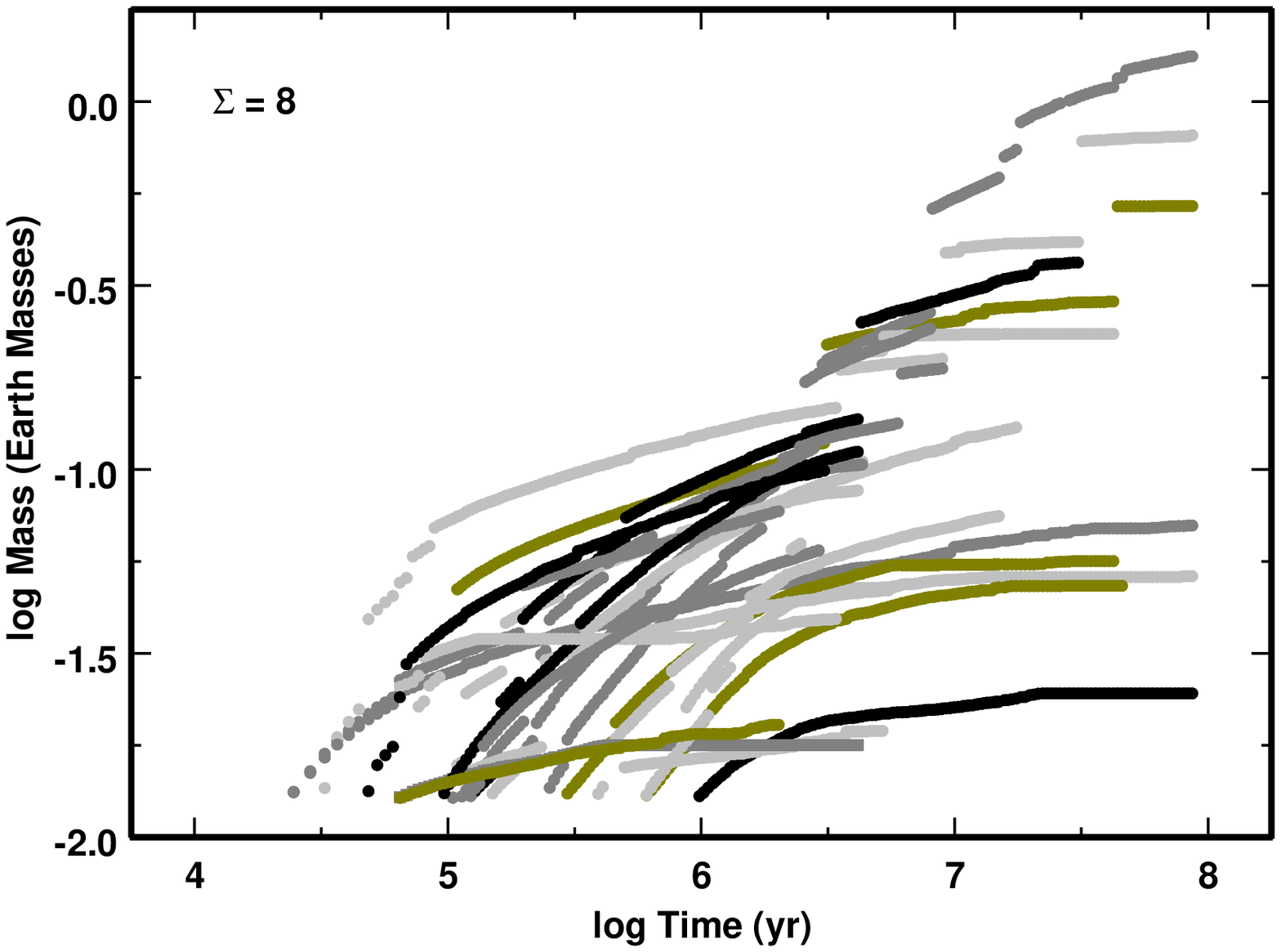}
\end{array}$
\end{center}
\caption{Evolution of oligarchs in the terrestrial zone. The calculation starts 
with 1 km planetesimals ($\rho_\bullet$ = 3 ~g~cm$^{-3}$) in a disk with 
$\Sigma_{\rm s}$ = 8~g~cm$^{-2}~(a / 1~{\rm AU})^{-1}$.  {\em Left panel:} The
time evolution of semimajor axis shows three phases that start at the inner edge 
of the grid and propagate outward: (i) after runaway growth, isolated oligarchs 
with $m \gtrsim$ $4 \times 10^{25}$ g form and continue to grow very rapidly;
(ii) oligarchs develop eccentric orbits, collide, and merge; and
(iii) a few massive oligarchs eventually contain most of the mass and
develop roughly circular orbits. The legend indicates masses (in
$M_{\oplus}$) for the largest oligarchs.  {\em Right panel:} The mass
evolution of oligarchs shows an early phase of runaway growth (steep
tracks) and a longer phase of oligarchic growth (flatter tracks), which 
culminates in a chaotic phase where oligarchs grow by captures of other 
oligarchs (steps in tracks). Despite the steeper appearance of some of the
mass tracks during runaway growth, protoplanets grow more rapidly during
oligarchic growth.
}
\label{fig:hybrid_evol}
\end{figure}

Fig. \ref{fig:hybrid_evol} shows the evolution of oligarchs in an evolutionary sequence starting
with an ensemble of 1~km planetesimals at 0.4--2 AU.  Following a short runaway growth phase, 
protoplanets with $m \gtrsim m_{\rm pro}$ appear in a wave that propagates out through the planetesimal 
grid. As these oligarchs continue to accrete planetesimals, dynamical friction maintains 
their circular orbits and they evolve into ``isolated" protoplanets. Eventually, large oligarchs 
start to interact dynamically at the inner edge of the grid; a wave of chaotic interactions then moves 
out through the disk until all oligarchs interact dynamically. Once a few large oligarchs contain 
most of the mass in the system, dynamical friction between the oligarchs and a few leftover 
planetesimals starts to circularize their orbits. This process excites the lower mass oligarchs and 
leftover planetesimals, which are slowly accreted by the largest oligarchs. At the end of the
calculation, the masses, semimajor axes, and orbital eccentricities of stable planets are similar
to those of the terrestrial planets in the Solar System.

Comparisons between the results of hybrid and $N$-body calculations show the importance of 
including planetesimals in the evolution. Both approaches produce a few terrestrial mass 
planets in roughly circular orbits. Because dynamical friction between leftover planetesimals 
and the largest oligarchs is significant, hybrid calculations produce planets with more 
circular orbits than traditional $N$-body calculations.  In most hybrid calculations, lower
mass planets have more eccentric orbits than the most massive planets, as observed in the 
Solar System.  In both approaches, the final masses of the planets grow with the initial 
surface density; the number of planets is inversely proportional to the initial surface 
density of solid planetesimals. However, the overall evolution is faster in hybrid calculations: 
oligarchs start to interact earlier and produce massive planets faster.

In hybrid calculations, the isolation mass and the number of oligarchs are more important as 
local quantities than as global quantities. As waves of runaway, oligarchic, and chaotic growth 
propagate from the inner disk to the outer disk, protoplanets growing in the inner disk become 
isolated at different times compared to protoplanets growing in the outer disk. Thus, the 
isolation mass in hybrid models is a function of heliocentric distance, initial surface density, 
and {\it time}, which differs from the classical definition (eq. [\ref{eq:miso}]).  

\begin{figure}[t]
\begin{center}$
\begin{array}{cc}
\includegraphics[width=3.0in]{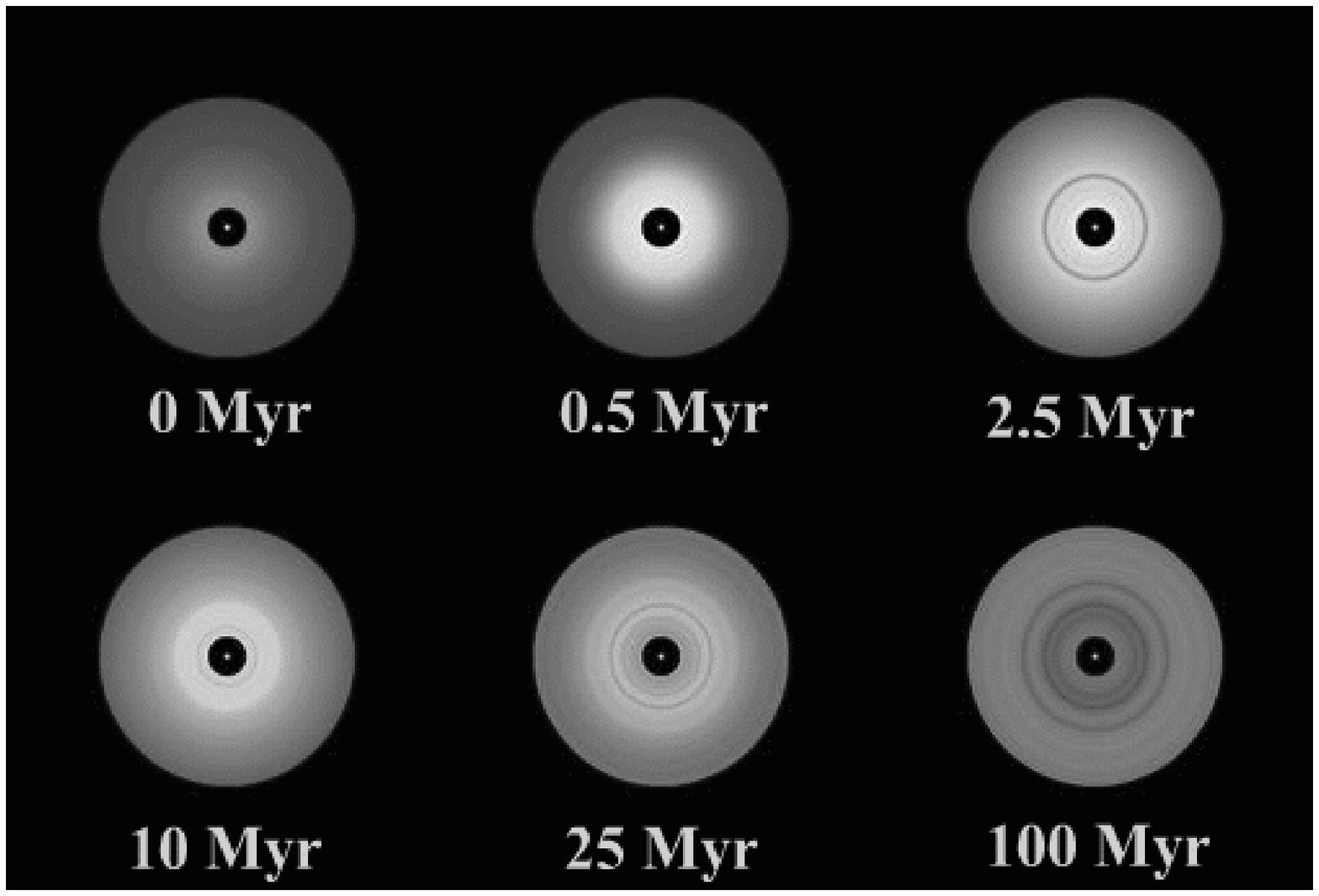} &
\includegraphics[width=3.0in]{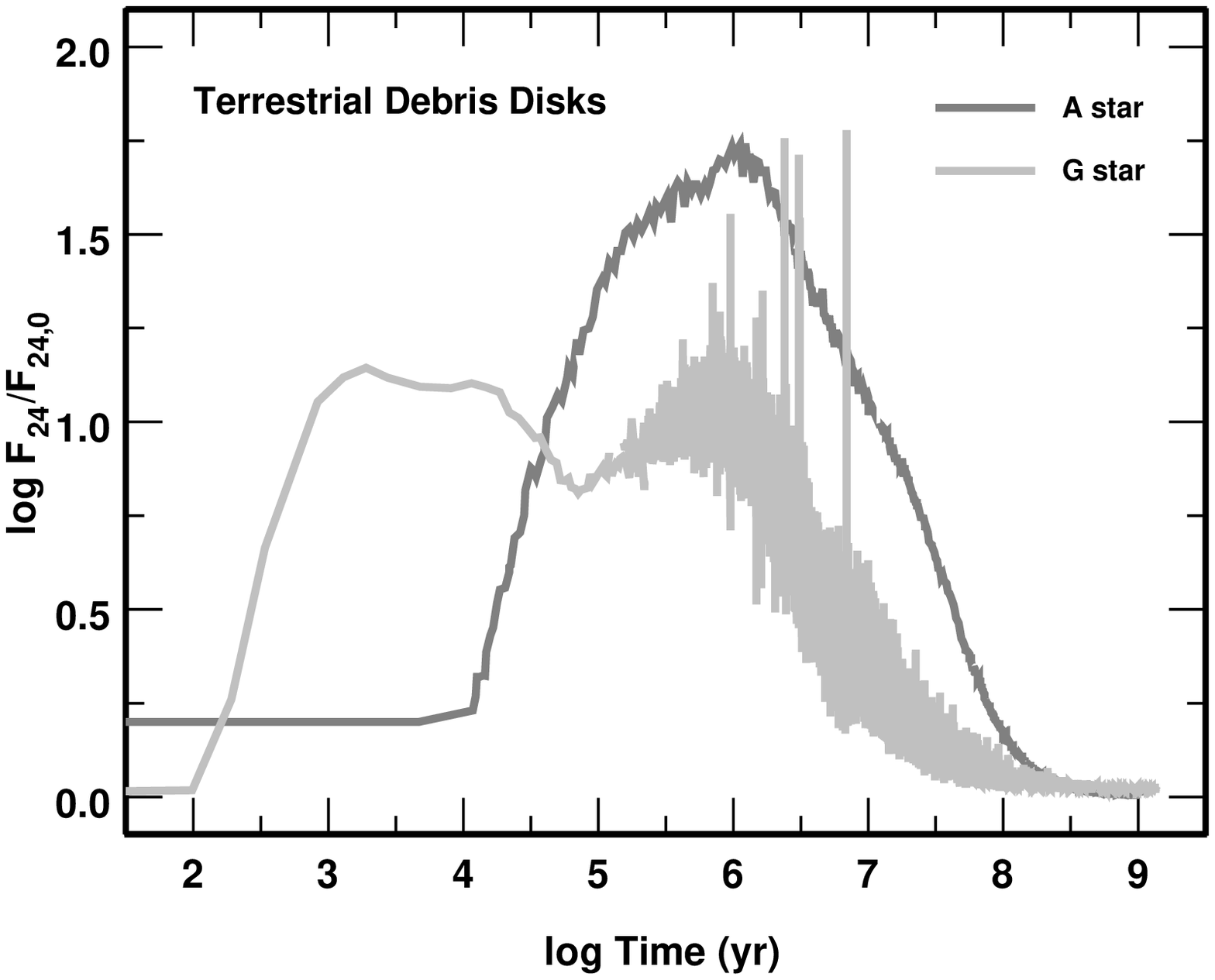}
\end{array}$
\end{center}
\caption{
Evolution of debris disks in the terrestrial zone \citep{kb2004b,kb2005}.
For an A-type star with a luminosity of $\sim$ 50 $L_{\odot}$,
the range in blackbody temperatures of planetesimals at 3--20 AU
(425--165 K) is similar to the range in the Solar System at
0.4--2 AU (440--200 K).
{\em Left panel:} Images of a disk extending from 3--20 AU around an 
A-type star. The intensity scale indicates the surface brightness of 
dust, with black the lowest intensity and white the highest intensity.
{\em Right panel:} Mid-IR excess for two debris disk models. The light 
grey line plots the ratio of the 24 $\mu$m flux from a debris disk at
0.4--2 AU disk relative to the mid-IR flux from a G-type star.
The dark grey line shows the evolution for the A-star disk shown
in the left panel.
\label{fig:debris1}
}
\end{figure}

During oligarchic growth of the simulation in Fig.  \ref{fig:hybrid_evol}, viscous stirring excites 
leftover planetesimals to the disruption velocity. A series of separate simulations demonstrates 
that the collisional cascade produces copious amounts of dust, which absorb and scatter radiation 
from the central star. Following the growth of protoplanets, the cascade begins at the inner edge
of the disk and moves outward. For calculations with a solar-type central star, it takes $\sim$ 
0.1 Myr for dust to form throughout the terrestrial zone (0.4--2 AU). The timescale is $\sim$ 1~Myr 
for the terrestrial zone of an A-type star (3--20 AU). As the collisional cascade proceeds, 
protoplanets impose structure on the disk (Fig. \ref{fig:debris1}, left panel). Bright rings form 
along the orbits of growing protoplanets; dark bands indicate where a large protoplanet has swept 
up dust along its orbit. In some calculations, the dark bands are shadows, where optically thick 
dust in the inner disk prevents starlight from shining on the outer disk 
\citep{grogin2001, kb2004a, durda2004}.

In the terrestrial zones of A-type and G-type stars, the dust emits mostly at mid-IR wavelengths. 
In calculations with G-type central stars, formation of a few lunar mass objects at 0.4--0.5 AU
leads to copious dust production in a few thousand years (Fig. \ref{fig:debris1}, right panel).  
As protoplanets form farther out in the disk, the disk becomes optically thick and the mid-IR 
excess saturates. Once the orbits of oligarchs start to overlap ($\sim$ 1 Myr), the largest 
objects sweep the disk clear of small planetesimals. The mid-IR excess fades. During this decline,
occasional large collisions generate large clouds of debris that produce remarkable spikes in 
the mid-IR excess \citep{kb2002b,kb2005}.

In A-type stars, the terrestrial zone lies at greater distances than in G-type stars. Thus, debris 
formation in calculations with A-type stars begins later and lasts longer than in models with G-type 
stars (Fig. \ref{fig:debris1}, right panel).  Because the disks in A-type stars contain more mass, 
they produce larger mid-IR excesses. At later times, individual collisions play a smaller role, 
which leads to a smoother evolution in the mid-IR excess with time. Although the statistics for 
G-type stars is incomplete, current observations suggest that mid-IR excesses are larger 
for A-type stars than for G-type stars (see Chapter by Moro-Martin).

Collisional cascades and debris disk formation may impact the final masses of terrestrial planets. 
Throughout oligarchic growth, roughly $\sim$ 25\% to 50\% of the initial mass in planetesimals is converted into
debris. For solar-type stars, the disk is optically thick, so oligarchs probably accrete the debris 
before some combination of gas drag, Poynting-Robertson drag, and radiation pressure remove it. In the
disks of A-type stars, the debris is more optically thin. Thus, these systems may form lower mass 
planets per unit surface density than disks surrounding less massive stars. Both of these assertions
require tests with detailed numerical calculations.

\subsection{Accretion of Atmospheres}
\label{sec:atmos}

Protoplanetary atmospheres have a rather different character than the mature planetary atmospheres of Solar System planets and exoplanets.  The crucial distinction is that protoplanets orbit within a gas disk.  The disk supplies the atmosphere's gas and provides an external binding pressure until the planet opens a clean gap.   The protoplanetary atmospheres inherits its composition from the disk.  However, the fraction of heavy elements that wind up in the core -- versus the dust and ablated planetesimals that remain in the atmosphere -- is a key uncertainty.  This uncertainty  crucially affects the mean molecular weight,  $\mu$, and opacity,  $\kappa$,  of the atmosphere.


\subsubsection{Static Protoplanet Atmospheres}
\label{sec:atmos-struc}

As protoplanets grow, they become massive enough to bind a gaseous atmosphere.  The atmosphere is significantly denser than the surrounding disk gas when the core's gravitational binding energy exceeds the thermal energy of the gas.   For a solid core of mass $m_c$ and radius $r_c$, this occurs when $r_B > r_c$, where the Bondi radius
\begin{equation} \label{eq:bondi-rad}
r_B = {G m/c_{\rm s}^2} ~ ,
\end{equation} 
where total planet mass $m = m_c + m_a$, including the gravitationally bound atmosphere's mass, $m_a$.
Equivalently a core with an atmosphere must exceed the Bondi mass
\begin{equation} 
m_c > m_B \equiv \sqrt{3 \over 4 \pi \rho_\bullet} {c_{\rm s}^3 \over G^{3/2}} \simeq 10^{-3} \left(a \over {\rm AU}\right)^{-9/14} {l_\star^{3/8} \over \tilde{\mu}^{3/2}} ~~ \mearth
\end{equation} 
where we use the gas temperature in an irradiated disk (eq.\ [\ref{eq:irtemp-2}]) and normalize the 
stellar luminosity as $l_\star = \lstar/\lsun$ and the gas mean molecular weight as 
$\tilde{\mu} = \mu/(2.4~m_H)$.

As the core mass increases beyond $m_B$ the atmosphere becomes denser and more massive.    The outer boundary of the atmosphere, $r_{out} = \min(r_B, R_H)$, is set by the Bondi radius until the atmosphere fills the Hill sphere.  This transition occurs for massive protoplanets, as $r_B > R_H$ requires
\begin{equation}
\label{eq:mtrans}
m > m_{\rm trans} = {c_s^3 \over \sqrt{3} G \Omega} \simeq 3 \left(a \over {\rm AU}\right)^{6/7} {l_\star^{3/8} \over \sqrt{m_\star} \tilde{\mu}^{3/2}} ~~ \mearth ~.\end{equation}  
Comparing to the disk's gas scale-height $H_g = c_s/\Omega$, the criteria $r_B > H_g$ and $R_H > H_g$ also reproduce \Eq{eq:mtrans} within order unity constants ($m_{\rm trans}$ increases by $3^{3/2}$ for the $R_H > H_g$ criterion).  Thus when $m > m_{\rm trans}$ the protoplanet is no longer uniformly embedded in the disk midplane.  It can feel the top and bottom of the disk and start to open a gap (see chapter by Morbidelli).   Outside $\sim 5$ AU, the core accretion instability generally occurs for $m < m_{\rm trans}$.  Thus the 3D structure of the gas disk can usually be ignored when describing the onset of core accretion (\S\ref{sec:atmos-CI}) but not its final evolution (\S\ref{sec:atmos-accr}).


The structure of an proplanetary atmosphere obeys the equations (which also govern stellar structure) of hydrostatic balance
\begin{equation}
{dP \over dr} = - {G m \over r^2} \rho = -\rho g ~ ,
\label{eq:hyd-eq}
\end{equation}
mass conservation
\begin{equation}
{dm \over dr} = 4 \pi r^2 \rho ~ ,
\label{eq:mass-con}
\end{equation}
and energy transport by optically thick, $\tau \sim \kappa P/g > 1$, radiative diffusion,
\begin{equation}
{16 \sigsb T^3 \over 3 \kappa \rho} {dT \over dr} = -{L \over 4 \pi r^2} ~ .
\label{eq:rad-tran}
\end{equation}
Wherever radiative diffusion would satisfy the Schwarzchild criterion, $d \ln T / d \ln P > \nabla_{\rm ad}$, the energy transport becomes convective.  The adiabatic index $\nabla_{\rm ad} = 2/7$ for an ideal diatomic gas, but in general must be determined from detailed equation of state calculations \citep{SauCha95, SauGui04}.    Because convective transport is efficient, the temperature profile follows an adiabat, $T\propto P^{\nabla_{\rm ad}}$ instead of \Eq{eq:rad-tran}  in convective regions.
For the ideal gas equation of state, $P = \rho \mathcal{R}  T$, with the (specific) 
gas constant $\mathcal{R} = k_B/\mu m_H$.

The masses of stable atmospheres (that do not undergo the core accretion instability) place interesting constraints on planet formation. In the
terrestrial zone, typical isolation masses (eq.\ [\ref{eq:miso}]) are roughly 0.1 \mearth. For almost any accretion time, these
planets have stable atmospheres with masses much smaller than the planet's mass. Icy planets formed at tens of AU, however,
have much larger isolation masses of several \mearth, and can support much more massive atmospheres. In 
the Solar System, the dichotomy between terrestrial planets with thin atmospheres and icy planets with massive
atmospheres is consistent with our estimates. Once we have a large
sample of rocky/icy exoplanets with well characterized atmospheres, we will see if the same dichotomy persists.

\subsubsection{Enhanced Planetesimal Accretion}\label{sec:atmos-planetesimal}
For low mass protoplanets with $r_{out} = r_B$ the size of the atmosphere relative to the core is
\begin{equation} \label{eq:chi}
{r_{out} \over r_c}  \simeq 800 \left(m_c \over 10 ~\mearth\right)^{2/3} \left(a \over 5 \AU\right)^{3/7} {\tilde{\mu} \over l_\star^{1/4}} ~ .  
\end{equation} 
For Mars-mass planets and larger, the radius of the atmosphere is 10 or more times larger than the radius of the core.  Extended atmospheres  can significantly enhance planetesimal accretion.

When a planetesimal encounters the atmosphere of a protoplanet, it experiences enhanced gas drag. Capture results if the planetesimal 
loses enough orbital energy. A thin atmosphere has little impact on very large 
planetesimals; the collisional cross-section is still $\pi r^2 f_g$. For sufficiently small planetesimals, any 
encounter with the atmosphere allows the planet to capture the planetesimal; the collisional cross-section 
is then $\pi r_{out}^2 f_g$. For intermediate sizes, the effective cross-section lies somewhere between 
$\pi r^2 f_g$ and $\pi r_{out}^2 f_g$. To address this regime, \citet{inaba2003} define an ``enhanced radius" 
$r_e$, where the collisional cross-section is $\pi r_e^2 f_g$. In this approach, $r_e = r$ for accreting
very large planetesimals and $r_e = r_{out}$ for accreting very small planetesimals.  However very small rocks and dust grains 
will be too tightly coupled to the gas to accrete.

Compared to the estimates for dispersion- and shear-dominated growth in \S5.1, atmospheres can enhance accretion
rates by 1--2 orders of magnitude \citep{chambers2008}. When leftover small planetesimals have 
typical radii of 0.1--10 km, an isolated terrestrial planet with a thin atmosphere has a small radius 
enhancement, $r_e \approx 1 - 2$. Thus, these isolated objects never experience rapid growth from an enhanced
radius.  Isolated icy planets are more massive and support massive atmospheres. These objects have $r_e \approx$
10 for accreting 0.1--10 km planetesimals and $r_e \approx$ 3 for accreting 100+ km planetesimals. Once they
develop atmospheres, icy isolated objects rapidly sweep up any leftover planetesimals along their orbits.

\subsubsection{The Core Accretion Instability}
\label{sec:atmos-CI}
Low mass protoplanets (near $m_B$) have low mass, optically thin atmospheres.  More massive cores bind thicker atmospheres, which trap the heat of planetesimal accretion.  As the heat is radiated away, the atmosphere becomes denser and more massive.  When the atmosphere's mass exceeds roughly the core mass, the atmosphere cannot maintain hydrostatic equilibrium and collapses \citep{Har78, miz1980}.  This collapse  --- referred to as the ``core accretion instability" --- leads to rapid gas accretion and the birth of a gas giant.  The critical core mass (sometimes called the ``crossover mass" because collapse occurs when the core and atmosphere masses roughly match) for this instability is often, but not always, $\sim 10 ~ \mearth$ \citep{IkoNak00, raf2006}.

Near the crtical core mass, real protoplanetary atmospheres are convective in the interior and (for low enough planetesimal accretion rates) radiative in the exterior region that matches onto the disk \citep{raf2006}.   However an illustrative, and historically important, calculation by \citet{stev1982} demonstrates the essential features of the core accretion instability by (incorrectly) assuming the atmosphere is completely radiative with constant opacity $\kappa$.  We summarize this calculation before comparing it to more detailed computations.

Following \citet{stev1982}, we calculate the structure and mass in the atmosphere by keeping the mass, $m$, constant in the hydrostatic balance equation (\ref{eq:hyd-eq}), i.e.\ neglecting the detailed variation from $m = m_c$ at the core to $m = m_c + m_a$ at the top. 
Equations (\ref{eq:hyd-eq}) and (\ref{eq:rad-tran}) then give (temporarily omitting order unity coefficients for clarity)
 \begin{equation} \label{eq:TPrad}
T^3 dT/dP \sim \kappa L / ( \sigsb G m)~.
\end{equation} 
To integrate this equation, we assume (correctly) that $P$ and $T$ are significantly higher at the base of the atmosphere than in the disk.  We further keep $L$ constant, appropriate if accreted planetesimals release their kinetic energy at the core's surface.  This assumption yields $L = G m_c \dot{m}/r_c$, where $\dot{m}$ is the planetesimal accretion rate.  The slowing of planetesimals as they fall through the atmosphere is a minor correction included in detailed numerical models.

With these approximations, \Eq{eq:TPrad} integrates to a simple $T-P$ profile
\begin{equation}
T \sim \left ( {  \kappa L \over \sigsb G m} P \right )^{1/4} ~ .
\label{eq:temp-atm}
\end{equation}
To obtain the density profile, we use the fact that for a barotropic relation $P \propto T^{\nabla_\infty}$ ($\nabla_\infty = 1/4$ for our example of a constant opacity), the hydrostatic balance equation for an ideal gas integrates to
\begin{equation} 
T = \nabla_\infty{G m \over \mathcal{R} r} ~ 
\end{equation} 
in the atmospheric interior.  The density profile follows as
\begin{equation}
\rho \sim {\sigsb \over \kappa L} \left ( { G m \over \mathcal{R}} \right )^4 {1 \over r^3} ~ .
\label{eq:rho-atm}
\end{equation}
More generally, $\rho \propto r^{1-1/\nabla_\infty}$, which is left as an exercise.

\begin{figure}[t]
\begin{center}
\includegraphics[width=6.5in]{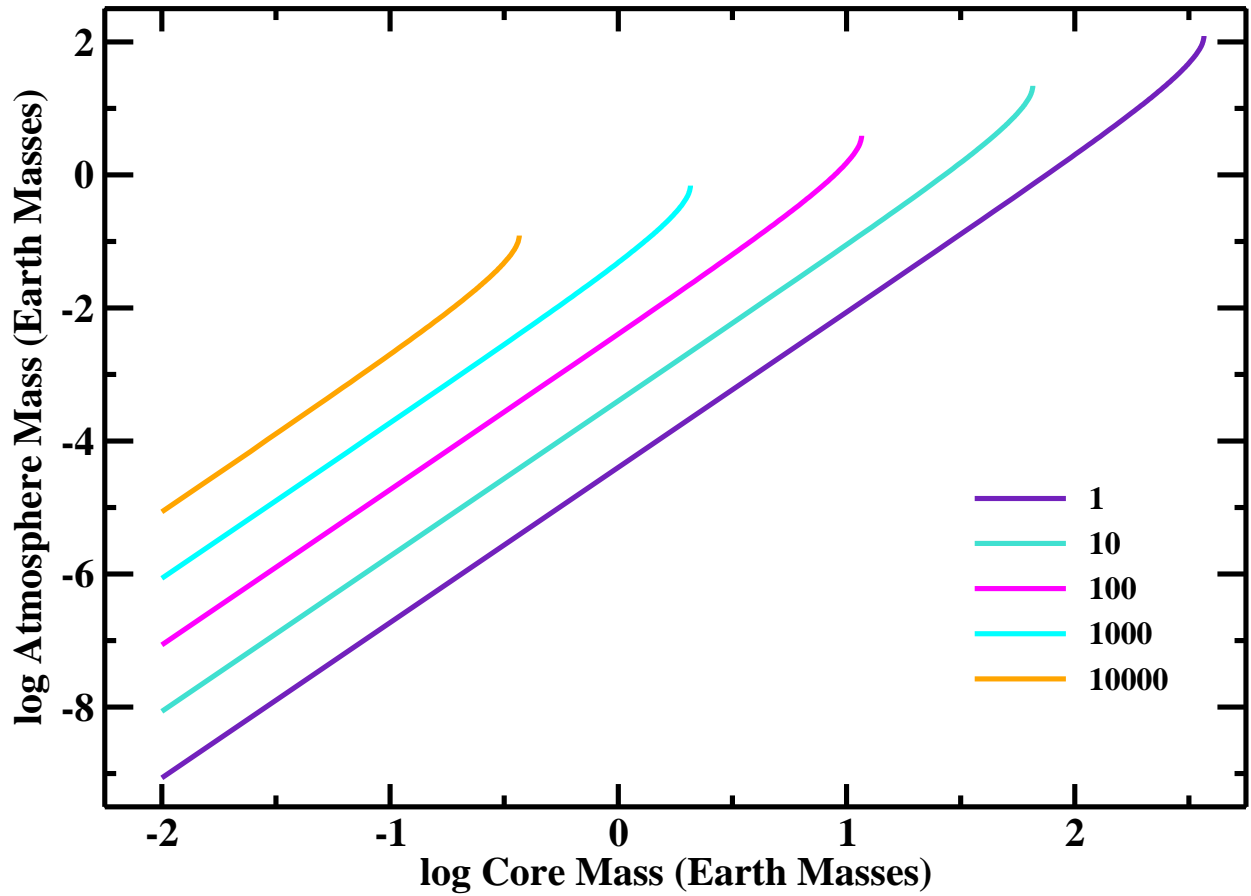}
\vskip 3ex
\caption{
Mass of planet atmosphere as a function of core mass and accretion time
(eq. [\ref{eq:mass-atm2}]). The legend indicates the accretion time in Myr.
At fixed core mass, the mass of the atmosphere grows with the accretion time.
Longer accretion times allow more massive cores to have hydrostatic atmospheres.
}
\label{fig:mass-atm}
\end{center}
\end{figure}

The atmosphere's mass follows by integrating eq. (\ref{eq:mass-con}) from $r_c$ to $r_{out}$:
\begin{equation}
m_a = {\pi^2 \over 3} \chi{ \sigsb \over  \kappa L} \left ( { G m \over  \mathcal{R} } \right )^4    \simeq 2.0 {\sigsb  \chi \over \kappa \mathcal{R}^4} {G^3 m^4 \over \rho_\bullet^{1/3} m_c^{5/3}} t_{acc}~ .
\label{eq:mass-atm1}
\end{equation}
with $\chi \equiv \ln(r_{out}/r_c)$ and order unity coefficients reinstated (despite the overall approximate nature of the calculation).  
The final expression relates the protoplanet luminosity to the (current) core growth timescale, $ t_{acc} = m_c/\dot{m} =  G m_c^2 / (r_c L)$.  Setting $m = m_c$ we can numerically evaluate 
\begin{equation}
m_a 
\approx  9.4  \left(m_c \over 10 ~\mearth\right)^{7/3} {\tilde{\mu}^4 \over \kappa_1}{t_{acc} \over {\rm Gyr}} ~~ \mearth ,
\label{eq:mass-atm2}
\end{equation}
with $\chi =  6.7$ (from eq.\ [\ref{eq:chi}]) and $\kappa_1 = \kappa/(1 \cm^2 \g^{-1})$.  For the chosen parameters, we are near the crossover mass, $m_a \sim m_c \sim 10~\mearth$.  Fig. \ref{fig:mass-atm} shows the behavior of this simple atmosphere model.  We emphasize that the numerical values of this simple model are only meant to be illustrative.  For instance the core must actually accrete in $t_{acc} \lesssim 10 ~{\rm Myr} \ll {\rm Gyr}$.

The existence of an instability arises from the non-linearities in $m_a$.  Expressing
\begin{equation}
m = m_c + k {m^4 \over m_c^{5/3}}
\end{equation} 
where $k$ incorporates all the constants in \Eq{eq:mass-atm1}, we can show that beyond a critical core mass the total mass (unphysically) declines as $m_c$ increases.  Using calculus, the turnover\footnote{Our values differ slightly from \citet{stev1982} because we assume constant accretion time instead of constant mass accretion rate.  Further the 3/4 exponent in his eq. (15) is a typo that should be 3/7.} occurs where $dm_c /dm = 0$ at $m =  m_c^{5/9}(4k)^{-1/3}$ and $m_c = 0.19 k^{-3/4}$.

This simple derivation at least qualitatively explains many  features of more detailed core accretion models.  The strong dependence of the atmospheric mass on $\mu$ is supported by studies showing that envelope pollution lowers the critical core mass \citep{HorIko11}.  The opacity is very sensitive to the amount  and sizes of dust grains  \citep{pollack1985}.  One popular way to speed up core accretion is to reduce the opacity \citep{HubBod05}.  The ``correct" choice of opacity likely varies between planets and is poorly constrained.  It is unclear how much dust from ablated planetesimals will remain in the atmosphere.  Small grains both contribute significantly to opacity and settle slowly.
 
High planetesimal accretion rates increase the critical core mass.  For very high accretion rates, especially in the inner disk, the protoplanet atmosphere will be fully convective \citep{raf2006}.  The atmosphere then matches onto the same adiabat (constant entropy curve) as the disk gas and thus has the lowest possible mass.   In this case the formation of gas giants is quite unlikely.
If planetesimal accretion stops (or becomes suitably small), the relevant luminosity comes from the Kelvin-Helmolz contraction of the atmosphere.  Models that omit planetesimal accretion (but correctly compute contraction) thus provide a meaningful lower limit on the critical core mass \citep{PapNel05}.


It has long been postulated that the cores of Solar System giants  might correspond to an isolation mass,
$\sim$ 5--20 \mearth, at 5--10 AU, even though this would require a massive planetesimal disk \citep{pollack1996}. To 
understand why this assumption is reasonable, we turn to Fig. \ref{fig:mass-atm}. Initially a low mass planet accretes 
planetesimals rapidly. With a short accretion time, the puffy atmosphere is well below the crossover mass. As the core grows in mass, two effects bring the atmosphere closer to instability, (i) extra compression of the gas and (ii) fewer and fewer planetesimals to accrete and heat the atmosphere.  As the atmosphere cools, the critical core mass drops until it reaches the actual core mass.  This cooling is most likely to happen after most planetesimals have been accreted, i.e.\ at the isolation mass.

\subsubsection{Direct Accretion of Disk Gas (and How it Stops)}
\label{sec:atmos-accr}

When the envelope collapses, the planet starts to accrete gas directly from the protostellar disk. 
This dynamical process is not amenable to a stellar structure calculation.
Direct accretion of gas is similar to accretion of planetesimals by oligarchs; the accretion rate is roughly the
product of the planet's cross-section (or impact parameter in the 2D limit), the local gas density (or surface density), and the encounter velocity, which is dominated by Keplerian shear for circular orbits.  

Shortly after the core accretion instability, the planet crosses the transition mass (eq.\ [\ref{eq:mtrans}]) and the relevant accretion radius is the Hill radius, $R_H$.   Since the transition mass also corresponds to the Hill radius exceeding the gas scale height (see discussion after eq.\ [\ref{eq:mtrans}]) accretion is not at the classic Bondi rate \citep[e.g.][]{shuv2}.   Instead the  two dimensional mass accretion rate applies,
\begin{equation} 
\dot{m}\gs \sim \varSigma\gs \varOmega R_H^2 \sim 10^5 {F \over m_\star^{1/6}} \left({m \over 60~\mearth}\right)^{2/3} \left(5~\AU \over a\right) ~{\mearth \over {\rm Myr}} ~ .
\end{equation} 
At $5 \times 10^{-7} \msunyr$, this rate exceeds the accretion rate onto most pre-main sequence T Tauri stars!

Clearly, something must stop this influx of gas.  The gaseous isolation mass is one natural stopping point.  Applying \Eq{eq:miso} to the gas disk gives
\begin{equation}
\label{eq:misog}
m_{\rm iso,g} = { (2 \pi B \varSigma\gs)^{3/2} \over (3 \mstar)^{1/2} } a^3 \approx 2 {F^{3/2} \over m_\star} \left ( { a \over 5~{\rm AU} } \right )^{3/4} ~ \mjup ~ .
\end{equation} 
While this result appears plausible for Jupiter, most disk models for Jupiter's core require $F \gtrsim 5$, requiring a second mechanism to halt gas accretion \citep{LisHub09}.

Opening a gap in the disk ---  which begins when \Eq{eq:mtrans} is satisfied --- can slow down accretion so that the disk dissipates before the planet reaches the gaseous isolation mass.   Since disk lifetimes are at least at least 1--3~Myr, accretion times must be at least this slow to halt growth (without invoking a fine-tuning of core accretion and disk dissipation timescales).
Only a wide and relatively clean gap can slow accretion enough to explain final planet masses \citep{DAnLub08}.  The effective viscosity of the disk must be low for a clean gap, which is an especially strong concern for self-gravitating disks \citep[\S\ref{sec:planets-bdgg}]{krat2010}.

\subsubsection{Numerical Simulations of Gas Giant Planet Formation}
\label{sec:atmos-GGsims}

\begin{figure}[p]
\begin{center}
\includegraphics[width=6in]{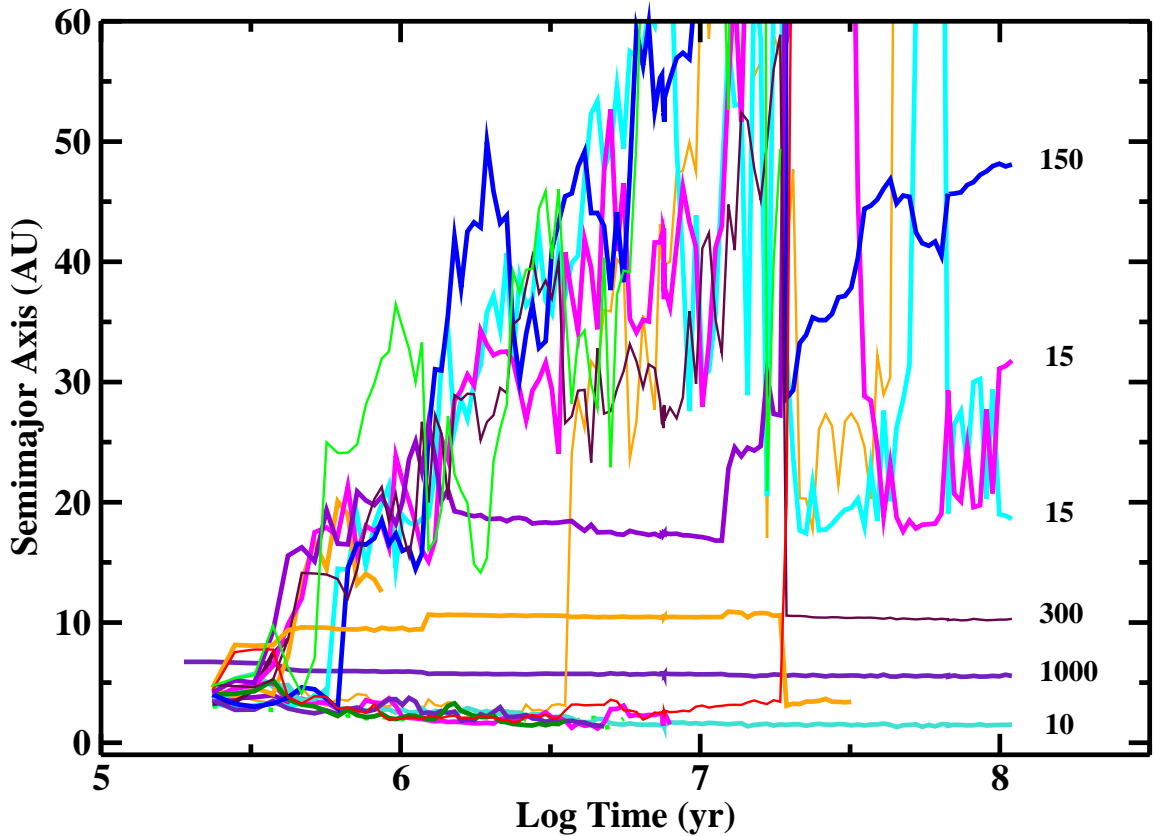}
\end{center}
\caption{Orbital evolution of icy oligarchs. The calculation starts with 1~cm and 1000 km 
planetesimals ($\rho_\bullet$ = 1.5 ~g~cm$^{-3}$) in a disk with initial
$\Sigma_{\rm s}$ = 14g~cm$^{-2}~(a / 3~{\rm AU})^{-1} e^{-a/{\rm 30~AU}}$.  
During the first $3 \times 10^5$~yr, $\sim$ 20 oligarchs with $m \approx$ 0.05 \mearth\ form 
in a relatively narrow range of semimajor axis, 3--7 AU.  As they grow, more massive oligarchs 
scatter lower mass oligarchs to large semimajor axes, $a \approx$ 1--20 AU.  At $\sim$ 1 Myr, 
some oligarchs begin to accrete gas. Over the next 10~Myr, continued growth and scattering 
leads to collisions and mergers of oligarchs. Eventually, only a few oligarchs remain. The
largest of these have masses comparable to the mass of Jupiter. 
The legend indicates masses (in $M_{\oplus}$) for the largest oligarchs in stable orbits
around the central star.
\label{fig:gasgiant_evol}}
\end{figure}

To conclude this section, Fig. \ref{fig:gasgiant_evol} illustrates the evolution of the semimajor axes of icy 
and gas giant planets from one simulation of material outside the terrestrial zone 
\citep[e.g.,][]{kb2011}. 
The calculation begins with a single 1000~km planetesimal and an ensemble of $\sim$ 1~cm planetesimals 
in each of 96 annuli from 3~AU to 30~AU. Because the system has large gravitational focusing factors
at $t = 0$, each large planetesimal rapidly sweeps up the small planetesimals along its orbit. With
growth times proportional to the orbital period (eq. [\ref{eq:t-slow}]), protoplanets at 3--7 AU grow 
much faster than those at larger $a$.  Growth produces many isolated mass objects packed closely 
together.  Early on, gravitational interactions among these objects jostle them around into overlapping
orbits. After $\sim$ 0.3 Myr, the most massive of these protoplanets begin to scatter lower mass 
protoplanets to smaller and larger $a$. Scattered protoplanets sweep up and scatter the large 
pre-existing planetesimals in these orbits, accelerating the growth of all large protoplanets. 
At $\sim$ 1 Myr, the largest protoplanets begin to accrete gas from the disk. As they grow, they
scatter lower mass protoplanets to larger and larger $a$; eventually, they eject some of these 
low mass protoplanets from the planetary system. 

At the end of the calculation at 100 Myr, six planets remain on stable orbits. Two gas giants,
with 1 and 3 Jupiter masses, have $a$ = 5 AU and 10 AU. Inside these gas giants, a
super-Earth with $m \approx 10~\mearth$ lies on a fairly circular orbit, $e \approx$ 0.02 at
$a \approx$ 1.5 AU. Outside the gas giants, two more super-Earths occupy orbits in a 2:1 
resonance. After many exchanges, the orbits of these two planets are likely stable. Finally,
a planet with roughly 1.5 times the mass of Saturn rests in an orbit with modest eccentricity,
$e \approx$ 0.1 at $a \approx$ 50~AU. The outcome of this simulation combines some properties 
of the Solar System -- four gas giants at 5--30~AU -- along with some properties of known 
exoplanets -- a super-Earth at 1--2 AU. 

This example illustrates several important differences between terrestrial and gas giant planet
formation. 

\begin{itemize}

\item Oligarchs form at 1 AU before they form at 5 AU. From eq. ([\ref{eq:t-slow}]), the growth time 
scales with the orbital period and the enhancement of the surface density at the snow line. Comparing
timescales at 0.4 AU and at 4 AU, we expect a factor of $\sim$ 300 from the orbital period and a factor 
of 1/2.4 from the snow line enhancement. The factor of $\sim$ 10 ratio of growth times at 0.4 AU and 
4 AU in the simulations agrees well with expectations.

\item Oligarchs reach chaotic growth faster at 5 AU than at 1 AU. From the discussion 
in \S\ref{sec:growth}, icy oligarchs at 5 AU have larger isolation masses than rocky oligarchs at 1 AU (eq. [\ref{eq:miso}]). 
Thus, their gravitational interactions are stronger and lead to chaotic growth sooner.

\item While rocky planets scatter low mass protoplanets a few tenths of an AU, gas giant planets scatter 
some low mass protoplanets close to the host star and eject many others from the planetary system.
Fortunately, the Solar System avoided either outcome.  However, many exoplanets close to their host 
stars have large $e$.  Although orbital migration probably accounts for exoplanets 
with small $a$ and nearly circular orbits (Chapter by Morbidelli), producing super-Earths with 
$a \lesssim$ 0.4 AU and $e \gtrsim$ 0.1 probably requires planet-planet scattering, as in Fig. 
\ref{fig:gasgiant_evol} \citep{juric2008,raymond2011}. In a few years, exoplanet statistics will
allow critical tests of the ability of migration and scattering to explain the observed ($a,e$) 
distribution.

\end{itemize}

\subsection{Direct Formation of Brown Dwarfs and Gas Giants}
\label{sec:planets-bdgg}

Though the core accretion is remarkably successful at explaining the diversity of planetary systems, it cannot be proved.  
It is thus useful to develop alternate theories.  Ideally each theory would make testable predictions, but consistency with physics and existing astronomical observations is also provides stringent constraints.  Here we consider the 
leading alternate theory, that a gravitationally unstable gas disk might fragment into bound objects that survive as
gas giant planets \citep{kuiper51, Cam78, BodGro80, boss2000}.  This theory should not be confused with the hypothesis that the solid component of the proplanetary disk gravitationally collapses into planetesimals (\citealp{saf1969, gold1973, youdin:2002}, \S\ref{sec:ptmlGI}).

Section \ref{sec:gravity} introduced the idea that disks are 
gravitationally unstable when the \citet{toom64} criterion,
\begin{equation} 
Q = {c_{\rm s} \varOmega \over \pi G \varSigma\gs} \lesssim 1~,
\end{equation} 
is satisfied for a surface mass density, $\varSigma$.  There, we outlined basic constraints on the accretion rate and temperature for stable 
disks.   For a fragment to survive it must cool quickly, so that it contracts  
on an orbital timescale.  Analytic theory suggests bound fragments are possible
but have the typical masses of brown dwarfs \citep{raf2005,krat2010}; so far, numerical 
simulations are inconclusive \citep{DAnDur10,CaiPic10}.

To outline the basic issues for gravitational instability, we consider a density perturbation in a
disk which satisfies the Toomre $Q$ criterion. For this fragment to continue to contract, it must 
cool on a sufficiently short timescale \citep{gamm2001},
\begin{equation}
t_c < \xi \Omega^{-1} ~ .
\label{eq:t-cool1}
\end{equation}
For likely conditions within a protoplanetary disk, numerical simulations suggest that the critical value of
$\xi \approx$ 3 \citep{gamm2001,rice2003}, with uncertainties due to the equation of state 
and the opacity.  Fragments that cannot cool on the orbital timescale will be sheared apart.
For the cooling time in a viscous disk with an $\alpha$-viscosity 
(eq. [\ref{eq:t-c1}]), this constraint places a limit on $\alpha$:
\begin{equation} 
\alpha \gtrsim 4/[9\xi \gamma (\gamma - 1)] \sim 0.3\, .
\end{equation} 
Fragmenting disks also have very high accretion rates. 

For the inner disk ($R <$ 10 AU), combining the Toomre ($Q \lesssim 1$) and cooling criteria yields
fragments only in hot, massive disks \citep{raf2005,MatLev05}.  With disk masses comparable to the stellar
mass, these conditions probably produce a bound stellar companion instead of a lower mass planet. 
Although colder, lower mass disks can have $Q \lesssim 1$, their slow cooling times do not allow the fragment to survive.

Forming planet-mass fragments in the outer disk, at 50 - 100 AU,  is more attractive. Irradiation (instead of viscous
transport) then dominates the energy budget of the disk \citep{DAless1998}.   Although estimates for the cooling time are more 
complicated, an irradiated disk is thought to fragment more readily once it reaches $Q \lesssim 1$ \citep{RicArm11, KraMur11}. 

In any part of the disk, the initial mass of a gravitational fragment is typically $\sim \varSigma\gs H\gs^2$ in terms of the disk scale-height $H\gs$.  
For the most optimistic assumptions about cooling -- specifically that the optical depth $\tau = 1$ -- 
fragment masses are $\gtrsim 5~\mjup$ at 100 AU \citep{raf2005}.  Less efficient cooling 
($\tau \neq 1$) and closer orbital separations increase the fragment mass.  These minimum masses require
very cold disks ($T \lesssim 10$ K), which might be achievable in the outer portions of 
disks in low mass star forming regions like Ophiuchus \citep{and2009}.

While making a fragment with a mass below $\sim 10 ~ \mjup$ is challenging, keeping the fragment 
mass low is an even more difficult problem.  Disks are likely to fragment before infall from the
surrounding molecular cloud ceases.  Preventing the disk fragment from accreting this infalling mass is a challenge.  
As explained in \S\ref{sec:atmos-accr}, stopping the flow of disk gas onto a planet requires a low mass 
and fairly quiescent disk, exactly opposite to the conditions required for an unstable disk.  
\citet{krat2010} quantifies these issues and concludes that disks in nearby star-forming regions are
more likely to produce 25--75 \mjup\ brown dwarfs than 1-10 \mjup\ planets.  The mass problem might be helped if fragments cool just rapidly enough to remain bound.  Then if they migrate inward they would overflow their Roche lobes and be ``tidally downsized" \citep{BolHay10, Nay10}.  This intriguing suggestion cannot yet be considered a solution.

Even if bound Jupiter-mass fragments form, understanding the time evolution 
of fragments within a $Q \sim 1$ disk requires sophisticated numerical calculations.  
In smooth particle hydrodynamics (SPH), an ensemble of particles represents the gas; each 
particle has a set of physical properties and responds to gravity, radiation pressure, and 
other forces \citep{benz1986,mona1992}.  Grid-based calculations lay out a set of 
points where the physical conditions of the gas are specified; solving a set of coupled 
differential equations yields the time evolution of the conditions at each point 
\citep{black1975,tohl1980,pick1998}.  

In both approaches, disks close to the stability limit develop multi-arm spiral structure
\citep{KraMat10,mayer2002}. Spiral modes generate turbulence throughout the disk 
\citep{nelson1998,gamm2001}. The large amplitudes of these modes create a rippled surface
above the disk midplane \citep{dur2001}. When the cooling rate is near the critical value $\xi$, 
the spiral structure maintains a rough equilibrium, where the amplitude of the spiral modes increases as 
the cooling rate declines \citep{mejia2005,cai2006}. 

When the cooling time is smaller than the local rotational period, as in equation 
(\ref{eq:t-cool1}), the disk fragments \citep{gamm2001,john2003,
rice2003}. For astrophysically relevant conditions, fragmentation requires the 
disk mass be at least 10\% of the stellar mass. As the disk mass grows, spiral waves 
propagate throughout the disk. Within a few rotation periods, fragments form in the densest 
portions of the spirals \citep{BolHay10}. If the fragments continue to cool rapidly, they grow in mass and 
become bound objects, otherwise they are sheared apart \citep{mayer2004}.

The long-term evolution of bound fragments in self-gravitating disks is unclear. Because 
the numerical calculations involve such a wide range of scales, none can evolve a bound 
fragment long enough to determine its final fate. If the cooling time is short, and accretion from the disk inefficient bound fragment could become gas
giant planets, or more likely brown dwarfs. Such objects might also grow to full-fledged stellar companions \citep{BonBat94}.  
Overcoming the numerical 
limits on the calculations requires faster computers and innovative techniques
to follow the evolution of fragments in a time-dependent disk. 

\section{SUMMARY}
\label{sec:summary}

In the last decade or two, observations have revolutionized our understanding of planetary 
astronomy. In the 1990's, the thrilling discovery of the Kuiper belt nearly doubled the 
empirical size of the disk of the Solar System \citep{jewitt:1993}.  A few years later, 
\citet{mayor1995} discovered an extraordinary exoplanet orbiting only 0.05 AU from the 
solar-type star 51~Peg.  Now, the number of known Kuiper belt objects easily exceeds
1000, including a grand variety of dynamical and taxonomic classes that place interesting
constraints on the origin and early evolution of the protosolar nebula \citep{barucci2008}.
Although there are only $\sim$ 1000 confirmed exoplanets, data from {\it Kepler} and many 
ground-based programs will certainly push the count past 10,000 in the next decade. 
Some of these will certainly challenge current ideas about planet formation.

Throughout this onslaught, theorists responded quickly with new ideas (dead zones in disks,
migration, symplectic $N$-body codes) and variants on old ideas (collisional cascades, disk 
instabilities, multiannulus coagulation codes). Rapid developments in computing hardware 
fueled many advances; new analytical approaches drove others. 

Today, we have a good basic theory of disk evolution. Despite uncertainties about the 
initial mass and temperature distribution and the origin of disk viscosity, analytic
and numerical disk models provide a good framework for interpreting observations and
for exploring the origin of planetary systems. Current research involves combining more 
elaborate versions of the basic theory (\S\ref{sec:disk}) with detailed models for the 
chemical evolution of disk material. Within the next decade, these investigations should
improve our insight into the overall evolution of the disk and the growth and composition 
of dust grains with sizes $\sim$ 1--10~mm.

We also understand how km-sized or larger planetesimals become planets (\S\ref{sec:planets}). 
Although there are major uncertainties about the onset and the end of gas accretion and the 
interactions of massive planets with the gas disk, analytic theory and numerical simulations 
demonstrate that -- on 1--10 Myr timescales -- ensembles of planetesimals can evolve into 
terrestrial and gas giant planets within $\sim$ 50 AU of the central star. Comparisons of
observations with the predicted masses and orbital properties of planets and the predicted 
dust masses and luminosity evolution of debris disks are promising and will eventually
produce stringent tests of the theory.

Despite these successes, we are still in search of a robust theory for planetesimal formation 
(\S\ref{sec:planetesimal}).  Excellent progress on the meter-size barrier isolates the 
importance of radial drift and the problems associated with direct coagulation models. Some
type of instability -- either by direct gravitational collapse or a concentration mechanism 
such as the streaming instability -- seems necessary to produce planetesimals on fast 
timescales. Larger numerical simulations will undoubtedly yield a better understanding of 
these instabilities. Exploring other physical mechanisms for particle growth and evolution 
is also essential.

Understanding fragmentation in protostellar disks promises to unify our understanding of star 
and planet formation.  Although many physical and numerical issues remain unresolved,  
fragmentation is a promising way to produce gas giants and brown dwarfs at $\gtrsim$ 50--100~AU
from the parent star. Despite the current lack of large samples of planets in this domain, 
direct imaging surveys are starting to discover 1--30~\mjup\ objects with $a \sim$ 10--100 AU
\citep{kalas2008,marois2008,lagrange2010,currie2010,KraIre11}. With large uncertainties in
model atmospheres, assessing the formation mechanism of these objects is difficult.  Once
large samples of planets and brown dwarfs at 10--100 AU are available, their properties will
allow a robust assessment of the core accretion and disk instability mechanisms.

\acknowledgements
We thank Ben Bromley, Margaret Geller, Paul Kalas, and Kaitlin Kratter for advice and comments 
on the manuscript.  Portions of this project were supported by 
{\it NASA's } {\it Astrophysics Theory Program} and the {\it Origin of Solar Systems Program} 
through grant NNX10AF35G and by Endowment Funds of the Smithsonian Institution.

\twocolumn

\bibliography{master}

\onecolumn
\clearpage

\clearpage

\clearpage

\clearpage

\clearpage

\clearpage

\clearpage

\clearpage 
\onecolumn
   \begin{longtable}{@{} lcp{10cm} @{}} 
    \caption[]{Frequently Used Symbols} \\
      \toprule
      Symbol    & Ref.\ & Meaning \\
      \midrule
      \endhead
      \multicolumn{2}{c}{General Physical Quantities}\\
      \cmidrule(l){1-2} 
      $\AU$ & \Eq{eq_sigmag}  & astronomical unit\\ 
      $c$ & \Eq{eq:tpr} & speed of light\\
      $c_s$ & \Eq{eq:disk-gasvel} & sound speed\\
      $C_D$ & \Eq{eq:turbdragforce} & drag coefficient\\
      $C_v$ &\S\ref{sec:disk-transport} & specific heat\\
      $P$ & \Eq{eq:disk-radmom} & gas pressure\\
      $t$ & & time or timescale, often subscripted\\
      $T~[T_{\rm eq}$] & \Eq{eq:t-eqbb} &[equilibrium] temperature\\
      $v_K$ & \Eq{eq:disk-gasvel} & Keplerian circular velocity\\
      $\gamma$ & \Eq{eq:disk-gasvel} & adiabatic index \\
      $\kappa$ & \Eq{eq:kappa} & radiative opacity\\
      $\lambda$ & \Eq{eq:rad-eff}, \S\ref{sec:gravity} & wavelength (light or other waves) \\
      $\lambda$ & \S\ref{sec:disk-transport}, \Eq{eq:ts} & gas mean free path \\
      $\mu$ & \S\ref{sec:disk-transport} & mean molecular weight\\
      $\sigma$ & \S\ref{sec:disk-transport} & cross section\\
      $\sigsb$ & \Eq{eq:vtemp1} & Stephan-Boltzmann constant\\
      $\rho$ & & density (mass per volume) of quantity, often subscripted\\
      $\rho_\bullet$ &\Eq{eq:tpr} & Internal density of solid grain or planetesimal \\
      $\varOmega$ & \S\ref{sec:disk-evol} & Keplerian orbital frequency\\
	\midrule
       \multicolumn{2}{c}{Stellar Quantities}\\
      \cmidrule(l){1-2}  
      \lstar ~[\lsun] & \Eq{eq:t-eqbb} & stellar [solar] luminosity\\
      $l_\star$ & \Eq{eq:mtrans} & \lstar/\lsun, dimensionless stellar luminosity\\
       \mstar ~[\msun] & \Eq{eq:fPMC} & stellar [solar] mass\\
       $m_\star$ & & \mstar/\msun, dimensionless stellar mass\\
       $Z_{\rm Fe}$ & \Eq{eq:fPMC} & stellar metallicity\\
      \midrule
       \multicolumn{2}{c}{Disk Quantities}\\
             \cmidrule(l){1-2}
        $D$ & \Eq{eq:disk-diss} & Dissipation (i.e.\ heating) rate, per area \\
       $F$ & \Eq{eq_sigmag} & Disk mass relative to MMSN \\
       $H$ & \Eq{eq:t-dyn} & disk scaleheight \\
       $J~[\dot{J}]$ & \Eq{eq:Mdisk-J} & angular momentum [torque]\\
       $L_{\rm acc}~[L_{\rm d}]$ & \Eq{eq:Lacc} ~[\Eq{eq:Ld}]& Accretion luminosity [part released in the disk]\\
       $\dot{M}$ & \Eq{eq:mdot} & accretion rate through disk\\
       $Q$ & \Eq{eq:Q} & dimensionless measure of gravitational stability \\
       $R$ &  \Eq{eq_sigmag} & distance from star (cylindrical radius) \\
       $R_{\rm in}$ &  \Eq{eq:Ld} & radius of disk inner edge\\
       $R_J$ & \Eq{eq:disk-nusigma}  & torque on inner disk edge, as length-scale\\
       $v_\phi~[v_R]$ & \Eq{eq:mdot} & orbital [radial] flow velocity\\
       $Z_{\rm disk}$ & \Eq{eq_sigmap} & Disk ``metallicity" as ratio of solids to gas \\
       $Z_{\rm rel}$ & \Eq{eq:Zrel}& ``Metallicity" relative to fiducial Solar value of 0.015\\
       $\alpha$ & \Eq{eq:t-c1} & dimensionless angular momentum transport coefficient\\ 
       $\alpha_D$ & \Eq{eq:halpha} & dimensionless turbulent diffiusion coefficient\\ 
       $\eta$ & \Eq{eq:etavK} & fraction by which rotation is slower than Keplerian\\
       $\theta$ & \Eq{eq:irtemp-1} & grazing angle of starlight on disk surface\\
       $\nu$ & \Eq{eq:disk-diffusion} & viscosity, usually ``anomalous" \\
       $\rho_{\rm R}$ & \Eq{eq:Roche} & Roche density for gravitational binding\\
        $\varSigma$ &  \Eq{eq_sigmag}, \Eq{eq:mdot} & Surface density (mass per area) of disk, e.g.\ gas (\S\ref{sec:disk}) or planetesimals  (\S\ref{sec:solidprotoplanets}).  Subscripted as needed.\\
        $\varSigma_g$~[$\varSigma_p$] & \Eq{eq_sigmag}~[\Eq{eq_sigmap}] & Surface density of gas [particle] disks.\\
        $\dot{\varSigma}$ & \Eq{eq:disk-diffusion} & Inflow or outflow of mass from disk, per area\\
      \midrule
       \multicolumn{2}{c}{Protoplanet Quantities}\\
        \cmidrule(l){1-2}
        $a$ & \Eq{eq:rhill} & semimajor axis (similar to disk $R$)\\
        $B$ & \Eq{eq:miso} & width of feeding zone in $R_H$\\
        $f_G$ & \Eq{eq:fg-disp} \& (\ref{eq:fg-shear})  & gravitational focusing factor for collisional cross section\\
       $m$~[$m_c$~,~$m_a$] & \S\ref{sec:solidprotoplanets} ~[\S\ref{sec:atmos-struc}] & protoplanet mass, total or [core, atmosphere]\\
       $m_s$~[$m_l$] & \S\ref{sec:solidprotoplanets} & mass of small protoplanets, i.e.\ planetesimals [larger protoplanets] \\
       $m_{\rm iso}$ & \Eq{eq:miso} & isolation mass\\
       $\dot{m}$ ~[$\dot{m}_l$] & \S\ref{sec:growth} & accretion rate, for [large] protoplanet's mass growth\\
       $Q_D^*$ & \Eq{eq:qdis} & energy (per mass) threshold for catastrophic collisional disruption \\
       $r$ ~[$r_s$~,~$r_l$~,~$r_c$] & \S\ref{sec:solidprotoplanets}~,~\S\ref{sec:atmos-struc} & radius of protoplanet [small, large, solid core]\\
       $r_{\rm B}$ & \Eq{eq:bondi-rad} & Bondi radius for planet gravity to exceed thermal energy of disk gas\\
       $R_H$ & \Eq{eq:rhill} & Hill radius for planet gravity to dominate stellar tides\\  
       $s$ & \Eq{eq:ts} & radius of small grain or planetesimal\\
       $\ts ~[\taus]$ & \Eq{eq:ts}  [\Eq{eq:taus}]  & [dimensionless] aerodynamic stopping timescale\\
       $v$ ~[$v_s$~,~$v_l$] & \S\ref{sec:solidprotoplanets} & velocity dispersion of [small, large] protoplanets\\
       $v_{esc} ~[v_H]$ & \Eq{eq:escvel-ratio} [\Eq{eq:vhill}] & escape speed from protoplanet's surface\\
       $\varSigma_s$~[$\varSigma_l$] & \S\ref{sec:solidprotoplanets} & surface mass density of small [large] protoplanets\\
       $\chi$ & \Eq{eq:mass-atm1} & log ratio of atmosphere to core radius\\
       $\psi$ & \Eq{eq:psi-star} & $r/R_H$, radius in Hill units\\
           \bottomrule
   \end{longtable}

\end{document}